\documentclass[aps,prd,onecolumn,groupedaddress,showpacs,nofootinbib,amssymb]{revtex4-2}
\usepackage[dvips]{graphicx}
\usepackage{amssymb}
\usepackage{amsmath}
\usepackage{graphicx,,color}
\usepackage{amsfonts}
\usepackage{bm}
\usepackage{cancel}
\usepackage{comment}
\usepackage{floatflt}
\usepackage{slashed}

\newcommand\be{\begin{equation}}
\newcommand\ee{\end{equation}}

\newcommand\e{\mathrm{e}}

\allowdisplaybreaks[4]

\begin{document}

\title{Recent Advances on Inflation}
\author{S.D.Odintsov$^{1,2}$}\,
\email{odintsov@ice.csic.es}
\author{V.K. Oikonomou$^{3}$}\,
\email{v.k.oikonomou1979@gmail.com;voikonomou@gapps.auth.gr}
\author{I. Giannakoudi$^{3,4,5}$}\,
\email{ifigeneiagiannakoudi@gmail.com}
\author{F.P. Fronimos$^{3}$}\,
\email{fotisfronimos@gmail.com}
\author{E.C. Lymperiadou$^{3,6}$}\,
\email{eilymper@auth.gr}

\affiliation{$^{1)}$ ICREA, Passeig Luis Companys, 23, 08010 Barcelona, Spain\\
$^{2)}$ Institute of Space Sciences (ICE,CSIC) C. Can Magrans s/n,
08193 Barcelona, Spain\\
$^{3)}$ Department of Physics, Aristotle University of
Thessaloniki, Thessaloniki 54124, Greece\\
$^{4)}$ University of Waterloo, Canada\\
$^{5)}$ Perimeter Institute, Canada\\
$^{6)}$ Department of Physics and Astronomy, University of Bonn,
Germany}

\tolerance=5000

\begin{abstract}
We review recent trends on inflationary dynamics in the context of
viable modified gravity theories. After providing a general
overview of the inflationary paradigm emphasizing on what problems
of hot Big Bang theory inflation solves, and a somewhat
introductory presentation of single field inflationary theories
with minimal and non-minimal couplings, we review how inflation
can be realized in terms of several string motivated models of
inflation, which involve Gauss-Bonnet couplings of the scalar
field, higher order derivatives of the scalar field, and some
subclasses of viable Horndeski theories. We also present and
analyze inflation in the context of Chern-Simons theories of
gravity, including various subcases and generalizations of string
corrected modified gravities which also contain Chern-Simons
correction terms, with the scalar field being identified with the
invisible axion, which is the most viable to date dark matter
candidate. We also provide a detailed account of vacuum $f(R)$
gravity inflation, and also inflation in $f(R,\phi)$ and
kinetic-corrected $f(R,\phi)$ theories of gravity. In the end of
the review we discuss the technique for calculating the overall
effect of modified gravity on the waveform of the standard general
relativistic gravitational wave form.
\end{abstract}


\maketitle

\section{Introduction}

Physicists of the current era are lucky to live in the era of
precision cosmology, in which a plethora of observational data are
available. Without exaggeration, nearly every 3 years a great
discovery takes place. It started in 2012 with the Higgs discovery
at the LHC \cite{CMS:2012qbp}, followed by the first direct
observation of gravitational waves in 2015
\cite{LIGOScientific:2016aoc}, followed by the one in a million
kilonova event in 2017, nowadays known as GW170817
\cite{LIGOScientific:2017vwq}. After that, in 2020 the NANOGrav
collaboration reported on the discovery of something that could be
either a pulsar red noise, or a gravitational wave
\cite{NANOGrav:2020bcs}. The latter was verified in 2023 in which
year the NANOGrav reported the first detection of a stochastic
gravitational wave background, verified by Hellings-Downs
correlations, so the signal is clearly a gravitational wave of
cosmological or astrophysical origin \cite{NANOGrav:2023gor}. The
chorus of observations will be further augmented by future
experiments, like the stage 4 Cosmic Microwave Background (CMB)
experiments \cite{CMB-S4:2016ple,SimonsObservatory:2019qwx},
expected to commence operations in 2027, and the future
gravitational wave experiments
\cite{Hild:2010id,Baker:2019nia,Smith:2019wny,Crowder:2005nr,Smith:2016jqs,Seto:2001qf,Kawamura:2020pcg,Bull:2018lat,LISACosmologyWorkingGroup:2022jok}
like LISA and the Einstein Telescope which will commence their
operation in 2035. All these new experiments will shed light to
fundamental problems in cosmology and astrophysics. The most
important of all, they will probe the primordial tensor modes in
our Universe and give a definitive answer on the question whether
inflation ever occurred. The inflationary scenario
\cite{Guth:1980zm,Linde:1983gd,Starobinsky:1982ee} is one of the
most viable theoretical proposals for the early time era of our
Universe. This is because inflation as a proposal solves all the
basic problems of the standard hot Big Bang model and in addition
it serves as a mechanism for generating matter structures at large
scales in our Universe. The primordial quantum fluctuations
generated during the inflationary era act as attractors on which
baryons and cold dark matter are accumulated, so the large scale
structure at large redshift up to $z=6$ may be explained by a
nearly scale invariant power spectrum of primordial scalar
perturbations. Thus from a theoretical point of view, the
inflationary scenario is the most appealing and vital for the
viable description of the primordial evolution of the Universe,
and for explaining the large scale structure of the Universe. In
principle many theoretical frameworks may generate an accelerating
era primordially, for a main stream of articles and reviews see
\cite{Guth:1980zm,Linde:1983gd,Starobinsky:1982ee,Linde:2007fr,Gorbunov:2011zzc,Lyth:1998xn,Linde:1985ub,Albrecht:1982wi,Sasaki:1995aw,Turok:2002yq,Linde:2005dd,
Kachru:2003sx,Brandenberger:2016uzh,Bamba:2015uma,Martin:2013tda,Martin:2013nzq,Baumann:2014nda,Baumann:2009ds,Linde:2014nna,Pajer:2013fsa,Yamaguchi:2011kg,
Byrnes:2010em,Kallosh:2013hoa,Kobayashi:2011nu,Bezrukov:2010jz,McAllister:2008hb,Cognola:2007zu,Cheung:2007st,Chen:2006nt,Nojiri:2005pu,Bartolo:2004if,Peebles:1998qn,
Kofman:1997yn,Berera:1995ie,Lidsey:1995np,Liddle:1994dx,Kofman:1994rk,Sasaki:1995aw,Lucchin:1984yf,Vagnozzi:2023lwo,Forconi:2021que,Benetti:2021uea,Casalino:2018wnc,Cicoli:2020bao,Visinelli:2018utg,Dutta:2017fjw,DiValentino:2016ucb,Gerbino:2016sgw,Vagnozzi:2022qmc,
Giare:2023wzl,Vagnozzi:2020gtf,Giare:2019snj,Kinney:2018nny,DiValentino:2018wum,Myrzakulov:2014hca,Luciano:2022viz,Renzi:2018dbq,DiValentino:2016nni,DiValentino:2016ziq,
DiValentino:2016ikp,DiValentino:2011zz,
vandeVis:2020qcp,Faraoni:1996rf,Bloomfield:2019rbs,Faraoni:2000wk,Nguyen:2019kbm,Faraoni:2000nt,Kaiser:2015usz,Faraoni:2006ik,Amin:2014eta,Elizalde:2008yf,Lahiri:2023oth,
Schutz:2013fua,Kaiser:2012ak,Greenwood:2012aj,Faraoni:2013igs,Kaiser:1995fb,Kaiser:1993bq,Nojiri:2020wmh}.
Traditionally, the inflationary scenario was firstly considered in
terms of some false vacuum decay of a scalar field
\cite{Guth:1980zm}, but it was then realized that a slow-rolling
canonical scalar field may appropriately describe the inflationary
era \cite{Linde:1983gd}. Usually inflation is formalized by the
use of a canonical minimally coupled scalar field or by a
non-minimally coupled scalar field, however there are many
shortcomings of the scalar field description of inflation that
make an alternative description rather compelling. The most
important shortcoming is that the single scalar field description
relies on a scalar field, which must couple to all the Standard
Model particles in order to reheat the Universe. Thus there are
too many unknown arbitrary couplings that must be explained and
also the inflaton itself must be identified or experimentally
verified. The only fundamental scalar field that has ever been
observed is the Higgs particle, so inflationary scenarios that use
the Higgs particle as the inflaton are well motivated
\cite{Bezrukov:2008ej}. There exist however models that assume
that the axion is the inflaton, with the axion being also a dark
matter candidate. However, this could be deemed problematic, since
if the axion is the inflaton, it must couple to Standard Model
particles in order to reheat the Universe. This is a rather
unwanted situation, since in most contexts the axion is a
non-thermal relic. Regardless the theoretical shortcomings, the
scalar field inflationary models are the most common and
frequently used descriptions of inflation. An alternative to
scalar field inflationary models comes from modifications of
Einstein-Hilbert gravity which contain higher order curvature
corrections. There are many modified gravity models and for some
important reviews in the field, see Refs.
\cite{reviews1,reviews2,reviews3,reviews4,reviews5}. The most
important and more common models of modified gravity use $f(R)$
gravity, with $R$ being the Ricci scalar, see Refs.
\cite{reviews1,reviews2,reviews3,reviews4,reviews5}. These models
are simple and mainstream models and can also have an Einstein
frame counterpart theory which is a minimally coupled scalar
field. One important feature of this $f(R)$ gravity framework is
that it is possible to achieve a unified description of inflation
with the dark energy era, see the pioneer work on this
\cite{Nojiri:2003ft} and several works thereafter
\cite{Nojiri:2007as,Nojiri:2007cq,Cognola:2007zu,Nojiri:2006gh,Appleby:2007vb,Elizalde:2010ts,Odintsov:2020nwm,Sa:2020fvn,Oikonomou:2020oex,Oikonomou:2020qah}.
Apart from $f(R)$ gravity inflation, modified Gauss-Bonnet
theories of gravity are also well studied in the literature of the
form $f(G)$
\cite{Cognola:2006sp,Nojiri:2005vv,Kanti:2015pda,Lidsey:2003sj,Oikonomou:2015qha}
or $f(R,G)$ \cite{Bamba:2010wfw} where $G$ is the Gauss-Bonnet
scalar. The $f(R,G)$ theories are plagued with ghost instabilities
and degrees of freedom, thus they are less frequently used and are
less appealing. In addition, quite popular theories are the
Einstein-Gauss-Bonnet theories of gravity
\cite{Hwang:2005hb,Nojiri:2006je,Nojiri:2005vv,Nojiri:2005jg,Satoh:2007gn,Bamba:2014zoa,Yi:2018gse,Guo:2009uk,Guo:2010jr,Jiang:2013gza,Kanti:2015pda,vandeBruck:2017voa,Koh:2014bka,Bayarsaikhan:2020jww,Kanti:1998jd,Pozdeeva:2020apf,Vernov:2021hxo,Pozdeeva:2021iwc,Fomin:2020hfh,DeLaurentis:2015fea,Chervon:2019sey,Nozari:2017rta,Odintsov:2018zhw,Kawai:1998ab,Yi:2018dhl,vandeBruck:2016xvt,Kleihaus:2019rbg,Bakopoulos:2019tvc,Maeda:2011zn,Bakopoulos:2020dfg,Ai:2020peo,Oikonomou:2020oil,Odintsov:2020xji,Oikonomou:2020sij,Odintsov:2020zkl,Odintsov:2020mkz,Venikoudis:2021irr,Easther:1996yd,Antoniadis:1993jc,Antoniadis:1990uu,Kanti:1995vq,Kanti:1997br,Odintsov:2020sqy,Oikonomou:2021kql,Kong:2021qiu}.,
but these theories are plagued with a non-trivial gravitational
wave speed which is distinct from the speed of light in vacuum.
These theories were severely constrained after the GW170817 event
since the latter indicated that the electromagnetic signal arrived
almost simultaneously with the gravitational waves. However a
theoretical solution for this problem was given recently in Refs.
\cite{Odintsov:2020sqy,Oikonomou:2021kql} in which case the theory
resulted to a constraint between the scalar field potential and
the Gauss-Bonnet coupling function. In addition, models of
$f(R,\phi)$ gravity are also used to produce inflation and also
Chern-Simons models of inflation are also used. In addition,
several $f(R,T)$ models can also describe the inflationary era,
where $T$ is the trace of the energy momentum tensor, see the
reviews \cite{reviews1,reviews2}. Also two scalar field models are
used for inflation see \cite{Kaiser:2015usz}.

The recent NANOGrav observation of the stochastic signal of
gravitational waves if interpreted as a cosmological signal,
indicates that the inflationary era must have had a blue-tilt in
its tensor spectral index, a fact that severely constrains the
inflationary models that can reproduce such a signal. This can be
achieved by Einstein-Gauss-Bonnet models, at the cost of having a
low-reheating temperature, see for example
\cite{Oikonomou:2023qfz,Vagnozzi:2023lwo}, while single scalar
field models cannot explain the 2023 NANOGrav signal. Thus the
current epoch puts severe constraints on theoretical frameworks,
which must explain the observations that come out at an incredible
speedy way nowadays. Thus to stand to the challenge, the modern
theoretical physicist must master the techniques of inflationary
cosmology and know how to judge whether a theoretical model is
viable or not. In this review we aim to provide a timely text
which contains all the recent trends and techniques on
inflationary dynamics, but also discussing the standard problems
of Big Bang cosmology and why inflation itself describes
successfully in a theoretical way the primordial era of our
Universe. We shall analyze inflationary dynamics in a quantitative
way for single scalar field inflation, both minimally and
non-minimally coupled, and for most mainstream modified gravity
theories, including those for which the tensor spectral index can
be blue tilted, a necessary ingredient in order for the models to
be compatible with the NANOGrav stochastic gravitational wave
observation, if the cosmological description is responsible for
the signal of course. Our analysis will be limited to providing
the necessary tools in order for someone to be able to produce
viable inflationary cosmologies, compatible with the most recent
(2018) Planck constraints on inflation \cite{Planck:2018jri}. The
scalar and the tensor power spectrum can be written in terms of a
certain set of dimensionless parameters which are known as
``slow-roll'' parameters. When these are smaller than unity, a
perturbation expansion can be performed on the scalar and tensor
power spectrum and the observable quantities which quantify the
inflationary era can be expressed in terms of the slow-roll
parameters. Regarding the observable quantities of inflation, we
shall focus on the most important ones, which are the spectral
index of the scalar primordial curvature perturbations, the tensor
spectral index of the primordial tensor perturbations and the
tensor-to-scalar ratio. Depending on the theoretical framework,
the number and the complexity of the slow-roll parameters varies,
so we shall emphasize on the calculation of the slow-roll
parameters for various mainstream theoretical frameworks and we
shall express the observational quantities of inflation in terms
of the slow-roll indices needed for each theory. We shall also
provide relations in closed form for all the observational
quantities and the corresponding slow-roll indices so that the
reader is able to reproduce the results quoted in each case. In
the end of the review we provide a concrete self-contained section
on the evolution of tensor perturbations in the context of
modified gravity and we quantify the effect of modified gravity in
terms of a single parameter. Then we analyze how the general
relativistic waveform of the tensor perturbation may acquire a
non-trivial multiplicative factor which contains the overall
modified gravity effect from the present day back to the redshift
corresponding to the mode that reentered the Hubble horizon back
in our Universe's past.

We need to note that the inflationary scenario is a theoretical
necessity in order to alleviate the problems of standard hot Big
Bang cosmology. Even if we have no direct sign that inflation ever
occurred, the observations ``cry for inflation'' since it is the
only consistent scenario that can be compatible with the
observational necessity of having a nearly scale invariant power
spectrum of primordial perturbations, and it is  the only
consistent answer to the question how large scale matter structure
was generated in the first place. Inflation and dark matter are
theoretical predictions that still wait to be revealed
observationally and experimentally. We may still be far away from
discovering those theoretical predictions and even if we did not
find them yet, it is almost certain in the minds of theorists that
both play an important role in the evolution of the Universe. The
situation here is the same as in the discovery of the Higgs, every
theorist believed that the elusive spinless particle gives mass to
the Standard Model particles, but it was never observed until
2012. We all knew it was there and only the proof of its existence
remained. And we knew because the Higgs particle and the
electroweak symmetry breaking was the only theoretical mechanism
that could yield a mass to the Standard Model particles. The same
applies with the inflationary paradigm. It must be the underlying
mechanism responsible for the CMB anisotropies and the reason that
large scale matter structure exists, and we need to find a proof
for its existence. The road might be long and thorny till we
unveil inflation, or the proof might be right at the corner of our
``local time frame of inertia''. Nobody knows and that is the
magic of nature.

This review is organized as follows: In section II we provide an
overview of the inflationary paradigm. We point out the
shortcomings of the standard hot Big Bang scenario, and we explain
how the inflationary paradigm theoretically solved these problems.
We emphasize how important is the inflationary paradigm
theoretically and why it eventually should be the correct
description of nature, since it is the only scenario which
provides a nearly scale invariant power spectrum of primordial
scalar curvature fluctuations, which are necessary in order to
explain large scale structure in our Universe as a whole. In
section III we analyze in detail how the inflationary era may be
generated by a single scalar field theory with minimal and
non-minimal couplings. We calculate the necessary slow-roll
indices, we provide and prove in detail several well-known
formulas regarding single scalar field inflation. In section IV we
provide a brief account of the swampland criteria, while in
section V we discuss in brief the constant-roll evolution as an
alternative to the standard slow-roll evolution. In section VI, we
study and analyze several string motivated models of inflation,
which also involve Gauss-Bonnet couplings of the scalar field,
higher order derivatives of the scalar field, and some subclasses
of viable Horndeski theories. In section VII, we present and
analyze inflation in the context of Chern-Simons theories of
gravity, presenting various subcases and generalizations of string
corrected modified gravities which also contain Chern-Simons
correction terms, with the scalar field being identified with the
invisible axion, which is the most viable to date dark matter
candidate. Section VIII is devoted to vacuum $f(R)$ gravity
inflation in its most general form, while section IX is devoted to
generalized $f(R,\phi)$ theories of gravity, in the form of scalar
field assisted $f(R)$ gravity inflation. Section X focuses on
k-essence $f(R)$ gravity theories, while section XI focuses on
kinetic-corrected $f(R,\phi)$ theories of gravity. Finally in
section XII we provide a concrete overview of the evolution
equations for the tensor perturbations in the context of modified
gravity and we review how to calculate the overall effect of
modified gravity on the general relativistic gravitational wave
waveform. Finally the conclusions follow at the end of the review.

\section{Brief Overview of Inflation}

In the early period of the 1970's, physicists started to realize
some problematic elements of the conventional Big Bang model and
thus the first aspects of inflationary cosmology started to
formulate. Different characteristics of the mechanics for
inflation were discovered and the first somewhat realistic model,
in which the early universe went through an inflationary de Sitter
era, was proposed by Starobinsky \cite{Starobinsky:1982ee}. Also
an equally important point in the historical development of
inflation was the model proposed by Guth \cite{Guth:1980zm}, in
which the inflationary era was a period when the Universe
exponentially expanded in a super-cooled false vacuum state (a
metastable state with no particles or fields, but large vacuum
energy density). However due to the possible outcomes of this
model, it was recognized that it is  not realistic and viable even
with improvements in order to explain several shortcomings of Big
Bang cosmology. The solution was given by Linde
\cite{Linde:1983gd} , who proposed the ``new inflationary theory''
\cite{Linde:1983gd}, in which inflation can start either in a
false vacuum or at the top of an effective potential in an
unstable state  and then the inflationary field slowly rolls down
to the minimum of that effective potential. Various models for
inflation have been constructed ever since. The main fundamental
idea for all of them is that:

\textit{-Inflation is an era of an abrupt accelerated expansion
that took place in a very early period after the beginning of the
Universe.-}

Some are following a canonical approach using scalar fields  and
others are using a description based on modified gravity for
inflation, with the most interesting and promising ones being the
$f(R)$-gravity models. Descriptions for a lot of these cases are
going to be presented in this text, however we start by presenting
the key problematic elements of the Big Bang model that were the
reason for introducing inflation as the optimal theoretical
description of our Universe's primordial era.

\noindent

\subsection{The Shortcomings of the Hot Big Bang Cosmology}

The Standard Big Bang (SBB) cosmology paradigm was a very
successful model, since there are various successful observations
about the properties of cosmic objects based on the SBB theory.
Also, along with the study of the cosmic microwave background
(CMB), it guided us to our first understanding of various
cosmological phenomena and subsequently how the Universe evolved
at the very early epochs of its existence, and how it evolved to
its subsequent large-scale form. However, even though the Standard
SBB scenario found success and was highly embraced, one cannot
ignore that some of its fundamental features can be problematic.

In the SBB, the early Universe is adiabatically expanded and
radiation-dominated. The model depends on the assumption of
homogeneity and isotropy on large scales (cosmological principle),
which lead to the ability to use the
Friedmann-Robertson-Walker(FRW) metric for the space-time of the
Universe, given by the following form,
\begin{equation}
    \label{FRW}
    ds^2= -c^2dt^2 + a^2(t)\left(\dfrac{dr^2}{1-Kr^2} + r^2 (d\theta^2 + \sin^2\theta d\phi^2)  \right) \,\, ,
\end{equation}
where $a(t)$ is the scale factor that is associated with the
spatial part of the metric and its evolution in time. Depending on
the value of constant $K$, we have three different case. For $K=0$
the space described by this metric is flat (flat Universe) , for
$K=+1$ is a closed Universe that could be described with a sphere
and for $K=-1$ the Universe is open, with the spatial part of the
spacetime being some hyperbolic hypersurface, like a saddle.

\begin{figure}[h!]
\label{lightcone}
    \centering
    \includegraphics[scale=0.65,]{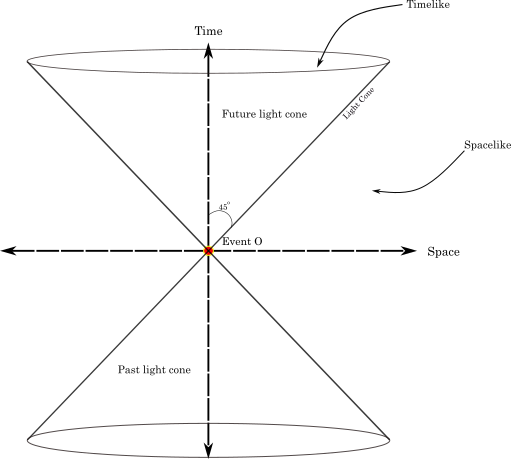}
    \caption{Light propagates along null geodesics $ds^2=0$ (world lines of
zero proper time). The straight lines (null geodesics) create a
light cone when they pass from a specific event O. Inside of the
light cone there are the time-like geodesics ($ds^2>0$) describing
the massive particles that travel along the world lines. The
interior is the causally connected region to the event O. Outside
is the spacelike regions which are causally disconnected and are
described by spacelike geodesics ($ds^2<0$).  }
    \label{fig.lightcone}
\end{figure}

In Eq. \eqref{FRW} the coordinates used are called the comoving
coordinates, meaning that as the Universe and thus space itself
expands, the coordinates $r, \theta$ and $\phi$ are not affected
by this expansion being expanded in the same way. To obtain a form
of the metric with the physical distance, the scalar factor is
multiplied by $r$, $R=a(t) r$. Therefore the metric is now written
in the following time-dependent form \cite{Baumann:2009ds}:
\begin{equation}
    \label{FRWt}
    ds^2=- c^2dt^2+a^2(t)\left[d\chi^2 + \Theta_k(\chi^2)(d\theta^2+\sin^2\theta d\phi^2) \right] \,\, ,
\end{equation}
where,
\begin{equation}
    \label{rTheta}
    r^2=\Theta_k(\chi^2) =\left\{ \begin{array}{rcl}
    \sinh^2\chi & \mbox{,} & K=-1 \\ \chi^2 & \mbox{,} & K=0 \\
    \sin^2\chi & \mbox{,} & K=+1
    \end{array}
    \right.  \,\,.
\end{equation}
Assuming the FRW metric and $c=1$ (in natural units for example),
the evolution of the Universe is mainly described by the form of
the scalar factor $a(t)$ and by transforming the form of the
metric with respect to the conformal time $\tau$ we have,
\begin{equation}
    \label{FRWtau}
    ds^2= a^2(\tau) \left[-c^2d\tau^2+ (d\chi^2 + \Theta_k(\chi^2)(d\theta^2+\sin^2\theta d\phi^2) )\right] \,\, ,
\end{equation}
where,
\begin{equation}
    \label{conftime}
    \tau=\int \dfrac{dt}{a(t)} \,\,.
\end{equation}
So by using this metric and these coordinates, the structure of
spacetime is more easily visualized. Specifically in an isotropic
Universe, the geodesics of propagating light are null $ds^2=0$.
Thus the radial null geodesics for propagating light in an
isotropic, homogeneous Universe in the conformal time frame are,
\begin{equation}
    \label{geodesics}
    \chi(\tau)=\pm \tau + constant   \,\,,
\end{equation}
which correspond to straight lines at angles of $\pm 45^{\circ}$
in the $\tau-\chi$ (conformal time-space) plane (see
\ref{fig.lightcone}).
At this point, two very important quantities must now be
introduced. The \textit{Hubble parameter H} expresses the rate of
expansion of the Universe and has units of inverse time (or mass
units in natural units). It is defined as,
\begin{equation}
    \label{Hhubble}
    H\equiv\dfrac{\dot{a}}{a}  \,\,,
\end{equation}
where $\dot{a}=\frac{\mathrm{d}a}{\mathrm{d} t}$. The Hubble
parameter is positive for an expanding Universe and negative for a
collapsing one. The other important quantity is the e-foldings (or
e-folds) number defined as,
\begin{equation}
    \label{Nefolds}
    N\equiv \ln{\dfrac{a_2}{a_1}} \,\,,
\end{equation}
which represents the number of Hubble times between two epochs
with scale factors $a_1$ and $a_2$. The Hubble time is $H^{-1}$ ,
which represents the time period that takes for the Universe to
expand substantially. The e-folds number can also be given as $N
=\int^{t_2}_{t_1} H dt$. The range of the e-folds number for a
successful inflationary model that can effectively solve the
flatness and horizon problems, which are going to be explained
later, is from $N \approx 45$ to $N^{tot}_{min}\simeq 60$, in the
context of most inflationary cosmologies, with the estimations
being based on single scalar field inflationary models. However
this value is model dependent and for some specific inflationary
models , it is possible that $N$ has a higher value than $60$
e-folds, a feature that mainly depends on the total equation of
state parameter of the Universe at the end of the inflationary era
\cite{Oikonomou:2023kqi}, and in others $N_{tot}\gg 60$, with
$N_{tot}$ having no upper bound that is specifically correlated
with the idea of ``eternal'' inflation \cite{Linde:1986fd} .[2pt]
\noindent

In general, any  physical system and its behavior can be
described, in the context of the fundamental laws of physics, with
the action,
\begin{equation*}
    \mathcal{S}[q]=\int dt\, L\left(q(t),\dot{q}(t),t \right)\,\,,
\end{equation*}
where $L$ is the Lagrangian of the system, with an allowed
explicit dependence on $t$, \cite{LythLiddle}.  According to
Hamilton's principle by taking the variation of the action $\delta
\mathcal{S}=0$, the path such as to maximize or minimize the
action integral can be derived as,
\begin{equation}
    \label{Hamiltons}
    \delta \mathcal{S}=\int dt \left(\dfrac{\partial L}{\partial q}\delta q + \dfrac{\partial L}{\partial\dot{q}}\delta{\dot q}\right)
    =0\, ,
\end{equation}
which leads to the Euler-Lagrange equations of motions for the
system,
\begin{equation}
    \label{EulerLagrange}
    \dfrac{\partial L}{\partial q}-\dfrac{\partial}{\partial t}\left(\dfrac{\partial L }{\partial \dot{q}} \right)=0\,\,.
\end{equation}
Now accordingly in the context of field theory, a Universe as a
physical system can be described by the scalar field action with
generic coordinates,
\begin{equation}
    \label{action}
    \mathcal{S}=\int d^4x\, \sqrt{-g} \mathcal{L}\,\,,
\end{equation}
where $g$ is the determinant of a metric $g_{\mu \nu}$ and
$\mathcal{L}=\mathcal{L}(\phi,\partial_{\mu}\phi)$ is the
Lagrangian density that has dimensions of $[m]^4$ in natural
units, it is Lorentz invariant and related to the Lagrangian of
the fields by $L=\int d^3x \, \mathcal{L}$. The equations of
motion and basically the evolution of the field can be determined
by the same variation principle for the action with respect to the
field $\phi$ and they are called the field equations . Relating to
the Lagrangian density $\mathcal{L}$, the energy-momentum tensor
$T^{\mu \nu}$ is defined as,
\begin{equation}
    \label{energymomentumtensor}
    T^{\mu \nu}=-\frac{2}{\sqrt{-g}}\frac{\delta\mathcal{L}}{\delta g_{\mu\nu}}
\end{equation}
and it is conserved with,
\begin{equation}
    \label{Tconservation}
    \nabla_{\mu}T^{\mu \nu}=0\,\,.
\end{equation}
Eq. \eqref{Tconservation} can be thought as expressing two
continuity equations,
\begin{align}
    \label{energycont}
    \nabla_\mu T^{\mu0} &=0\\[2pt]
    \label{momcont}
    \nabla_\mu T^{\mu i} &= 0\, ,
\end{align}

where in Eq. \eqref{energycont} corresponds to the energy
continuity equation and Eq. \eqref{momcont} corresponds to the
momentum continuity equation in curved spacetime, also known as
the Euler equation as well.

On another note, related to the Hubble rate $H$, by inserting Eq.
\eqref{FRWt} into the Einstein equations, we can derive the
following two equations also known as the Friedmann Equations,
\begin{equation}
    \label{Friedmann1}
    H^2 \equiv \left(\dfrac{\dot{a}}{a} \right)^2= \dfrac{\kappa^2}{3}\rho -\dfrac{K c^2}{a^2}+\dfrac{\Lambda}{3} \,\,,
\end{equation}
\begin{equation}
    \label{Friedmann2}
    \dot{H}+H^2 =\dfrac{\ddot{a}}{a}= -\dfrac{\kappa^2}{6}\left(\rho+\frac{3p}{c^2}\right)+\dfrac{\Lambda}{3} \,\,,
\end{equation}
where $\kappa^2=8\pi G/c^4=\frac{1}{M^2_{Pl}}$, $G$ is the
gravitational constant, $M_{Pl}$ is the reduced Planck mass and
$c$ the speed of light. When Eq. \eqref{Friedmann1} and
\eqref{Friedmann2} are combined they give rise to the continuity
equation, or the energy conservation equation,
\begin{equation}
    \label{continuity}
    \dfrac{d\rho}{dt}+3H(\rho+\frac{p}{c^2})=0 \,\,.
\end{equation}
By defining the equation-of-state parameter $\mathrm{w}$,
\begin{equation}
    \label{wparameter}
    \mathrm{w}\equiv\dfrac{p}{\rho c^2}  \,\,,
\end{equation}
and integrating Eq.\eqref{continuity} we get,
\begin{equation}
    \label{densityrho}
    \rho \propto \alpha^{-3(1+\mathrm{w})} \,\,.
\end{equation}
which describes how the energy density decays for a specific
epoch. So Eq. \eqref{densityrho} along with the Friedmann equation
Eq. \eqref{Friedmann1} leads to the following equation,
\begin{equation}
    \label{a(t)}
     \alpha(t) \propto  \left\{ \begin{array}{rcl}
    t^{2/3(1+\mathrm{w})} & , & \mathrm{w}\neq -1 \\ e^{Ht} & ,&  \mathrm{w}=-1
    \end{array}
    \right.  \,\,,
\end{equation}
 which indicates how the scale factor evolves with
time \cite{Baumann:2009ds}. The value of the equation-of-state
parameter depends on what perfect fluid energy density dominates
the scale factor of the flat Universe. Possible values of
$\mathrm{w}$ could be: $\mathrm{w}=0$ for non-relativistic
non-baryonic matter, $\mathrm{w}=-1$ for vacuum energy (for a
cosmological constant), $\mathrm{w}=1/3$ for radiation or
relativistic matter and $\mathrm{w}=+1$ for stiff matter.
Additionally, causality demands that $\mathrm{w}$ has to be
strictly less or at the very least equal to 1 , while no
lower-bound appears. For negative values
$\mathrm{w}<-\frac{1}{3}$,  the equation-of-state parameter
relates to the presence of a dark fluid. For a quintessence and
quintessential evolution of the Universe,  $\mathrm{w}\neq -1$ and
$-1<\mathrm{w}< -1/3$, while also data from Planck+WP, SNLS and
BAO suggest the limits of $-1.3<\mathrm{w}<-0.81$ (95$\%$C.L.)
\cite{Planck:2018vyg}. For $\mathrm{w}<-1$, relates to the
existence of hypothetical phantom dark energy, and it can lead to
a Big Rip singularity. While the equation-of-state parameter can
be taken as a constant, it can also be a function of redshift or
time. It is also possible to have significant contributions from
various kinds of matter to the total energy density $\rho$ and
pressure $p$, thus they can be given by the sum of all the
individual components as,
\begin{equation}
    \label{Sdensitypressure}
    \rho \equiv \sum_i \rho_i \hspace{0.5cm},  \hspace{0.5cm} p\equiv\sum_i p_i  \,\,.
\end{equation}
For each of the components there is also the ratio of the observed energy density to the critical density for the present time $t_0$ ,
\begin{equation}
    \label{Omegaratio}
    \Omega_{i0} \equiv \dfrac{\rho_{i0}}{\rho_{crit}}  \,\,,
\end{equation}
where $\rho_{crit}=3H_0^2/\kappa^2$ is the density required to
make the expansion of the Universe slow down, stop and reverse,
meaning the existence of a closed Universe. Additionally for the
curvature contribution at $t_0$,
\begin{equation}
    \label{OmegaK0}
    \Omega_{K0}=-\dfrac{Kc^2}{a^2(t_0) H_0^2} \,\,,
\end{equation}
and using Eq.\eqref{Friedmann1} and setting the scalar factor as $a(t_0)=a_0=1$,
\begin{equation}
    \label{FriedmanOmega}
    \left(\dfrac{H}{H_0}\right)^2 =\sum_i \Omega_i a^{-3(1+w_i)}+\Omega_K a^{-2} \,\,.
\end{equation}
Therefore, Eq.\eqref{FriedmanOmega} for the present time $t_0$ leads to,
\begin{equation}
    \label{AllOmega}
    1=\sum_i \Omega_{i0}+\Omega_{K0} \,\,.
\end{equation}
Generally as a function of the cosmic time $t$ the curvature
parameter is,
\begin{equation}
    \label{OmegaK}
    \Omega_K=-\dfrac{Kc^2}{(a(t)H)^2}=-\dfrac{Kc^2}{\dot{a}^2} \,\,,
\end{equation}
and by assuming the simplest case of the expansion being dominated
by a form of matter with an equation of state parameter
$\mathrm{w}$ and with the scale factor given by
$a(t)=t^{2/3(1+\mathrm{w})}$ and also by taking the first
derivative with respect to time, we obtain the following equation,
\begin{equation}
    \label{Omegaderivative}
    \dot{\Omega}_K=-Kc^2 \dfrac{d}{dt}\left( \dfrac{1}{\dot{a}^2}\right)=Kc^2\dfrac{2\dot{a}\ddot{a}}{\dot{a}^4} \rightarrow \hspace{0.2cm} \dot{\Omega}_K= \Omega_K H (1+3w) \,\,.
\end{equation}
From Eq. \eqref{Omegaderivative}, it is deduced that the value
$\mathrm{w}=-\dfrac{1}{3}$ is an equilibrium point and an
\textbf{unstable} one, more specifically. This indicates that if
the strong energy condition is satisfied, when $\mathrm{w}$
deviates a bit from that unstable equilibrium point, for
$\mathrm{w}>-1/3$ with $\Omega_K>0$, the parameter $\Omega_K$
keeps growing and with $\Omega_K<0$, $\Omega_K$ keeps decreasing
away from zero. From experimental data and observations of
large-scale structures and the CMB it is concluded that $-0.0179 <
\Omega_K < 0.0081$ (95$\%$CL). Meaning that the presently observed
Universe is very flat and therefore it used to be even more flat
in the past with the value of parameter $\Omega_K$ reaching
exceptionally small values, $\Omega_K \thicksim 0$. This could be
the case, if we assumed that $K=0$ precisely, for the Universe in
its very early initial state. However, it seems quite peculiar to
have such a precise value of $K$ and no explanation why this would
be the case.

And this is where the first signs of how some characteristics of
the SBB could be problematic and a new approach is needed. The
conventional SBB model describes the Universe as very homogeneous,
isotropic and flat, which are features that do not emerge from a
fundamental mechanism that explain how the Universe came to be or
evolved in that fine-tuned way, but only from the initial
assumptions of the construction of the model. Apart from the
flatness problem, there exist several other shortcomings of the
SBB model, and inflation, provides concrete theoretical
explanations for these problems \cite{guth,lindeproblems} and
there are presented one by one in the following subsections.

\subsubsection{The Horizon problem}

Before continuing, we will provide definitions of some quantities
which are important for better understanding of the concepts
described here. In Eq. \eqref{Hhubble}, the Hubble parameter was
defined. There are also the quantities of the Hubble distance, or
horizon, given as $H^{-1}(t)$ and the \textbf{particle horizon
$\mathcal{R}$}, which is the distance that light could have
travelled since the beginning when $a=0$ and regions separated by
distance more that the particle horizon are ``causally
disconnected'', meaning they can \textit{never} communicate with
each other. The particle horizon is also called the
\textbf{comoving horizon} and is given by the following relation,
 \begin{equation}
     \label{Comovinghorizon}
     \mathcal{R}=\int_{t}^{0}\dfrac{dt}{a(t)}=\int_{0}^{a} \dfrac{da}{Ha^2} = \int_0^{a} d\ln a \left(a H \right) ^{-1} \,\,,
 \end{equation}
where $\left(a H \right)^{-1}$ is the \textbf{comoving Hubble
radius} and if some regions are separated by a distance greater
that the coming Hubble radius, they cannot communicate with each
other at that time. By this definition, for a Universe dominated
with a fluid with an equation of state parameter $w$, the Hubble
radius is,
\begin{equation}
    \label{Hubbleradius}
    \left(a H\right)^{-1}= H_0^{-1} a^{(1+3w)/2} \,\,.
\end{equation}
So by the conventional SBB, where $w\geq 0$, the Hubble radius
grows monotonically and the comoving horizon $\mathcal{R}$
increases with time. This means that comoving regions that are
entering the horizon in the present time, and thus becoming
causally connected now, used to be outside of the limits of the
horizon and therefore causally disconnected in the past and
especially in the primordial era of the Universe. Consequently,
the high homogeneity and thermal equilibrium of far distant
regions of the Universe observed in the CMB, that were supposed to
be causally disconnected in the early times, is quite problematic.
This is also sometimes mentioned as the \textit{Horizon problem}.
So if homogeneity is not assumed as an initial feature, there must
be some mechanism that made the Universe to evolve that way.
However the SBB model cannot provide an explanation on how the
Universe could be that homogeneous without fine-tuning the initial
conditions.

Inflation solves this problem by introducing a period with a
decreasing comoving Hubble radius $\left(a H \right)^{-1}$. The
exponential growth of the scale factor $a$ during inflation and
the relatively constant $H$ allows the Hubble radius to be
decreasing while Inflation is happening. So in the very early
period, the comoving Hubble radius was much larger than the
comoving horizon $\mathcal{R}$ and all the scales-regions of the
Universe (relevant to cosmological observation) used to be small
enough to be inside the Hubble radius and thus they were in fact
causally connected. Then during inflation, the radius started to
decrease, became smaller than the horizon while the volume of the
horizon itself expanded. After Inflation ended, the Hubble radius
started increasing its size again and in consequence, at  present
time, the horizon is larger than the Hubble radius and regions
that were causally connected, now seem that they are causally
disconnected, see Fig. (2). According to this mechanism, different
regions of the Universe, which used to be closer together and
inside the horizon in the early epoch, were able to communicate,
become homogeneous with respect to each other and established
thermal equilibrium. When the Hubble radius decreased and the
Universe rapidly expanded they became causally disconnected, as we
observe them to be today.
\begin{figure}[h!]
    \centering
    \includegraphics[width=0.8\textwidth]{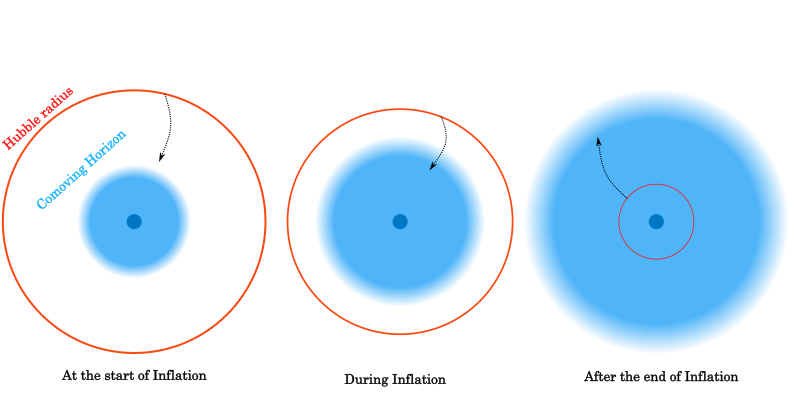}
    \caption{At the beginning, all the Universe, whose ends are represented by
the edge of the comoving horizon, is inside the Hubble radius.
Thus all the regions of the Universe are causally connected and
during this period there was enough time for them to become
homogeneous and isotropic. As inflation begins the Hubble radius
(red line) decreases in size and it keeps decreasing as inflation
is in process. On the contrary, the comoving horizon (blue patch)
is increasing in size, as the Universe is expanding. At some point
the comoving horizon becomes greater in size than the Hubble
radius and thus some regions of the Universe become causally
disconnected from the regions of the Universe that are still
inside the Hubble radius, therefore impossible for those different
parts to communicate with each other. After inflation the Hubble
radius starts to increase again and therefore those regions
re-enter the Hubble radius and become visible again.}
    \label{horizonradius}
\end{figure}

\subsubsection{The Flatness Problem}

We briefly discussed the Flatness Problem, now we shall further
elaborate on this. The second issue that arises is the Flatness
problem, where the conventional Big Bang model points to the fact
that the Universe is very flat, with,
\begin{equation}
    \label{OmegaKflat}
    \Omega_{K0} \sim 0 \xrightarrow[]{Eq.\eqref{AllOmega}} \Omega_0=\sum_i \Omega_{i0} \sim 1 \,\,,
\end{equation}
which is verified by observational data. However, in the
conventional model, this value is an unstable fixed point of the
equations and there is no significant reason why the Universe
should be at that unstable point, except in the case in which the
initial conditions were fine-tuned that way.

Combining Eqs. \eqref{AllOmega} and \eqref{OmegaK}, we obtain,
\begin{equation}
    \label{Omegaflat}
    1-\Omega=\Omega_K=-\dfrac{Kc^2}{(a H)^{2}} \,\,.
\end{equation}
Therefore in the duration of the inflationary period, since there
is the decrease of the Hubble radius $(a H)^{-1}$, the parameter
$\Omega_K$ decreases towards zero as well and thus the Universe
evolves towards its flatness naturally, which is in accordance
with the experimental observations. It also justifies the
disregard of the cases of a closed or open Universe since if the
Universe was not flat, and for example we had a closed Universe
with an intrinsic spatial curvature $K=+1$, no phase transition
can occur and even if inflation washes out the effect of curvature
such that the density parameter $\Omega_K\simeq0$, $K$ can never
transition from $+1$ to 0, or in the case of an open Universe from
$-1$ to 0.

\subsubsection{The Primordial Relics Problem}

In a very early period after the ``Big Bang'', the Universe can be
described by \textbf{Grand Unified Theories} (GUT) or string
theory or some Standard Model extensions. In a GUT scenario the
Universe used to have a temperature of the order of the GUT
temperature, which is about $T_{GUT}\sim 10^{28} K$ and the
electromagnetic, weak and strong forces were unified. According to
GUT, the Universe went through a phase transition when the
temperature of the Universe dropped below the $T_{GUT}$, and
during that transition there was a production of primordial relics
(e.g. domain walls, magnetic monopoles or other topological
defects etc.), which are described as point-like topological
defects in the scheme of GUT. So at the time of their creation,
the number and energy density of primordial relics like magnetic
monopoles should had been large, but still smaller than the ones
for radiation during the GUT period and thus the Universe in that
early stage is still radiation-dominated. Later on, as those
relics should have been quite large, they would have quickly
become non-relativistic, since $\rho_M \propto a^{-3}$ for
magnetic monopoles (massive particles) and $\rho_r \propto a^{-4}$
for radiation, thus they would have dominated over radiation and
ordinary matter until the present time. However, from observations
there is no evidence for the existence of such primordial relics
and definitely no signs that they dominate the Universe, with
research setting an upper limit to their number density today of
$n_M(t_0)\sim {10^{-19}}{cm^{-3}}$. This very small number density
and the lack of this kind of observations today in the Universe
compared to the predictions of the Standard Big Bang scenario
along with particle physics, is the \textit{Primordial Relics
Problem}.

In the case of inflation, once the monopoles were created before
or during the inflationary period, thereafter their number density
would have decreased significantly during the rapid exponential
expansion of the Universe. After the Universe expanded, the
magnetic monopoles were basically so outspread in space that their
number density $n_M(t_0)$ today would have reached such a small
value, which renders them nearly impossible to detect.

\subsection{Conditions for Inflation}

Having presented the issues that motivated the construction of the
inflationary paradigm, it is time to introduce the conditions
under which inflation takes place. During the inflationary period
the Hubble radius decreases with time, as mentioned in the
previous paragraphs, therefore,
\begin{equation}
    \label{Hradiuscondition}
    \dfrac{d}{dt}\left(\dfrac{1}{a H}\right) <0 \,\,.
\end{equation}
Since the Hubble rate is given by Eq. \eqref{Hhubble}, two more
equivalent conditions can be deduced for inflation,
\begin{align}
    \label{infcon1}
    \ddot{a}> 0 \,\,,\\
    \label{infcon2}
    -\dfrac{\dot{H}}{H^2}<1 \,\,.
\end{align}
From condition Eq. \eqref{infcon2}, if there is a strong
inequality, $|\dot{H}|\ll H^2$, then the Hubble rate $H$ is almost
constant for many Hubble times and there is approximately an
exponential expansion with $a(t) \propto e^{Ht}$. For $H$ exactly
constant over many Hubble times, inflation is described by a
 de Sitter expansion. If however the Hubble rate
 contains linear or higher order time dependencies, like for
 example $H(t)\sim H_0-H_i t$, then this evolution is called
 quasi-de Sitter evolution. Thus inflation is described as an early era of rapid nearly exponential
expansion of the Universe. From the conditions of Eq.
\eqref{infcon1}, it is also derived that during inflation the
scale factor $a$ increases fast as a function of the cosmic time
$t$. This condition is also achieved by an expansion with a
power-law scale factor of $a(t)\sim t^n$ with $n>1$, however the
tensor-to-scalar ratio for these power-law inflation models is
larger than the limits set by \textit{Plank} data, thus they are
not appropriate to describe the inflationary era. Furthermore,
from Eq. \eqref{a(t)} for an exponential form of the scale factor,
the equation-of-state parameter is $\mathrm{w}=-1$, which
corresponds to a cosmological constant and from Eq.
\eqref{densityrho}, the density is $\rho \sim $ constant, which
can also be approximately taken for a dominant dark energy era
($\mathrm{w}<-1$) with $\mathrm{w}\sim -1$. There is also an
inflationary condition for the pressure that arises if Eq.
\eqref{Friedmann2} and Eq. \eqref{infcon2} are combined,
\begin{equation}
    \label{infcon3a}
    \left(\rho+\dfrac{3 p}{c^2}\right)< \dfrac{2\Lambda}{\kappa^2}
\end{equation}
where $\Lambda$ is the cosmological constant. For $\Lambda=0$ or absorbed into $\rho$ and $p$,
\begin{equation}
\label{infcon3b}
\left(\rho+\dfrac{3p}{c^2} \right)<0\,\,.
\end{equation}
Again here, it can be seen that for a dark energy dominated era
$\mathrm{w}<-1$ or for $\mathrm{w}\sim -1$, from the definition of
the equation-of-state parameter Eq. \eqref{wparameter}, this
condition if fulfilled.

In the next sections, we shall present the standard models that
are used in the literature to describe inflation, namely scalar
field inflation models and modified gravity models. While scalar
field inflationary models are more customary, nowadays it seems
that single scalar field inflation might be insufficient for
describing the primordial era of our Universe. The reason is
two-fold, firstly it is theoretically unappealing to have to
explain the couplings of the inflaton scalar field to all the
Standard Model particles in order to reheat the Universe and
secondly, after the recent NANOGrav detection of the stochastic
gravitational wave background \cite{NANOGrav:2023gor}, the
cosmological perspective of such a background complexifies the
single scalar field description of inflation, while leaving room
for modified gravity descriptions \cite{Oikonomou:2023qfz}.

\section{Canonical-Scalar field Inflation}

\subsection{Minimally-coupled Scalar field Inflation}

For the scalar field models, the Lagrangian density for $N$ real fields is assumed to have the form,
\begin{equation}
    \label{NsfLagrangian}
    \mathcal{L}=-\dfrac{1}{2}\sum_{i=1}^{N} \partial^{\mu}\phi_i \partial_{\mu}\phi_i-V(\phi_1,\dots ,
    \phi_n)\, ,
\end{equation}
where $V$ is a function of all the fields and the fields with this
Lagrangian are said to be canonically normalized. For the simplest
case of a single scalar field in flat spacetime, the Lagrangian is
given as,
\begin{equation}
    \label{Lscalarfield}
    \mathcal{L}=-\dfrac{1}{2}\partial^{\mu}\phi \partial_{\mu}\phi
    -V(\phi)\, ,
\end{equation}
where the first is a kinematic term, $V(\phi)$ is the scalar field
potential and both terms separately have mass dimensions of
$[m]^4$ in natural units. In this section we will focus on the
single scalar field inflation models with minimal coupling to
gravity. In this case the inflationary period is associated with a
single scalar field, dubbed \textit{inflaton} and the minimal
coupling to gravity conveys that the action for the inflationary
field $\phi$ is not coupled with the scalar curvature in any way
but through a term of the Lorentz invariant $\sqrt{-g}d^4x$ from
the metric. The energy density of the inflaton is dominant
compared to the rest of the matter fields for this period and
thus, no additional field emerges. We also consider a flat
Universe, $K=0$, described by a FRW metric. In the literature
there are many models that may consider the effect of both a
scalar and gauge fields simultaneously or model with intrinsic
curvature, however this is not relevant to the context of this
text.

So the action for minimally-coupled Single scalar field inflation
considering Eq.\eqref{FRW} is,
\begin{equation}
    \label{sscAction}
    \mathcal{S}= \int d^4x \sqrt{-g} \left(\dfrac{R}{2 \kappa^2}-\dfrac{1}{2}g^{\mu \nu} \partial_{\mu}\phi\partial_{\nu} \phi -V(\phi) \right)  \,\,,
\end{equation}
where $\kappa^2=8\pi G/c^4 = \frac{1}{M_{Pl}^2}$ and $g$ is the
determinant of the FRW metric $g^{\mu \nu}$ for $K=0$
\cite{reviews1}. The first term is the gravitational
Einstein-Hilbert action including the Ricci scalar $R$, which in
terms of the metric connection it has a form of $R=g^{\mu
\nu}R_{\mu\nu}$ with $R_{\mu\nu}$ being the Ricci tensor. The last
two terms in the action of the scalar field are the canonical
kinetic term and the scalar field potential $V(\phi)$ which takes
into account the self-interactions of the scalar field. It is
worth mentioning that quantum fluctuations can have as a result
perturbations in the inflaton field and in the metric tensor with,
\begin{equation}
\label{quantumfluc1}
    \phi \rightarrow \phi_0 +\delta \phi \,\,\,,\,\,\, g_{\mu \nu} \rightarrow \bar{g}_{\mu\nu} + \delta g_{\mu\nu}
\end{equation}
where $\bar{g}_{\mu\nu}$ is the FRW metric and $\phi_0$ the
classical solution for homogeneous, isotropic evolution in the
inflationary era. For the scalar field action of Eq.
\eqref{sscAction}, through its variation, the scalar field
equation is,
\begin{equation}
\label{preEoM}
 \ddot{\phi}-\alpha^{-2}\nabla^2\phi +3H\dot{\phi} +V'(\phi)=0
\end{equation}
where $\nabla^2$ is determined with comoving coordinates and for a homogeneous field, it takes the form,
\begin{equation}
    \label{sscFE}
    \ddot{\phi}+3H\dot{\phi}+V'(\phi)=0
\end{equation}
where $V'(\phi)$ is the derivative of the potential with respect to the field. Accordingly, the energy-momentum tensor here is defined as,
\begin{equation}
    \label{emTensor}
    T_{\mu \nu}=\partial_{\mu}\phi \partial_{\nu} \phi -g_{\mu \nu}\left(\frac{1}{2}g^{\rho\sigma}\partial_{\rho}\phi \partial_{\sigma}\phi -V(\phi)  \right) \,\,.
\end{equation}
and therefore the pressure and the energy density of the Universe
can be obtained from $T_{j}^{i}=-P\delta_{j}^{i}$ and $T_0^0$
respectively,
\begin{equation}
    \label{ssfPressure}
    p = \dfrac{1}{2}\dot{\phi}^2-V(\phi) \,\,,
\end{equation}
and
\begin{equation}
    \label{ssfDensity}
    \rho=\dfrac{1}{2}\dot{\phi}^2+V(\phi) \,\,.
\end{equation}
In order for a scalar field to be able to produce a viable model
for inflation, it needs to slow-roll under the influence of the
inflationary potential $V(\phi)$ as it evolves towards the minimum
of the potential, and also in a sufficiently slow manner, in order
for the scale factor to increase enough to be able to resolve the
standard Big Bang cosmology problems. The condition that is
imposed for this effect to be secured is called the slow-roll
assumption,
\begin{equation}
    \label{SRassumption}
    \frac{1}{2}\dot\phi^2\ll V    \,\,,
\end{equation}
which indicates that the kinetic term is sub-leading and becomes
really small compared to the potential. Also under this condition
the Friedmann equation \eqref{Friedmann1} takes the form,
\begin{equation}
    \label{FriedmannSR}
    H^2\simeq \dfrac{\kappa^2}{3} V \,\,\,.
\end{equation}
and from the derivative of the condition \eqref{SRassumption}, it
is indicated that $\ddot\phi\ll V'$, and in effect the scalar
field equation of \eqref{sscFE} results in,
\begin{equation}
    \label{feSR}
    3H\dot{\phi} \simeq -V'
\end{equation}
and from the Raychaudhuri equation \eqref{Friedmann2}, this
condition, implies that $\dot H\ll H^2$, which is in agreement
with the condition for the Hubble radius for homogeneity. In order
to quantify the consequences of this assumption, the slow-roll
indices are introduced and for the single scalar field inflation,
they can be expressed with respect to the Hubble parameter as
follows,
\begin{equation}
    \label{e1e2}
    \epsilon_1 =-\dfrac{\dot{H}}{H^2} \hspace{0.2cm},\hspace{0.2cm} \epsilon_2=\dfrac{\ddot{H}}{2H\dot{H}} \,\,\,.
\end{equation}
and therefore the results of the slow-roll assumption for the
canonical field of \eqref{SRassumption} are translated as the
following conditions for the slow-roll indices,
\begin{equation}
\label{e1e2COND}
  \epsilon_1,\epsilon_2 \ll 1 \, ,
\end{equation}
for the time when inflation is taking place \cite{reviews1}, with
$\epsilon_1 \ll 1$ ensuring the occurrence of the inflationary era
and $\epsilon_2 \ll 1$ ensuring that inflation lasts a sufficient
amount of time so that the scalar field slowly evolves with
respect to cosmic time $t$ for a large number of e-folds, and the
density parameter for curvature vanishes from the background
equations. It is worth mentioning that the condition on the
$\epsilon$ slow-roll parameter is not just a mathematical boundary
of a specific mechanism in an inflationary model, but it is linked
to the natural process of inflation. By elaborating on the
mathematical form of \eqref{e1e2} we get that $\epsilon
=-\frac{\dot{H}}{H^2}= 1-\frac{\ddot{a}a}{\dot{a}^2}$.
Independently of any specific mathematical formulation of a
particular model, generally a key feature of inflation is that
during the inflationary era the Hubble radius decreases and so
$\frac{d}{dt}\left(\frac{1}{a H}\right)<0 \rightarrow
-\frac{\ddot{a}}{\dot{a}^2} <0$ and thus
$-\frac{\ddot{a}a}{\dot{a}^2} <0 \rightarrow
1-\frac{\ddot{a}a}{\dot{a}^2} <1 \rightarrow \epsilon_1<1$. The
same applies to the conditions for the end of inflation, since the
Hubble radius rate with respect to time is equal to zero the
moment that inflation ends. Another more familiar expression for
the slow-roll parameters is the one with respect to the canonical
scalar field potential $V(\phi)$,
\begin{equation}
    \label{SRwithV}
    \epsilon \simeq \dfrac{1}{2\kappa^2} \left(\dfrac{V'(\phi)}{V(\phi)} \right)^2  \hspace{0.2cm} , \hspace{0.2cm} |\eta| \simeq \dfrac{1}{\kappa^2}\left|\dfrac{V''(\phi)}{V(\phi)}\right| \,\,,
\end{equation}
where $V'$, $V''$ are the first and second derivatives of the
potential with respect to the field $\phi$ and the form of the
potential $V(\phi)$ can take various forms depending on the model.
Also in \eqref{SRwithV}, the sign of $\eta$ is insignificant and
only its order of magnitude matters. The connection between the
two representations of the slow-roll parameters is
\cite{reviews1},
\begin{equation}
    \label{SRequiv}
    \epsilon=\epsilon_1 \hspace{0.2cm},\hspace{0.2cm} \eta=\epsilon_1-\epsilon_2
\end{equation}
where $\epsilon_1=V'(\phi)^2/6H^2V(\phi)$ and
$\epsilon_2=\ddot{\phi}/H\dot{\phi}$, with the derivative of
\eqref{feSR}. The end of the inflationary era occurs when
basically the inflationary condition is violated, and the
slow-roll perturbative expansion for the power spectrum breaks
down, which is equivalent to the condition,
\begin{equation}
    \label{ordernsr}
    \epsilon, \eta \sim  \mathcal{O}(1)\,\,.
\end{equation}
After the end of inflation, the reheating era follows and the
field oscillates about the minimum value of its potential. The
quantum fluctuations $\delta \phi$ of the scalar field basically
generate the CMB fluctuations nearly 60 e-folds before the end of
inflation. Lastly, two of the most important observational
quantities, the spectral index of the primordial scalar curvature
perturbations $n_\mathcal{S}$ and the tensor-to-scalar ratio $r$,
can be also expressed with respect to these slow-roll parameters
as,
\begin{equation}
    \label{nsr1}
    n_\mathcal{S}\simeq 1-6\epsilon +2\eta \,\,\,,\,\,\,r\simeq 16 \epsilon
\end{equation}
or as $n_\mathcal{S}=1-4\epsilon_1-2\epsilon_2$, $r=16\epsilon_1$
for the slow-roll indices of \eqref{e1e2}. The are various models
that consider different forms for the inflationary potential
$V(\phi)$. Some examples are presented in the following
subsections.

\subsubsection{Massive Scalar Field}
\noindent

This is a simple case of single scalar field inflation, called the
\textit{Chaotic Inflation}, driven by a mass term. The field
starts at a large value and rolls down towards the origin of the
scalar potential and the form of the potential is,
\begin{equation}
    \label{sscM1}
    V(\phi)= \dfrac{1}{2}m^2\phi^2 \,\,.
\end{equation}
Since the potential is axial symmetric it is expected that either
positive or negative values for the scalar potential could work
and therefore the results are independent of the sign of $\phi_k$,
which is the value of the field at the horizon crossing, something
which is hinted by the slow-roll indices as well. The choice of
sign can however in principle affect the evolution of the scalar
field with respect to time. From \eqref{SRwithV} the slow-roll
parameters are,
\begin{equation}
    \label{M1.sroll}
    \epsilon=\dfrac{2}{\kappa^2 \phi^2} \hspace{0.1cm},\hspace{0.1cm} \eta=\dfrac{2}{\kappa^2 \phi^2} \,\,
\end{equation}
and the end of inflation occurs when,
\begin{equation*}
   \epsilon,\eta \sim \mathcal{O}(1)\rightarrow \dfrac{2}{\kappa^2 \phi^2}\simeq 1 \,\,,
\end{equation*}
\begin{equation}
    \label{fendM1}
    \phi_{end}= \dfrac{\sqrt{2}}{\kappa}=\sqrt{2} M_{Pl} \,\,.
\end{equation}
The integral definition of the e-folds number $N$ with respect to
the filed $\phi$ is,
\begin{equation}
    \label{Nintegral}    N=\int_{\phi_k}^{\phi_{end}}\frac{H}{\dot\phi}d\phi\, .
\end{equation}
By taking the slow-roll approximation, with \eqref{FriedmannSR} and \eqref{feSR}, the integral for this specific potential form is,
\begin{equation}
\label{fkM1}    N=\int_{\phi_k}^{\phi_{end}}\dfrac{H}{\dot{\phi}} d\phi =-\dfrac{\kappa^2}{6}\int^{\phi_{end}}_{\phi_k} \phi \, d\phi \rightarrow N=-\dfrac{\kappa^2}{12}\left(\phi^2_{end}-\phi^2_k\right) \,\,.
\end{equation}
By replacing the $\phi_{end}$, in \eqref{fkM1}, the value of the field at horizon crossing is determined as,
\begin{equation}
    \phi_k\simeq \sqrt{2}M_{Pl}\cdot\sqrt{6N+1}
\end{equation}
and inserting this value into \eqref{M1.sroll} and then in \eqref{nsr1}, the spectral index and the tensor-to-scalar ration can be determined in the horizon crossing as,
\begin{equation}
    \label{nsrinfM2}
    n_\mathcal{S}\simeq \dfrac{6N-3}{6N+1} \,\,\,\,,\,\,\,\, r\simeq \dfrac{16}{6N+1}
\end{equation}
According to the relations of Eq.\eqref{nsrfendM2}, for a number
of e-folds of $N\simeq60$, the spectral index and the
tensor-to-scalar ratio have values of $n_s\simeq 0.9889$ and
$r\simeq 0.044$ respectively. From observations of the CMB by the
Plank collaboration, the constrains on the values of the spectral
index and of the tensor-to-scalar ratio are determined to be,
\begin{equation}
\label{nsrPlank}
n_\mathcal{S}=0.9649\pm 0.0042 \,\,(68\% \mathrm{CL})\hspace{0.2cm} \mathrm{and} \hspace{0.2cm} r<0.064 \,\, (95\% \mathrm{CL}).
\end{equation}
Therefore we see that the values of $n_\mathcal{S}$ and $r$
deviate enough from the one determined by observation, thus this
potential of the power-law form of Eq. \eqref{sscM1} cannot be
viable to generate an inflationary period. It is noted, as
previously mentioned, that ``power-law'' models like the examples
1, 2 and 3 are not viable due to the range of values for the
quantities $n_\mathcal{S}$ and $r$, simultaneously compared to the
limits from observational data from Planck. However this is
specifically for the case of minimally-coupled scalar field
inflation model and for the case of non-minimally coupled theories
those models could possibly be proven viable under certain
circumstances.

\subsubsection{Self Interacting Scalar Field}
\noindent

In this case, the potential has a quadratic form with respect to the field,
\begin{equation}
    \label{ssc.M2}
    V(\phi)=\lambda \phi^4 \,\,,
\end{equation}
which has mass dimension of $[m]^4$, since $\lambda$ is a
dimensionless coupling constant and $\phi^4$ has a dimension of
$[m]^4$. The slow-roll parameters here are,
\begin{equation}
    \label{M2.sroll}
    \epsilon =\dfrac{8}{\kappa^2 \phi^2} \hspace{0.1cm},\hspace{0.1cm} \eta=\dfrac{12}{\kappa^2 \phi^2} \,\,,
\end{equation}
therefore for the end of inflation,
\begin{equation*}
    \epsilon=1 \rightarrow \dfrac{8}{\kappa^2 \phi^2}=1\,\,,
\end{equation*}
\begin{equation}
    \label{fendM2}
    \phi_{end}=2\sqrt{2}M_{Pl} \,\,.
\end{equation}
Similarly by taking the integral form of $N$ and the slow-roll
approximation, with \eqref{FriedmannSR} and \eqref{feSR}, the
integral for this specific potential form is,
\begin{equation}
\label{fkM2}
N \simeq -\dfrac{\kappa^2}{4}\int_{\phi_k}^{\phi{end}}\phi \,d\phi  \rightarrow N=-\dfrac{\kappa^2}{8}\left(\phi^2_{end}-\phi^2_k \right)
\end{equation}
where $\kappa^2=\frac{1}{M^2_{Pl}}$. So by replacing the
$\phi_{end}$ in \eqref{fkM2}, the value of the scalar field
$\phi_k$ can be determined as,
\begin{equation}    \phi_k=2\sqrt{2}M_{Pl}\cdot\sqrt{1+N}
\end{equation}
where $M_{Pl}$ is the reduced Planck mass and $N$ the number of
e-folds. By inserting this value in \eqref{M1.sroll} and then in
\eqref{nsr1}, the scalar spectral index and the tensor-to-scalar
ratio can be determined at the horizon crossing as,
\begin{equation}
    \label{nsrfendM2}
    n_\mathcal{S}\simeq \dfrac{N-2}{N+1} \,\,\,\,, \,\,\,\, r\simeq \dfrac{16}{N+1}
\end{equation}
According to the relations of Eq. \eqref{nsrfendM2}, for a number
of e-folds of $N\simeq60$, the spectral index and the
tensor-to-scalar ratio have values of $n_\mathcal{S}\simeq 0.95$
and $r\simeq 0.26$ respectively. So based on these values in
comparison again with the Planck values of \eqref{nsrPlank}, a
self interacting scalar field with the potential of \eqref{ssc.M2}
of a power-law form, cannot be viable and is not fit to generate
an inflationary era according to observational data.

\subsubsection{Natural Inflation}

Let us now consider the natural inflation model, usually
considered in axion field contexts. In this case the inflationary
field, inflaton, is represented by a pseudo-Nambu-Goldstone
boson(PNGB), which could be e.g. an axion . The potential of the
inflaton is,
\begin{equation}
    \label{52}
    V(\phi)=L^4\left(1\pm \cos\left(\frac{\phi}{f}\right) \right) \,\,
\end{equation}
where $L$ and $f$ are two mass scales related to the height and
the width of the potential respectively and they are of order
$f\sim M_{Pl} [GeV]$ and $L\sim M_{GUT} [GeV]$. The inflationary
era corresponds in the region of $0<\phi<\pi f$ and it occurs as
the inflaton evolves towards the potential minimum at $\phi= \pi
f$. The slow-roll parameters for this potential are,
\begin{equation}
    \label{eq53}
    \epsilon=\dfrac{1}{2(\kappa f)^2}\dfrac{\sin^2\left(\frac{\phi}{f}\right)}{\left(1 \pm \cos\left(\frac{\phi}{f}\right)\right)^2} \hspace{0.1cm}, \hspace{0.1cm} \eta=\dfrac{1}{(\kappa f)^2}\dfrac{\cos\left(\frac{\phi}{f}\right)}{\left(\pm 1 -\cos\left(\frac{\phi}{f}\right) \right)} \,\,.
\end{equation}
as seen, they depend solely on $f\sim M_{Pl}$ and not on $L$.
Following the same process as before, it can be determined that,
\begin{equation}
    \sin{\left(\dfrac{\phi_k}{f}\right)}\simeq \sin{\left(\dfrac{\phi_{end}}{f}\right)}\cdot exp\left\{-\dfrac{NM^2_{Pl}}{16\pi
    f^2}\right\}\, .
\end{equation}
In order for this model to be viable for inflation and to obtain a
number of e-folds $N \gtrsim 70$, it is required that the initial
value of the inflaton is $\phi_k  \lesssim 0.1M_{Pl}$. It is also
worth mentioning that the Natural inflation model, or axion model,
can safely produce the power-law models examined before by simply
assuming that $\frac{\phi}{f}\ll 1$ and performing a Taylor
expansion, which is connected to the kinetic axion model.

\noindent

\subsection{Observable quantities in the inflationary paradigm}

In the previous sections, two important observable quantities,
that of the spectral index of the scalar perturbations
$n_\mathcal{S}$ and the tensor-to-scalar ratio $r$, have already
been introduced in the context of the single scalar field
inflation. This section, concentrates more on the origin of these
quantities and on the introduction of the tensor spectral index
$n_\mathcal{T}$ as well.

Even though the primordial Universe is considered to be
homogeneous, CMB observations have proved that this is not
entirely the case and it has anisotropies of lower order $\sim
10^{-5}$ than the homogeneous background. Inflation can explain
sufficiently these anisotropies, with the existence of quantum
fluctuations in sub-horizon scales during the early periods of the
inflationary epoch. So during inflation, perturbations are defined
around the homogeneous background solutions $\bar{\phi}(t)$ of the
inflaton and the metric $\bar{g}_{\mu\nu}$, as also similarly seen
in \eqref{quantumfluc1},
\begin{equation}
    \label{quantumfluc2}
    \phi(t,\textbf{x})=\bar{\phi}(t)+\delta\phi(t,\textbf{x}) \hspace{0.15cm},\hspace{0.15cm} g_{\mu\nu}(t,\textbf{x})= \bar{g}_{\mu\nu}(t)+\delta g_{\mu\nu}(t,\textbf{x}) \,\,.
\end{equation}
Specifically, during the inflationary era, the comoving Hubble
radius decreases as the Universe expands and it becomes smaller
than the comoving wavelength (horizon), \ref{horizonradius}. So
when these fluctuations exit the horizon, they become causally
disconnected and they remain frozen until the end of inflation,
when the physical horizon expands again and they gradually reenter
as classical density perturbations. During the time of inflation,
the stress-energy tensor contributions are heavily dominated by
the energy of the inflaton and therefore the perturbations of the
inflationary field have some effect on the geometry of the
spacetime through the field equations. Also, since the background
spacetime is considered fairly symmetric, justified by being
spatially flat, homogeneous and isotropic, the decomposition of
the metric and stress-energy perturbations into independent
scalar, vector and tensor components is possible. This approach is
called the SVT decomposition, it can be described in the Fourier
space and each type is able to evolve independently and treated
separately. For Vector perturbations, it can be seen from the
decomposition of metric perturbations that they are not created by
inflation and nevertheless they are resolved while the Universe
expands. So the focus is going to be on the scalar and tensor
perturbations, which are observable as density fluctuations and
gravitational waves. Depending on the comoving wavelength $k$ of a
mode it can be characterized as super-horizon when $k< \alpha H$
and sub-horizon for $k>\alpha H$, while the sub-horizon modes also
satisfy $k\gg \alpha H$ when inflation is considered to be in its
vacuum state and thus the fluctuations are produced at very scales
inside the horizon. So after a mode has exited the horizon during
its contraction, they can be described by a classical probability
distribution, whose invariance is determined by the power spectrum
at horizon crossing. Typically the condition reads
$c_\mathcal{S}k=a H$ however the model at hand predicts a sound
wave velocity equal to the speed of light. $k=\alpha H$. This is
true for a single scalar field theory in a homogeneous flat
background like the FRW spacetime. For scalar perturbations the
power spectrum is expressed as,
\begin{equation}
    \label{scalarPower}    \mathcal{P}_\mathcal{S}=\left.\dfrac{H^2}{2k^3} \dfrac{H^2}{\dot{\phi}^2}  \right|_{k=\alpha H}
\end{equation}
which relates to the primordial scalar curvature perturbations and for tensor perturbations,
\begin{equation}
    \label{tensorPower}    \mathcal{P}_\mathcal{T}=\left.\dfrac{4}{k^3} \dfrac{H^2}{M^2_{Pl}}  \right|_{k=\alpha H}
\end{equation}
which corresponds to the power spectrum of primordial
gravitational waves. The dependence of the power spectra on the
scale is described through the scalar and tensor spectral indices
respectively with,
\begin{equation}
    \label{scalarspectralindex}
    n_\mathcal{S}-1=\dfrac{d \ln{\mathcal{P}_\mathcal{S}}}{d \ln{k}}
\end{equation}
and
\begin{equation}
    \label{tensorspectralindex}
    n_\mathcal{T}=\dfrac{d \ln{\mathcal{P}_\mathcal{T}}}{d \ln{k}} \,\,.
\end{equation}
Additionally, the tensor-to-scalar ratio is defined as,
\begin{equation}
    \label{tensortoscalarratio}
    r=\dfrac{\mathcal{P}_\mathcal{T}(k)}{\mathcal{P}_\mathcal{S}(k)}
\end{equation}
in order to correlate the amplitudes of scalar and tensor
fluctuations and be able to compare them. It can be noted that
these quantities can be considered scale invariant, since their
values remain essentially unaffected under the change of scale
$k$. Under the assumption of the slow-roll approximation, for
canonical-scalar field models the power spectra can be expressed
solely with respect to the potential of the field $V(\phi)$  and
considering the definitions of the slow-roll parameters in
\eqref{SRwithV} and the relations between the two representations
\eqref{SRequiv}, we arrive in the relations of \eqref{nsr1}
\cite{reviews1},
\begin{equation}
    n_\mathcal{S}= 1-6\epsilon +2\eta \,\,\,,\,\,\,n_\mathcal{T}=-2\epsilon\,\,\,,\,\,\,r= 16 \epsilon  \,\,.
\end{equation}
For more complicated models than the ones presented previously,
there is the introduction of extra slow-roll parameters, which are
going to be included in the presentation of each case in later
sections. Also for modified gravity models, the sound speed and
the propagation speed of primordial gravitational waves is
non-trivial too.

Observations that can contribute to the evaluation of these
quantities play a detrimental role to obtaining valuable insight
for physics in the primordial Universe. The tensor-to-scalar ratio
$r$ is an auxiliary parameter and as previously mentioned, it
quantifies the ratio of the amplitude of tensor over scalar
perturbations and it is evaluated at the CMB pivot scale $k=0.002$
Mpc$^{-1}$. Some of the parameters may be scale dependent, for
example in some models the scalar spectral index may have a
non-trivial scale dependence, called ``running'', but we shall not
consider such issues here. For the scalar spectral index
$n_\mathcal{S}$, in principle it is predicted that for a
completely homogeneous Universe $n_\mathcal{S}=1$. However
perturbations, which are, quantified by the power spectrum of
\eqref{scalarPower}, result in an apparent deviation observed in
the CMB as mentioned in \eqref{nsrPlank}.  Lastly, in contrast to
the scalar spectral index $n_\mathcal{S}$, the tensor spectral
index $n_\mathcal{T}$ has not been computed yet, due to the lack
of B-modes (curl modes) in the CMB. B-modes, which are a specific
mode of polarization, can arise from the conversion of the E-mode
polarization modes to B-modes that occurs at late times or on
small angular scales, or from primordial tensor perturbations
which are the inflationary tensor modes. So a detection of such
B-modes directly give a verification for the existence of the
inflationary era.

\subsection{Non-minimally coupled Scalar field Inflation}

\noindent

There are a lot of models that can be used to describe the
inflationary era, which have a more complicated theoretical
background than the canonical minimally-coupled single scalar
field that was mentioned previously. Some models may include
further curvature correction terms with respect to the Ricci
scalar, for the coupling to gravity described as $f(R)$ gravity
corrected canonical scalar field models or include multiple
fields. A more general class of inflationary models can be
described by the following action \cite{reviews1},
\begin{equation}
    \label{sGeneral}
    \mathcal{S}= \int d^4x \sqrt{-g} \left(\dfrac{f(R,\phi)}{2 \kappa^2}-\dfrac{1}{2}\omega(\phi) g^{\mu \nu} \partial_{\mu}\phi\partial_{\nu} \phi -V(\phi) \right)  \,\,.
\end{equation}
where $f(R,\phi)$ is a smooth function of $R$ and $\phi$
indicating the non-minimal coupling to gravity and the kinetic
term $\omega(\phi)$, for which if $\omega(\phi) \neq 1$ it refers
to a non-canonical scalar field.

In this section, we focus on the canonical non-minimal coupled
model for scalar field inflation. In this case, the scalar
curvature is no longer coupled with gravity only through the
Lorentz invariant term $\sqrt{-g} d^4x$, but there is also another
term that couples the field with the scalar curvature of the form
$f(R,\phi)$, $f(R,\phi)=f(\phi)R$. This is a sub-case of the more
general class from \eqref{sGeneral}, described by the following
action \cite{reviews1},
\begin{equation}
    \label{sCNMS}
    \mathcal{S}= \int d^4x \sqrt{-g} \left(\dfrac{f(\phi)R}{2 \kappa^2}-\dfrac{1}{2}g^{\mu \nu} \partial_{\mu}\phi\partial_{\nu} \phi -V(\phi) \right)
    \,\,,
\end{equation}
with $f(\phi)$ being a dimensionless scalar coupling function. In
principle when $\phi$ reaches its vacuum expectation value, it can
become equal to unity and generate Einstein's gravity at late
times. It has to be specified that the action of \eqref{sCNMS} is
in the Jordan frame and one may choose to work in the Einstein
frame, which means rewriting the action so that a linear term of
the Ricci scalar appears as the sole contribution of curvature
while higher order curvature corrections are described by means of
a scalar field by performing a conformal transformation.
Performing a conformal transformation in order to change the scale
is not forbidden since general relativity has not an exclusive
scale. Nevertheless, the description should in essence be the same
between the two frames, when conformal invariant quantities are
considered, while also taking into consideration the differences
when imposing the slow-roll conditions.

By varying the action \eqref{sCNMS} with respect to the metric and
the scalar field $\phi$, assuming the flat FRW metric, we get the
equations of motion \cite{reviews1},
\begin{equation}
    \label{nonminEoM1}
    \dfrac{\dot{\phi}^2}{2}-\dfrac{3\dot f}{\kappa^2} H +V= \dfrac{3f}{\kappa^2} H^2\,\,,
\end{equation}
\begin{equation}
    \label{nonminEoM2}
    \dot{\phi}^2 +\dfrac{\ddot f}{\kappa^2}-H\dfrac{\dot f}{\kappa^2}+\dfrac{2f}{\kappa^2}\dot{H}=0 \,\,,
\end{equation}
\begin{equation}
    \label{nonminEoM3}
    \ddot{\phi}+3H\dot{\phi}+\dfrac{d V}{d \phi }-\dfrac{R}{2\kappa^2}\dfrac{d f}{d \phi}=0
\end{equation}
The slow roll indices in this case are defined as \cite{reviews1},
\begin{equation}
    \label{NMparameters}
    \epsilon_1 =-\dfrac{\dot{H}}{H^2} \,\,,\,\,\epsilon_2=\dfrac{\ddot{\phi}}{H\dot{\phi}} \,\,,\,\, \epsilon_3=\dfrac{\dot f}{{2Hf}} \,\,,\,\, \epsilon_4=\dfrac{\dot{E}}{2HE} \,\,,
\end{equation}
where $E$ is a function defined as
$$E=f+\frac{3 \dot f^2}{2\kappa^2 \dot{\phi}^2}\,\,.$$
The parameters $\epsilon_1$ and $\epsilon_2$ are the slow-roll
parameters also used previously in the minimally coupled scalar
field and the two new parameters $\epsilon_3$ and $\epsilon_4$
were added in light of the additional functional degree of freedom
$f(\phi)$ introduced in \eqref{sCNMS} for this case. By taking the
slow-roll assumption holds true and the slow-roll condition that
$\epsilon_i \ll 1$, $i=1,2,3,4$, then the observational quantities
can be expressed with respect to these parameters as
\cite{reviews1},
\begin{equation}
    \label{nonminNsr}
    n_\mathcal{S}\simeq 1-4\epsilon_1 -2\epsilon_2 +2\epsilon_3 -2\epsilon_4 \,\,\,, \,\,\, r=\dfrac{8\kappa^2 Q_s}{f} \,\,,
\end{equation}
where $Q_s$ is defined as the function of,
\begin{equation}
    \label{Qs}
    Q_s=\dot{\phi}^2 \dfrac{E}{f H^2 (1+\epsilon_3)^2}\,\,.
\end{equation}
Also from the imposed slow-roll condition to the parameters, the
equations of motion from \eqref{nonminEoM1},\eqref{nonminEoM2} and
\eqref{nonminEoM3} take the following form,
\begin{align}
   \label{newNMeom1}
   &\dfrac{3 f H^2}{\kappa^2}\simeq V \,\,,\\
   \label{newNMeom2}
   &3H\dot{\phi}-\dfrac{6H^2}{\kappa^2}f'+V'\simeq 0\,\,, \\
    \label{newNMeom3}
    &\dfrac{H\dot{f}}{\kappa^2}-\dfrac{2f\dot{H}}{\kappa^2}\simeq \dot{\phi}^2 \,\,.
\end{align}
Taking the slow-roll condition and \eqref{newNMeom1},\eqref{newNMeom2} and \eqref{newNMeom3} into consideration, the parameter $Q_s$ can also be approximated as,
\begin{equation}
    \label{newQs}
    Q_s\simeq \dfrac{H\dot{f}}{H^2\kappa^2}-\dfrac{2f\dot{H}}{H^2\kappa^2}\,\,,
\end{equation}
therefore from \eqref{nonminNsr}, the tensor-to-scalar ratio $r$ and the scalar spectral index $n_s$ during the slow-roll era are,
\begin{equation}
    \label{newNMr}
    r\simeq 16(\epsilon_1+\epsilon_3)\,\,,
\end{equation}
\begin{equation}
    \label{newNMns}
    n_\mathcal{S}\simeq 1-2\epsilon_1\left(\dfrac{3H\dot{f}}{\dot{\phi}^2}+2\right)-2\epsilon_2-6\epsilon_3\left(\dfrac{H\dot{f}}{\dot{\phi}^2}-1\right) \,\,\,.
\end{equation}
Now since the observational quantities of $r$ and $n_\mathcal{S}$
are expressed with the forms of \eqref{newNMr} and \eqref{newNMns}
respectively, specifically under the slow-roll condition, we are
able to analytically compute them for the duration of the
slow-roll era for any given function $f(\phi)$ \cite{reviews1}.
For the impact of the non-minimal coupling to the quantity of
tensor spectral index $n_\mathcal{T}$, included also in the
context of string corrections.

\subsubsection{Example for Specific Form of the Function $f(\phi)$}

To implement the above formalism, a specific example, which can
also be found in \cite{Odintsov:2016jwr}, for the form of the
function $f(\phi)$ is considered as follows
\cite{Odintsov:2016jwr},
\begin{equation}
    \label{exampleF}
    f(\phi)=\dfrac{1+\xi \left(e^{-\beta n \phi}+e^{-\frac{n}{\beta}\phi} \right)}{2}\,\,\,,
\end{equation}
where $\beta$ is a constant. This form of $f(\phi)$ has a special
symmetry and $\beta\rightarrow \frac{1}{\beta}$ for $\beta>1$ and
for further simplification $\xi=1$, $f(\phi)$ can be approximated
to the form of,
\begin{equation}
    \label{aproxexampleF}
    f(\phi)\simeq \dfrac{1+e^{-\frac{n}{\beta} \phi}}{2}\,\,.
\end{equation}
Additionally, the most simple form for the potential $V(\phi)$ is assumed with $V(\phi)=\Lambda$, where $\Lambda$ is a positive constant parameter.

Thus by considering the slow-roll approximation and $\kappa^2=1$ for simplicity, from \eqref{newNMeom2} we can derive the expression for $\dot{\phi}$ as,
\begin{equation}
    \label{exampleEQ1}
    \dot{\phi}\simeq -H \dfrac{n}{\beta} e^{-\frac{n}{\beta}\phi}
\end{equation}
and also for $\dot{f}$ as,
\begin{equation}
    \label{exampledotF}
    \dot{f}\simeq \dfrac{n^2 e^{-2\frac{n}{\beta}\phi}}{2}\,\,\,.
\end{equation}
Thus by taking the formulas for the slow-roll parameters from
\eqref{NMparameters} and considering the relations of
\eqref{exampleEQ1} and \eqref{exampledotF}, it is determined that,
\begin{equation}
    \label{NMnewe1}
    \epsilon_1 \simeq \dfrac{n^2 e^{-2\frac{n}{\beta}\phi}}{2 \beta^2} \,\,\,,
\end{equation}
\begin{equation}
    \label{NMnewe2}
    \epsilon_2 \simeq \dfrac{n^2}{\beta ^2} e^{-\frac{n}{\beta}\phi} +\epsilon_1 \,\,\,,
\end{equation}
\begin{equation}
    \label{NMnewe3}
    \epsilon_3\simeq \epsilon_1\,\,\,.
\end{equation}
Additionally, by substituting these relations into \eqref{newNMns}
and \eqref{newNMr}, the scalar spectral index $n_\mathcal{S}$ and
the tensor-to-scalar ratio $r$ can be computed as,
\begin{equation}
    \label{NMnewnsr}
    n_\mathcal{S}\simeq 1-2\dfrac{n^2}{\beta^2}e^{-\frac{n}{\beta}\phi} \hspace{0.1cm}, \hspace{0.1cm} r\simeq 16 \dfrac{n^2}{\beta^2}e^{-2\frac{n}{\beta}\phi}
\end{equation}
Now by using the integral \eqref{Nintegral}, the number of e-folds $N$ can be determined as,
\begin{equation}
    \label{NMefoldsN}    N=\int^{\phi_{end}}_{\phi_k}\dfrac{H}{\dot{\phi}}d\phi \simeq \dfrac{\beta^2}{n^2}e^{\frac{n}{\beta}\phi_k}
\end{equation}
where $\phi_{end}$ is the value of the field when inflation ends
and $\phi_k$ at horizon crossing. Also for the result of
\eqref{NMefoldsN} with respect to $\phi$, the approximation of
$\phi_k\gg \phi_{end}$ was used, which is justified for the
duration of the slow-roll era. So lastly, the observable
quantities of $n_\mathcal{S}$ and $r$ take the following forms
with respect to the e-folds number $N$,
\begin{equation}
    \label{NMnsrwithN}
    n_\mathcal{S}\simeq 1-\dfrac{2}{N} \hspace{0.1cm},\hspace{0.1cm} r \simeq\dfrac{16\beta^2 }{n^2 N^2} \,\,\,.
\end{equation}
An interesting result of this case, is when we select the value
$n=\frac{2}{\sqrt{3}}$, in which case the observational quantities
become $n_\mathcal{S}=1-\frac{2}{N}$ and $r\simeq \frac{12
\beta^2}{N^2}$, which is the same relations derived by
$\alpha$-attractor models \cite{Kallosh:2013yoa}. The difference
is that in the $\alpha$-attractor models $\beta\ll 1$ and so the
same attractor behavior can be exhibited by a non-minimal theory
with correctly chosen parameters. It is also worth noting that
here we have a different formalism than that of a strongly coupled
non-minimal theory and in that case the parameter $\xi$ is large
and $f(\phi)$ is chosen under a different criteria.

\section{A Brief Account of the Swampland Criteria}

As a sidenote, we shall briefly discuss, but shall not cover
explicitly in every example, the completeness of models through
the prism of the Swampland criteria. The interested reader may
check for example \cite{Vafa:2005ui,Palti:2020qlc}. for a wide
range of applications. In short, the gravitational action that one
may chose to work with can in principle be regarded as a
low-energy effective model. In order to distinguish theories based
on their UV completeness and therefore ascertain whether they
serve indeed as effective modes or not, a set of criteria can be
investigated. Let us showcase them explicitly.

The first criterion is the Swampland distance conjecture,
\begin{equation}
\centering
\label{swamp1}
|\kappa\Delta\phi|\leq\mathcal{O}(1)\, .
\end{equation}
This condition states that the field range of the scalar field
must not be arbitrary during inflation but in principle is smaller
than or equal to the Planck mass. As shown, the condition is the
same irrespective of the sign therefore the scalar field could
increase in value as time flows by. The second criterion is the de
Sitter conjecture,
\begin{equation}
\centering
\label{swamp2}
\bigg|\frac{V'}{\kappa V}\bigg|\geq\mathcal{O}(1)\, .
\end{equation}
It is applied at the start of inflation and suggests that the
slope of the scalar potential has a lower bound. The same
criterion can be written in a different form as
$-\frac{V''}{\kappa^2V}\geq\mathcal{O}(1)$. This in turn implies
that for a positive scalar potential, its form is specific with
$-V''<0$ and it also has a lower bound. These conditions can be
applied to several models. The reader should also keep in mind
that the aforementioned criteria can in principle be satisfied as
separate conditions and not simultaneously. In fact, if one can
ensure that a single criterion is satisfied then in consequence
the model belongs to the Swampland and serves an effective model
which is UV incomplete in the high energy regime. A characteristic
example is the power-law model where it was shown that while the
de Sitter conjecture is indeed satisfied, the equivalent condition
$-\frac{V''}{\kappa^2V}\geq\mathcal{O}(1)$ is not. A similar
example will be covered subsequently in the following sections
however the Swampland criteria shall not be covered in this review
but are nonetheless mentioned here for the sake of completeness.

\section{Evading the Slow-roll Evolution: The Constant-roll Evolution}

Here we shall briefly discuss a different approach on the scalar
field evolution that has interesting phenomenological implications
for the inflationary era. Previously, it was shown that under the
slow-roll assumption, the scalar field evolves slowly and
therefore issues like the apparent flatness and the horizon
problem can be explained properly. In this approach, for potential
driven inflation, it is shown that the necessary condition is the
dominance of the scalar potential over the kinetic term, i.e.
$\frac{1}{2}\dot\phi^2\ll V$. This condition, along with the
continuity equation, can be used in order to derive the additional
slow-roll condition $\dot H\ll H^2$ and $\ddot\phi\ll H\dot\phi$,
where it should be stated that these inequalities are indicative
of the order of magnitude of the respective object and not its
sign. These two conditions are not postulated but are derived from
the assumption, or from a different perspective, the necessity of
the dominance of the scalar potential. In this approach, the
scalar field is said to slowly evolve with respect to time, or the
e-foldings number.

Another assumption that can be made about the dynamics of the
scalar field is known as the constant roll condition. In this
case, the scalar field evolves approximately under the condition,
\begin{equation}
\centering
\label{constantroll}
\ddot\phi=\beta H\dot\phi\, ,
\end{equation}
where $\beta$ is an auxiliary dimensionless parameter which is not
necessarily constant for a $\phi$ dependent constant roll
condition. This evolution rate can be used as an approximation
during the inflationary era and in principle can be used along
with the slow-roll condition for $\beta\ll 1$, however it is not
necessary. One of the advantages of the constant-roll condition is
that the contribution of the second order derivative of the scalar
field is now considered in the continuity equation of the scalar
field however now the degrees of freedom are increased. In
addition, the value of the second slow-roll index
$\epsilon_2=\frac{\ddot\phi}{H\dot\phi}$ is specified completely
by the aforementioned parameter and it is constant in the
constant-roll case.

The evolution of the scalar field under the constant-roll
condition could be a leading factor in the production of
primordial scalar non-Gaussianities in the CMB, however it is not
the only factor that leaves a non-Gaussian imprints, as subsequent
cosmological eras could leave an imprint as well. In short,
anisotropies in the CMB are not only expected, but they are also
quantified by the scalar spectral index. Such anisotropies must
obey a specific distribution pattern. Information about such
pattern can be derived by examining curvature perturbations of
uniform density hypersurfaces. Theoretically, the origin of
curvature perturbations can be quantum fluctuations in cold
inflationary models or thermal fluctuations in warm inflationary
models, if not both, on superhorizon scales. Regardless of their
origin, the distribution pattern of CMB anisotropies can be
studied through the bispectrum, meaning the Fourier transform of
the three-point correlation function. As it is known, a Gaussian
distribution implies that even correlations can be written as
combinations of lower but nonetheless even correlations, i.e. the
four point correlation can be written through combinations of the
two point correlation function and so on. It is therefore expected
that the bispectrum is zero in usual Gaussian distributions. Since
secondary anisotropies are not always linear in physical systems,
one can introduce a nonlinear parameter $f_{NL}$ in order to
quantify the deviation of CMB anisotropies from a Gaussian
distribution. In principle, the distribution pattern, or
equivalently the numerical value of such nonlinear parameter,
differs if the observer used different wavelengths in order to
perform the measurement, for instance the equilateral nonlinear
term $f_{NL}^{eq}$, in which the wavelengths in momentum space are
equal. In consequence, the three point correlation function is
dominated by the scalar field dynamics primordially and is
connected to the numerical value of auxiliary parameters such as
the slow-roll indices during the first horizon crossing. In the
literature, there exist several studies that analyze such patterns
for several scalar-tensor models of gravity. The main result is
that the amount of non-Gaussianities in the CMB, or in other words
the deviation of the distribution of the CMB anisotropies from a
Gaussian distribution, is negligible as
$f_{NL}\sim\mathcal{O}(10^{-2})$ and only more involved models
that predict a propagation velocity of scalar perturbations that
clearly deviates from the speed of light can produce a larger
value, closer to the upper bounds currently available. Obviously,
the same applies to tensor perturbations as well where information
can be extracted by examining the three-point correlation function
for gravitons.

\section{String Inspired Models Of Gravity}

In this section we expand on the previously presented canonical
scalar field theory by including additional terms related to the
scalar field, originating from string corrections of the scalar
field Lagrangian. In general, the four dimensional scalar field
Lagrangian which contains at most two derivatives has the
following form,
\begin{equation}\label{generalscalarfieldaction}
\mathcal{S}_{\varphi}=\int
\mathrm{d}^4x\sqrt{-g}\left(\frac{1}{2}Z(\varphi)g^{\mu
\nu}\partial_{\mu}\varphi
\partial_{\nu}\varphi+\mathcal{V}(\varphi)+h(\varphi)\mathcal{R}
\right)\, .
\end{equation}
Note that when the scalar fields are considered in their vacuum
configuration, the scalar field has to be either conformally or
minimally coupled. When quantum corrections of the local effective
action are considered, with the quantum corrections being
consistent with diffeomorphism invariance of the action and also
contain up to fourth order derivatives, the scalar field action is
generalized to \cite{Codello:2015mba},
\begin{align}\label{quantumaction}
&\mathcal{S}_{eff}=\int
\mathrm{d}^4x\sqrt{-g}\Big{(}\Lambda_1+\Lambda_2
\mathcal{R}+\Lambda_3\mathcal{R}^2+\Lambda_4 \mathcal{R}_{\mu
\nu}\mathcal{R}^{\mu \nu}+\Lambda_5 \mathcal{R}_{\mu \nu \alpha
\beta}\mathcal{R}^{\mu \nu \alpha \beta}+\Lambda_6 \square
\mathcal{R}\\ \notag &
+\Lambda_7\mathcal{R}\square\mathcal{R}+\Lambda_8 \mathcal{R}_{\mu
\nu}\square \mathcal{R}^{\mu
\nu}+\Lambda_9\mathcal{R}^3+\mathcal{O}(\partial^8)+...\Big{)}\, ,
\end{align}
with the parameters $\Lambda_i$, $i=1,2,...,6$ being appropriate
dimensionful constants. For the purposes of this section, the
gravitational action of the model is defined as
\cite{Hwang:2005hb},
\begin{equation}
\centering
\label{Stringaction}
\mathcal{S}=\int d^4x\sqrt{-g}\left(\frac{f(\phi)R}{2\kappa^2}-\frac{1}{2}\omega(\phi)g^{\mu\nu}\nabla_\mu\phi\nabla_\nu\phi-V(\phi)-\xi(\phi)\bigg[c_1\mathcal{G}+c_2G^{\mu\nu}\nabla_\mu\phi\nabla_\nu\phi+c_3\Box\phi g^{\mu\nu}\nabla_\mu\phi\nabla_\nu\phi+c_4\bigg(g^{\mu\nu}\nabla_\mu\phi\nabla_\nu\phi\bigg)^2\bigg]\right)\, ,
\end{equation}
where $f(\phi)$, similar to the non-minimal case, is an arbitrary
dimensionless function depending on the scalar field $\phi$ while
$\xi(\phi)$ is an arbitrary function of the scalar field with
parameters $c_i$'s being auxiliary parameters with mass dimensions
of eV$^{-i+1}$ for the sake of consistency. For generality, an
additional dimensionless parameter $\omega$ is introduced so that
one can distinguish between the canonical ($\omega=1$) and phantom
($\omega=-1$) case however for the time being it shall be
considered as a dynamical variable depending solely on the scalar
field. This is the most general string inspired model that can be
introduced where for simplicity the same coupling function
$\xi(\phi)$ is considered however this is not mandatory. Indeed,
the user may feel free to change the coupling function that
accompanies each $c_i$ factor in subsequent computations.
Referring to the contributions themselves, the first term
describes the Gauss-Bonnet density $\mathcal{G}$ which serves as a
non minimal coupling between curvature and the scalar field. The
introduction of the coupling $\xi(\phi)$ is in fact quite
important because, due to the nature of the Gauss-Bonnet density,
it does not participate in the background equations as a total
derivative if it is introduced linearly in the gravitational
action, therefore the coupling function $\xi(\phi)$ keeps in place
the Gauss-Bonnet density. Of course in the literature there exist
several extensions that do not require an arbitrary coupling such
as $f(\mathcal{G})$ gravity, and even a linear model
$\alpha\mathcal{G}$ in $D$ dimensions which, upon rescaling the
auxiliary parameter $\alpha$ as $\alpha\to\frac{\alpha}{D-4}$ and
taking the limit $D\to4$, the Gauss-Bonnet density indeed
participates in the equations of motion, however we shall not
consider these examples in this brief review. The Gauss-Bonnet
model is commonly known as a low-energy effective string model,
and in an essence, when introduced, it affects accordingly not
only the background equations but also the behavior of scalar and
tensor perturbations respectively for as long as the scalar
coupling function evolves dynamically. Now the second term that is
introduced in the gravitational action (\ref{Stringaction}) is the
kinetic coupling. As the name stands, it serves as a coupling
between curvature and the kinetic term of the scalar field and
serves as an effective corrective term. It should be stated that
the inclusion of only these two terms manages to affect the
propagation velocity of tensor perturbations, therefore the model
may be at variance with recent observations such as the GW170817
event, however as we shall showcase subsequently, there exists a
way that the model can be rectified. The kinetic coupling manages
to effective shift the contribution of the kinetic term of the
scalar field and in principle it does not require a dynamical
coupling function in front, in order to participate but is
nonetheless introduced for the sake of generality. In fact, as
long as the scalar field evolves dynamically then the kinetic
coupling has an active role on the field equations. The third
contribution in (\ref{Stringaction}) is the Galilean model
\cite{Kobayashi:2010cm,Burrage:2010cu} which is a type of higher
order coupling between the scalar field and its kinetic term and
finally the last term can be interpreted as a coupling between the
scalar field with the square of its kinetic term. These
corrections refer to the kinetic term as well, however they are
more involved as they are more intricate. The Galilean term serves
as a non-minimal coupling between first and second order
variations of the scalar field and are introduced in a nonlinear
way while the final term, reminiscing of the k-essence models,
serves exactly as a higher power of the kinetic term and can be
treated either as an important contribution or higher order
corrections based on the occasion. The reason why this action is
considered to be a general case is due to the fact that all these
additions in the action, even though they result in several
inclusions in the background equations which are not only
nontrivial but also nonlinear, the continuity equation of the
scalar field still remains a second order differential equation.
Let us show this explicitly by working on the background
equations. By performing a similar work to the previous sections,
one can easily see that the field equations for gravity read
\cite{Hwang:2005hb},
\begin{equation}
\centering
\label{Stringfieldeq}
\frac{f}{\kappa^2}G_{\mu\nu}=(\nabla_\mu\nabla_\nu-g_{\mu\nu}\Box)\frac{f}{\kappa^2}+\omega\nabla_\mu\phi\nabla_\nu\phi-\bigg(\frac{1}{2}\omega g^{\alpha\beta}\nabla_\alpha\phi\nabla_\beta\phi+V\bigg)g_{\mu\nu}+T^{(string)}_{\mu\nu}\, ,
\end{equation}
where as usual the energy-stress tensor due to the presence of
string corrections is defined as
$T^{(string)}_{\mu\nu}=\frac{2}{\sqrt{-g}}\frac{\delta(\sqrt{-g}\mathcal{L}_{string})}{\delta
g^{\mu\nu}}$ where $\mathcal{L}_{string}$ is the Lagrangian
density of the $\xi$ dependent part in action
(\ref{Stringaction}). Also, the continuity equation of the scalar
field reads,
\begin{equation}
\centering
\label{conteqstring}
-\omega\Box\phi+V' -\frac{f_{,\phi}}{2\kappa^2}-\frac{1}{2}\omega_{,\phi}\nabla_\mu\phi\nabla^\mu\phi+\frac{T^{(string)}}{2}=0\, ,
\end{equation}
which as mentioned before is a second order differential equation
with respect to the scalar field. In principle, the inclusion of
additional scalar terms, either minimally or non-minimally coupled
to curvature, is not forbidden however the model at hand is the
most general case that yields second order nonlinear differential
equations. Now in this context, the contribution of the string
correction terms included in (\ref{Stringaction}) is quite lengthy
and is showcased below,
\begin{align}
\centering
\label{Tabstring}
T^{(string)}_{\mu\nu}&=2c_1\bigg\{-\xi\bigg[\frac{1}{2}\mathcal{G}g_{\mu\nu}+4R_{\mu\alpha}R^{\alpha}_\nu+4R^{\alpha\beta}R_{\mu\alpha\nu\beta}-2R_{\mu}^{\,\,\alpha\beta\gamma}R_{\nu\alpha\beta\gamma}-2RR_{\mu\nu}\bigg]-2g_{\mu\nu}(2\xi_{;\alpha\beta} R^{\alpha\beta}-\Box\xi R)\nonumber\\&+4\bigg(\xi_{;\alpha\beta} R_{\mu\alpha\nu\beta}-\Box\xi R_{\mu\nu}+2\xi_{;\alpha(\nu}R^\alpha_{\mu)}\bigg)\bigg\}+2c_2\bigg\{ \xi\bigg(2R^{\alpha}_{(\mu}\phi_{\nu)}\phi_{,\alpha}-\frac{1}{2}R_{\mu\nu}\phi^{,\alpha}\phi_{,\alpha}-\frac{1}{2}R\phi_{,\alpha}\phi_{,\beta}\bigg)+\frac{1}{2}\Box(\xi\phi_{,\mu}\phi_{,\nu})\nonumber\\&+\frac{1}{2}(\xi\phi^{,\alpha}\phi_{,\alpha})_{,\mu;\nu}-(\xi\phi^{,\alpha}\phi_{,(\mu})_{;\nu)\alpha}\bigg\}+2c_3\bigg\{-(\xi\phi^{,\alpha}\phi_{,\alpha})_{,(\mu}\phi_{,\nu)}+\xi\Box\phi\phi_{,\mu}\phi_{,\nu}+\frac{1}{2}(\xi\phi^{,\alpha}\phi_{,\alpha})_{,\beta}\phi^{,\beta}g_{\mu\nu}\bigg\}\nonumber\\&+2c_4\xi\phi^{,\alpha}\phi_{,\alpha}\bigg\{2\phi_{,\mu}\phi_{,\nu}-\frac{1}{2}\phi^{,\beta}\phi_{,\beta}g_{\mu\nu}\bigg\}\, ,
\end{align}
where for simplicity, $\phi_{,\mu}=\nabla_\mu\phi$, while the
contribution in the continuity equation is,
\begin{align}
\centering
\label{Tabstringtrace}
T^{(string)}&=2c_1\xi_{,\phi}\mathcal{G}-2c_2G^{\mu\nu}\bigg[\xi_{,\phi}\phi_{,\mu}\phi_{,\nu}+2\xi\phi_{,\mu;\nu}\bigg]+2c_3\bigg[\xi_{,\phi}\Box\phi\phi^{,\mu}\phi_{,\mu}+\Box(\xi\phi^{,\mu}\phi_{,\mu})-2(\xi\Box\phi\phi^{,\mu})_{;\mu}\bigg]\nonumber\\&+2c_4\bigg[\xi_{,\phi}(\phi^{,\mu}\phi_{,\mu})^2-4(\xi\phi^{,\mu}\phi^{,\nu}\phi_{,\nu})_{;\mu}\bigg]\, .
\end{align}
Here, it becomes abundantly clear that their contribution is
important only if the scalar field evolves dynamically, therefore
de Sitter solutions are not affected by the inclusion of string
corrections. At this point it should also be stated that the above
expressions are valid even for the case of a nonzero spatial
curvature $K$ however hereafter we shall limit our work to only
the flat case for the sake of simplicity. The generalization to
nonzero spatial curvature is relatively straightforward at the
level of background equations however it becomes tedious when
linear perturbations are considered. Let us now focus on the
equations of motion for the case of vanishing curvature. In this
case, the temporal and spatial components of the field equations
are written as \cite{Hwang:2005hb},
\begin{equation}
\centering
\label{Friedmannstring}
\frac{3fH^2}{\kappa^2}=\frac{1}{2}\dot\phi^2+V-\frac{3H\dot f}{\kappa^2}-T^{(string)0}_{\,\,\,\,\,\,\,\,\,\,\,\,\,\,\,\,\,\,\,\,\,\,0}\, ,
\end{equation}
\begin{equation}
\centering
\label{Raychaudhuristring}
-\frac{2f\dot H}{\kappa^2}=\dot\phi^2+\frac{\ddot f-H\dot f}{\kappa^2}+\frac{1}{3}T^{(string)i}_{\,\,\,\,\,\,\,\,\,\,\,\,\,\,\,\,\,\,\,\,\,\,i}\, ,
\end{equation}
where according to the results of (\ref{Tabstring}),
\begin{align}
\centering
\label{Tstring00}
T^{(string)0}_{\,\,\,\,\,\,\,\,\,\,\,\,\,\,\,\,\,\,\,\,\,\,0}&=-24c_1\dot\xi H^3+9c_2H^2\xi\dot\phi^2-c_3(\dot\xi-6H\xi)\dot\phi^3+3c_4\xi\dot\phi^4\, ,
\end{align}
\begin{align}
\centering
\label{Tstringii}
T^{(string)i}_{\,\,\,\,\,\,\,\,\,\,\,\,\,\,\,\,\,\,\,\,\,\,i}&=-24c_1(\ddot\xi H^2+2H\dot\xi(\dot H+H^2))+3c_2\dot\phi((2\dot H+3H^2)\xi\dot\phi+4H\xi\ddot\phi+2H\dot\xi\dot\phi)+3c_3\dot\phi^2(2\xi\ddot\phi+\dot\xi\dot\phi)-3c_4\xi\dot\phi^4\, ,
\end{align}
whereas (\ref{Tabstringtrace}) is identically equal to,
\begin{align}
\centering
\label{Tstring}
T^{(string)}&=2c_1\xi_{,\phi}\mathcal{G}-6c_2\bigg[H^2(\dot\xi\dot\phi+2\xi\ddot\phi)+2H\xi\dot\phi(2\dot H+3H^2)\bigg]+2c_3\dot\phi\bigg[\ddot\xi\dot\phi+3\dot\xi\ddot\phi-6\xi(\dot H\dot\phi+2H\ddot\phi+3H^2\dot\phi)\bigg]\nonumber\\&-6c_4\dot\phi^2\bigg[\dot\xi\dot\phi+4\xi\ddot\phi+4\xi\dot\phi H\bigg]\, .
\end{align}
In this approach, a homogeneous scalar field was once again
considered. Hence, it becomes clear that the inclusion of
additional string corrective terms, apart from introducing a new
degree of freedom if the non-minimal coupling function $\xi(\phi)$
is indeed dynamical, something which is mandatory for the
Gauss-Bonnet model at the very least, it results in the appearance
of several terms that evolve dynamically with respect to the
scalar field. In particular, each $c_i$ factor generates a term
proportional to $\dot\phi^i$ therefore string corrective terms
seem to introduce corrections to the kinetic term of the canonical
scalar field which in the context of higher order gravity are well
motivated. Their contribution as showcased appears in a highly
nonlinear manner as now $\ddot\phi$ is also coupled to not only
$\xi$ and its derivatives but also appropriate powers of
$\dot\phi$. Obviously the same thing applies to the case of higher
powers of the kinetic term, not just quadratic, that may be
inserted in the gravitational action. The same term also results
in the appearance of the second time derivative of the scalar
field in continuity equation, as stated by (\ref{Tstring}) hence
the reason why they were selected. In the literature, their
contribution has been thoroughly investigated, mainly in separated
models however it is not obligatory to consider only one
correction at a time. As an example, one could consider the
k-essence model that was presented previously, with or without the
inclusion of the scalar potential and combine it with a kinetic
coupling, meaning that $c_2$ is the only nonzero parameter in
(\ref{Stringaction}). In this approach, the kinetic coupling
introduces an additional $\dot\phi^2$ in the background equations
therefore phenomenologically speaking, it acts as a shift in the
kinetic term. String corrections, although they have a similar
behavior in the continuity equation of the scalar field, meaning
that they actively affect the evolution of the scalar field, their
contribution is quite different at the level of perturbations as
we shall showcase explicitly.

Let us now see how the inflationary era is described in this
context. In order to do so, we shall follow similar steps as in
the previous sections given that the previous results are
obviously subcases. Firstly, the slow-roll indices are defined as
\cite{Hwang:2005hb},
\begin{align}
\centering
\label{slowrollstring}
\epsilon_1&=-\frac{\dot H}{H^2}\, ,&\epsilon_2&=\frac{\ddot\phi}{H\dot\phi}\, ,&\epsilon_3&=\frac{\dot f}{2Hf}\, ,&\epsilon_4&=\frac{\dot E}{2HE}\, ,&\epsilon_5&=\frac{\dot Q_t}{2HQ_t}\, ,
\end{align}
where in this case a total of five indices are introduced. Here,
it should be stated that each index describes a specific part. In
particular, the first slow-roll index is specified by all
components of the gravitational action (\ref{Stringaction}) and is
completely specified by the background equations
(\ref{Friedmannstring})-(\ref{Raychaudhuristring}). It is mainly
used in order to quantify the duration of inflation and in
consequence extract information that is crucial for the
computation of the observables. The second index can be extracted
from the continuity equation of the scalar field
(\ref{conteqstring}) and, while it is affected by a possible
non-minimal coupling $f(\phi)$, it mostly captures the
contribution of the potential and the string corrective terms
however every scalar coupling function participates in principle.
In an essence, it carries information about the dynamical
evolution of the scalar field which is important for the study of
scalar perturbations. The third index addresses the contribution
of the non-minimal coupling, if existent, while the last two
indices contain both contributions. Index $\epsilon_3$, similar to
the previous non-minimal case, is indicative of the dynamical
evolution of the respective non-minimal scalar coupling function
and is used in both scalar and tensor perturbations. In addition,
$\epsilon_4$ shows the apparent dominance of the scalar functions
of the model in a compact manner and is mainly used in order to
address scalar perturbations but in principle it can be written
with respect to the rest indices. Finally the last index in
indicative of the running Planck mass, where $Q_t$ is the modified
Planck mass squared due to the presence of string corrections. It
should be stated that the fifth index is identical to the third in
the limit of vanishing string corrections therefore $\epsilon_3$
can carry information about the running of the Planck mass but
only for the non-minimal coupling. The auxiliary parameters in
this context read,
\begin{align}
\centering
\label{auxiliarystringterms}
Q_a&=-8c_1\dot\xi H+4c_2\xi\dot\phi^2H+2c_3\xi\dot\phi^3\, ,\nonumber \\
Q_b&=-16c_1\dot\xi H+2c_2\xi\dot\phi^2\, ,\nonumber\\
Q_c&=-6c_2\xi\dot\phi^2H^2+4c_3\dot\phi^3(\dot\xi-4\xi H)-12c_4\xi\dot\phi^4\, ,\nonumber\\
Q_d&=-4c_2\xi\dot\phi^2\dot H-4c_3\dot\phi^2\bigg(\dot\xi\dot\phi+\xi\ddot\phi-\xi\dot\phi H\bigg)+8c_4\xi\dot\phi^4\, \nonumber\\
Q_e&=-32c_1\dot\xi\dot H+4c_2\dot\phi\bigg(\dot\xi\dot\phi+2\xi\ddot\phi-2\xi\dot\phi H\bigg)-8c_3\xi\dot\phi^3\, ,\nonumber\\
Q_f&=16c_1(\ddot\xi-H\dot\xi)+4c_2\xi\dot\phi^2\, ,\nonumber\\
E&=\frac{f}{\kappa^2\dot\phi^2}\bigg(\omega\dot\phi^2+\frac{3(\dot f+\kappa^2Q_a)^2}{2\kappa^4Q_t}+Q_c\bigg)\, ,
\nonumber\\
Q_t&=\frac{f}{\kappa^2}+\frac{Q_b}{2}\, ,
\end{align}
which can be used in order to study scalar and tensor
perturbations. As shown, dynamical evolution of the scalar field
is crucial otherwise a finite value, at best nonzero, is obtained
and it thus results in vanishing indices. Note that the previously
defined indices, similar to the quite simpler canonical scalar
field model, can be specified and in the end, both the scalar and
tensor spectral indices as well as the tensor to scalar ratio can
be extracted. Before we proceed however with the specification of
such indices, let us study briefly the impact that string
corrections have on scalar and tensor perturbations.

Firstly, consider that the FRW metric is perturbed as,
\begin{equation}
\centering
\label{perturbedmetric}
ds^2=-a^2(1+2\alpha)d\eta^2-2a^2\beta_{,j}d\eta dx^j+a^2(g_{ij}(1+2\varphi)+2\gamma_{,i|j}+2h_{ij})dx^idx^j\, ,
\end{equation}
where $\alpha(\eta)$ is the cosmic scale factor depending on
conformal time $\eta$ specified by the condition $c\,dt=\alpha
d\eta$. Here, the conformal time is used for the sake of
simplicity. Furthermore, $\varphi$,$\alpha$, $\beta$ and $\gamma$
describe scalar perturbations whereas $h_{ij}$ tensor
perturbations and in particular traceless $g^{ij}h_{ij}=0$ and
transverse, $\partial_ih^{ij}=0$. The latter is connected to
gravitational waves and in the presence of certain string
corrections, the propagation velocity of tensor perturbations is
affected as we shall showcase subsequently. Typically, $\alpha$ is
the lapse function which specifies the connection between proper
time $\tau$ and conformal time $\eta$ while $\beta_{,j}$ is the
shift function, indicative of the apparent velocity between the
threading and the worldlines which are orthogonal to the chosen
slicing. By studying scalar perturbations using the above
perturbed metric, one can see that scalar modes propagate with a
nontrivial velocity \cite{Hwang:2005hb},
\begin{equation}
\centering
\label{cAstring}
\bigg(\frac{c_\mathcal{S}}{c}\bigg)^2=1+\frac{Q_d+\frac{(\dot f+\kappa^2Q_a)Q_e}{2\kappa^2Q_t}+\bigg(\frac{\dot f+\kappa^2Q_a}{2\kappa^2Q_t}\bigg)^2Q_f}{\omega\dot\phi^2+3\frac{(\dot f+\kappa^2Q_a)^2}{2\kappa^4Q_t}+Q_c}\, ,
\end{equation}
which obviously differs from the sound wave velocity of a perfect
fluid $c_s^2=\frac{\dot P}{\dot\rho}$. This is the most general
expression that the sound wave velocity has and as shown, the
inclusion of string corrections in the gravitational action has a
major impact on the propagation velocity of scalar perturbations.
In the end, the above expression should be well behaved for a
viable inflationary model, meaning that its numerical value during
the first horizon crossing where modes start becoming superhorizon
should be in the range $0< c_\mathcal{S}\leq c$. The fact that it
is strictly real suggests that no ghost instabilities are
initially present while the upper bound ensures that the model is
in agreement with causality. Physically speaking, while the sound
wave velocity can be equal to the speed of light, consider for
instance the canonical scalar field that was presented initially
or a non-minimal coupling between the scalar field and the Ricci
scalar, the lower-bound is purely mathematical since in a sense it
implies no propagation. Now for the tensor mode, once can show
that it is affected by string corrections as,
\begin{equation}
\centering
\label{tensormodestring}
\ddot h_{ij}+(3+a_M)H\dot h_{ij}-\bigg(\frac{c_\mathcal{T}}{c}\bigg)^2\frac{\nabla^2}{a^2}h_{ij}=0\, ,
\end{equation}
where $a_M=\frac{\dot Q_t}{HQ_t}$ is an auxiliary dimensionless
parameter which specifies the running Planck mass
\cite{Linder:2021pek} and is introduced for simplicity since in
subsequent sections a brief comment on the energy spectrum of
primordial gravitational waves shall be made. In this case it
becomes apparent that the non-minimal coupling $f(\phi)$ and
string corrections have a major impact on the behavior of tensor
perturbations. In particular, the Gauss-Bonnet density
$\xi(\phi)\mathcal{G}$ and the kinetic coupling
$\xi(\phi)G^{\mu\nu}\nabla_\mu\phi\nabla_\nu\phi$ are the only
string corrective terms in the gravitational action
(\ref{Stringaction}) that influence tensor perturbations. This can
be seen not only from the definition of $a_M$ but also from the
propagation velocity of tensor perturbations which is defined as,
\begin{equation}
\centering
\label{cTstring}
c_\mathcal{T}^2=c^2\bigg[1-\frac{Q_f}{2Q_t}\bigg]\, ,
\end{equation}
where similar to the case of the sound wave velocity
$c_\mathcal{S}$ previously discussed, it must also satisfy the
relation $0< c_\mathcal{T}\leq1$. This is a really interesting
result and is worth discussing further. Recently, the GW170817
event which was a signal from the merging of two Neutron Stars
(NS) has made it abundantly clear that gravitational waves
propagate through spacetime with the velocity of light. As a
result, the above subclasses are in peril since they predict a
propagation velocity for tensor perturbations which deviates from
that of light's. However, imposing the condition $Q_f=0$ restores
compatibility and the models can in fact be salvaged. This
condition imposes a rather strong constraint on the Gauss-Bonnet
scalar coupling function and has a significant implication on the
inflationary phenomenology. For instance, the degrees of freedom
are in fact decreased and one scalar coupling function can be
extracted from the continuity equation (\ref{conteqstring}) once
the rest have been specified. In turn, the continuity equation can
be treated either as a first order differential equation with
respect to the scalar potential or a second order differential
equation with respect to the coupling function $\xi(\phi)$, this
will become more clear in the following. Note also that this
constraint is discussed as a plausible scenario even in the early
era due to the fact that a primordial propagation velocity
$c_\mathcal{T}$ which deviates from the speed of light predicts
massive gravitons which should not be the case for this subclass
of models since they can be considered as low-energy effective
string theory models. Nevertheless, for the sake of generality,
several results shall be presented where the aforementioned
constraint is either implemented or not. The distinction will be
clear.

At this stage, the observed indices can be computed by making use
of (\ref{slowrollstring}). It should be stated that these indices
are not necessarily slow-roll indices, meaning that it is not
mandatory for their values to be approximately of order
$\mathcal{O}(10^{-3})$ and lesser even if the slow-roll assumption
is imposed but their combination should be quite small in order to
obtain results compatible with the latest (2018) Planck data. The
only conditions that need to be respected are,
\begin{align}
\centering
\label{viabilityofobservables}
-\epsilon_1-2\epsilon_2+2\epsilon_3-2\epsilon_4&\leq3&\epsilon_1-2\epsilon_3&\leq3\, ,
\end{align}
for the extraction of the scalar and tensor spectral index
respectively. These conditions do not require that
(\ref{slowrollstring}) are actually slow-roll conditions in order
to obtain a viable inflationary era, where the aforementioned
indices are of order $\mathcal{O}(1)$ and greater. In addition, in
order to extract the expressions for the spectral indices, it is
common to use the condition $\dot\epsilon_i=0$ however in Ref.
\cite{Oikonomou:2020krq} it was shown that the same results are
indeed extracted for slow-varying variables, that is
$\frac{\dot\epsilon_i}{H\epsilon_i}\leq\mathcal{O}(10^{-3})$
approximately. In the end, the spectral indices and the
tensor-to-scalar ratio obtain the following forms
\cite{Hwang:2005hb},
\begin{align}
\centering
\label{observablesstring}
n_\mathcal{S}&=1-2\frac{2\epsilon_1+\epsilon_2-\epsilon_3+\epsilon_4}{1-\epsilon_1}\, ,&n_\mathcal{T}&=-2\frac{\epsilon_1+\epsilon_5}{1-\epsilon_1}\, ,&r&=16\bigg|\bigg(\epsilon_1+\epsilon_3+\frac{\kappa^2}{4f}\bigg[\frac{2Q_c+Q_d}{H^2}-\frac{Q_e}{H}+Q_f\bigg]\bigg)\frac{f}{\kappa^2Q_t}\bigg(\frac{c_\mathcal{S}}{c_{\mathcal{T}}}\bigg)^3\bigg|\, ,
\end{align}
where everything is written as a function of the auxiliary
variables. It should be stated that this is the most inclusive
expression in terms of string corrections as it contains all the
known expressions in the literature for the observed quantities.
As an example, for the case of a minimally coupled canonical
scalar field, only indices $\epsilon_1$ and $\epsilon_2$ are
needed, which are showed are connected to the usual indices
$\epsilon$ and $\eta$ and therefore
$n_\mathcal{S}=1-6\epsilon+2\eta$, $n_\mathcal{T}=-2\epsilon$ and
$r=16|\epsilon|$ as shown before. As another example, consider the
k-essence models obtained for $c_i=0$ for $i=1,2,3$ and
$\xi(\phi)c_4=\frac{c_4}{2}$, for which after a few calculations,
it becomes clear that the tensor to scalar ratio as usual reads
$r=16c_\mathcal{S}\epsilon_1$. These forms also showcase the
previous statements made on the influence of indices $\epsilon_i$
on perturbations as now indices $\epsilon_2$-$\epsilon_4$ affect
scalar perturbations while $\epsilon_5$ tensor perturbations. On
the other hand, all auxiliary parameters needed participate in the
tensor-to-scalar ratio. Its form is general however under the
slow-roll assumption one can obtain a more simplified expression.
At this point, it should also be stated that the analysis of
scalar and tensor perturbations suggests that the respective
power-spectra obtain the following forms,
\begin{equation}
\centering
\label{stringpowerspectra1}
\mathcal{P}_{\mathcal{S}}=\frac{\kappa^2H^2}{4\pi^2}\bigg[1-\epsilon_1-(2\epsilon_1+\epsilon_2-\epsilon_3+\epsilon_4)\bigg(\gamma_E+\ln\frac{2}{1-\epsilon_1}-2\bigg)\bigg]^2\frac{H^2(1+\varepsilon)^2}{c_{\mathcal{S}}^{2\nu}(\kappa\dot\phi)^2}\frac{f}{\kappa^2E}\, ,
\end{equation}
\begin{equation}
\centering
\label{stringpowerspectra2}
\mathcal{P}_{\mathcal{T}}=\frac{\kappa^2H^2}{2\pi^2}\bigg[1-\epsilon_1-(\epsilon_1+\epsilon_4)\bigg(\gamma_E+\ln\frac{2}{1-\epsilon_1}-2\bigg)\bigg]^2\frac{1}{c_{\mathcal{T}}^{2\nu_\mathcal{T}}\kappa^2Q_t}\, ,
\end{equation}
where
$\varepsilon=\epsilon_3\frac{f}{\kappa^2Q_t}+\frac{Q_a}{2HQ_t}$,
$\nu=\frac{4-n_{\mathcal{S}}}{2}$ and
$\nu_\mathcal{T}=\frac{3-n_{\mathcal{T}}}{2}$. These are the
expressions from which the spectral indices are essentially
derived however they are presented here not only for the sake of
generality but also in order to impose constraints on the free
parameters of given models given that the amplitude of scalar
curvature perturbations $\mathcal{A}_{\mathcal{S}}$ is known.
Since string corrections affect tensor perturbations as well, it
stands to reason that the amplitude of tensor perturbations
$\mathcal{A}_{\mathcal{T}}$ could also impose further constraints
provided that it is observed in the future.

Here, it is worth making a comment on the impact of string
corrections and in particular on couplings between gravity and the
scalar field. Previously, it was shown that the Gauss-Bonnet term
$\xi(\phi)\mathcal{G}$ and the kinetic coupling
$\xi(\phi)G^{\mu\nu}\nabla_\mu\phi\nabla_\nu\phi$ manage to affect
the propagation velocity of tensor perturbations. Recall that the
latter does not require a dynamical coupling $\xi(\phi)$, in
contrast to the first. These string corrections are the only
additions in Eq. (\ref{Stringaction}) that result in a nontrivial
running Planck mass $a_M=\frac{\dot Q_t}{HQ_t}$, something which
can easily be seen from the fact that the slow-roll index
$\epsilon_5$ does not coincide with $\epsilon_3$ in those cases.
In turn, the tensor spectral index is affected by string
corrections not only indirectly through the first slow-roll index
$\epsilon_1$, but also directly thought the additional index. This
realization has interesting phenomenological implication on string
inspired models since a blue spectrum can easily be manifested
provided that the condition $\epsilon_5<-\epsilon_1$ is now
satisfied. This is an rather interesting statement since the
canonical scalar field model is incapable of producing a
blue-tilted tensor spectral index however the inclusion of
well-motivated string corrections can in fact affect tensor
perturbations to quite an extend. Indeed, in Ref.
\cite{Oikonomou:2021kql} it was shown that constrained
Gauss-Bonnet models that satisfy the condition $\ddot\xi=H\dot\xi$
can manifest a blue spectrum if certain criteria are met. As a
result, the blue spectrum leaves a major impact on the energy
spectrum of gravitational waves and in particular, in the high
frequency regime. This is because high frequency modes are the
first to re-enter the horizon, therefore they become sub-horizon
in the early Universe and can thus carry information about such
mysterious eras. The fact that a blue-tilted spectrum is a
possibility implies that the energy spectrum of high frequency
modes is amplified, compared to the general relativistic
description, and therefore it becomes easier to spot them. Indeed,
this will be shown explicitly in subsequent sections but it is
worth mentioning here since string corrections can leave a major
imprint on the energy spectrum of primordial gravitational waves.

Let us now see a few examples of string inspired models that have
been presented in the literature. For simplicity, in order to
avoid presenting a plethora of models, we shall restrict the
analysis on three models, one unconstrained and two constrained
models, with the first two revolving around the
Einstein-Gauss-Bonnet model while the last containing in addition
a kinetic coupling. This does not mean however that the rest
string corrections are not important of course.

\subsection{Gauss Bonnet Model}

For the first example, let us consider the dilaton model presented
in \cite{Odintsov:2018zhw},
\begin{align}
\centering
\label{example1EGBcouplings}
V(\phi)&=V_0\e^{-\frac{\phi}{f}}\, ,&\xi(\phi)&=\xi_0\e^{\frac{\phi}{f}}\, ,
\end{align}
where $f$ is an auxiliary parameter with mass dimensions of eV
whereas $V_0$ and $\xi_0$ serve as the potential amplitude and the
Gauss-Bonnet scalar coupling function amplitude respectively with
the former having mass dimensions of eV$^4$ while the latter is
dimensionless for the sake of consistency. In this approach, both
functions are exponential however they behave differently in the
limit $\frac{\phi}{f}\ll1$ since in this case the potential
vanishes while the Gauss-Bonnet coupling dominates. Note that the
product $V(\phi)\xi(\phi)=V_0\xi_0$ has a constant value, that is
$V(\phi)\xi(\phi)=V_0\xi_0$ and is specified completely by the
amplitudes. These types of models in which the Gauss-Bonnet scales
inverse with the potential amplitude are used in order to unify
early and late-time era, therefore it is interesting to present
the inflationary phenomenology. Note that in this approach the
propagation velocity of tensor perturbations is assumed to be
unconstrained. Another reason that makes the exponential potential
an interesting model is the fact that the first two slow-roll
index in Eq. (\ref{slowrollstring}) is now $\phi$ dependent in
contrast to the canonical scalar field case where $\epsilon_1$ and
$\epsilon_2$ are constant and cannot describe a graceful exit from
inflation. For the case at hand, $\epsilon_4$ can be used as an
index that quantifies inflation if the slow-roll assumption is
imposed therefore the inflationary era can be properly studied. In
consequence, by using index $\epsilon_4$, one can speculate that
the inflationary era ceases when the condition $\epsilon_4=1$ is
satisfied. In this case, apart from the slow-roll conditions, the
following assumptions are also made,
\begin{align}
\centering
\label{EGBexample1approx}
\ddot\xi&\ll H\dot\xi\, ,&\kappa^2\dot\xi H\ll1\, ,
\end{align}
which are motivated by the slow evolution of the scalar field
$\ddot\phi\ll H\dot\phi$. In principle, the slow-roll condition is
not mandatory in order to study the inflationary era however a
slow-varying scalar field, as mentioned before, manages to solve
both the flatness and horizon issues as a large duration can be
obtained. As a result, the scalar spectral index and the
tensor-to-scalar ratio are given by the following expressions,
\begin{align}
\centering
\label{EGBexampleobservables}
n_\mathcal{S}&=1-\frac{8\xi'^3\sqrt{V}(4(\kappa^4V)^2\xi'-3\kappa^4V')(4(\kappa^4V)^2\xi'+\kappa^4V')}{9\kappa^3\sqrt{3}}-\frac{4V''}{\kappa^2V}-\bigg(\frac{V'}{\kappa V}\bigg)^2\, ,\nonumber\\
r&=\bigg|\frac{32}{9}\kappa^6V^2\xi'^2+\frac{8}{3}\kappa^2\xi'V'+2\bigg(\frac{V'}{\kappa V}\bigg)^2\bigg|\, ,
\end{align}

\subsection{Constrained Einstein-Gauss-Bonnet Model}

For the second example we shall consider that the propagation
velocity of tensor perturbations is in fact identical to that of
light's. This condition can easily be satisfied if the
Gauss-Bonnet scalar coupling function satisfies the differential
equation,
\begin{equation}
\centering
\label{xiconstraint}
\ddot\xi=H\dot\xi\, ,
\end{equation}
regardless of the cosmological era that is studied. Therefore,
even in primordial eras, a concrete description that predicts
primordial massless gravitons can be realized. Now for simplicity
we shall assume that the scalar coupling functions of the model
are given by the following expressions,
\begin{align}
\centering
\label{example2EGBcouplings}
f(\phi)&=f_0(\kappa\phi)^n&\xi(\phi)&=\e^{-\frac{\phi}{\varphi}}\, ,
\end{align}
where $f_0$ is an auxiliary dimensionless parameter, $n$ is the
power-law exponent which is not necessarily integer and $\varphi$
is a free parameter of the model with mass dimensions of
$[\varphi]=$eV. These models were initially considered for two
reasons. Firstly, due to their respective forms, one can easily
see that $\frac{f'}{f}=\frac{n}{\phi}$,
$\frac{f''}{f}=\frac{n(n-1)}{\phi^2}$ and
$\frac{\xi''}{\xi'}=-\frac{1}{\varphi}$ therefore the inflationary
phenomenology should be quite straightforward since the
aforementioned rations participate in the slow-roll indices. The
second reason that this model is considered is because the
exponential Gauss-Bonnet scalar coupling function on its own was
proven to be inconsistent with observations under the assumption
(\ref{xiconstraint}) since, depending on the magnitude of
$\varphi$, it may lead to eternal or no inflation
\cite{Odintsov:2020sqy}. As a result, it is intriguing to examine
whether it can actually become viable if it is paired with the
simplest choice of a Ricci scalar coupling. Furthermore, the
constraint in the propagation velocity of primordial tensor
perturbations manages to decrease the degrees of freedom. In
particular, assuming that the second slow-roll index $\epsilon_2$
is actually quite lesser than unity, Eq. (\ref{xiconstraint})
yields,
\begin{equation}
\centering
\label{constraineddotphi}
\dot\phi\simeq H\frac{\xi'}{\xi''}\, ,
\end{equation}
hence the reason why the exponential Gauss-Bonnet scalar coupling
function is regarded as a convenient replacement, since $\dot\phi$
and in consequence every time derivative is simplified to a great
extend. Note that Eq. (\ref{constraineddotphi}) is valid only for
$\epsilon_2\ll1$. In turn, the $\phi$ dependence of $\dot\phi$ is
provided solely from the Hubble rate expansion and since we assume
a potential driven inflation, the extraction of the potential is
provided by the continuity equation of the scalar field
(\ref{conteqstring}). As a result, the scalar potential reads,
\begin{equation}
\centering
\label{stringexample2potential}
V(\phi)=V_0(\kappa\phi)^{2n}\e^{-\frac{\kappa\varphi}{f_0(n-1)}(\kappa\phi)^{1-n}}\, ,
\end{equation}
with $[V_0]=$eV$^4$ being the potential amplitude. As shown, the
potential is now a combination of a power-law and an exponential
function. For a specific exponent $n$ and small exponent values,
it stands to reason that the potential is in fact the linear
combination of power-laws however their exponents is not
necessarily integer as well. Non integer exponents are not a new
inclusion but they have been considered in cosmology. Now choosing
the values $\varphi=0.01M_P$, $n=\frac{1}{2}$ and $f_0=100$ for
$N=60$, implies that the exponential model accompanied by a
power-law non minimal coupling is in fact a viable inflationary
model as now $n_\mathcal{S}=0.967045$, $r=0.000551$ and
$n_\mathcal{T}=0.000069$ which are in agreement with the latest
observations. Interestingly enough, the inclusion of string
corrections can in fact result in the manifestation of a blue
tilted tensor spectral index. Obviously, this is not limited to
the non-minimal case, see for instance Ref.
\cite{Odintsov:2020sqy} where it becomes clear that having a
positive parameter $\lambda<1$ where
$\lambda=\frac{4\kappa^2\xi''V}{3}$ suffices. The results also
suggest that during the first horizon crossing, $\phi_k=0.6025M_P$
and therefore the exponential factor in the potential can be
expanded. This leads in the inclusion of additional power-law
factors such as a linear and $\phi^{\frac{3}{2}}$. This is not
limited to the above designation of free parameters and in fact
several other choices can lead to a viable inflationary era.

\subsection{Constrained Viable Horndeski Theories}

For the final model we shall consider that the action for the
model contains both terms that affect the propagation velocity of
tensor perturbations as shown below,
\begin{equation}
\centering
\label{actionexample3}
\mathcal{S}=\int d^4x\sqrt{-g}\bigg(\frac{R}{2\kappa^2}-\frac{1}{2}g^{\mu\nu}\nabla_\mu\phi\nabla_\mu\phi-V(\phi)-\xi(\phi)\bigg(\mathcal{G}+cG^{\mu\nu}\nabla_\mu\phi\nabla_\nu\phi\bigg)\bigg)\, ,
\end{equation}
where $c$ is an auxiliary parameter with mass dimensions of
eV$^{-2}$. Here, we consider both terms from Eq. (\ref{Tabstring})
that affect the propagation velocity of primordial tensor
perturbations, see equations (\ref{auxiliarystringterms}) and
(\ref{cTstring}). In principle one can include only the kinetic
coupling however the theory cannot become compatible with the
GW170817 event in that case. For the sake of consistency we shall
assume that $Q_f=0$ in this case. This in turn implies that the
differential equation that $\dot\phi$ must satisfy is given by the
following expression,
\begin{equation}
\centering
\label{thirdstringconstraint}
4(\ddot\xi-H\dot\xi)+c\xi\dot\phi^2=0\, ,
\end{equation}
where upon implementing the constant-roll condition
$\epsilon_2=\beta$, it generates the following solution,
\begin{equation}
\centering
\label{dotphistringmodel3}
\dot\phi\simeq\frac{4H\xi'(1-\beta)}{4\xi''+c\xi}\, .
\end{equation}
Obviously, in the limit $c,\beta\to0$ it becomes abundantly clear
that the previous form in Eq. (\ref{constraineddotphi}) is
recovered, as it should. Here we shall consider the constant-roll
condition in order to avoid presenting the same phenomenology over
and over again. Let us also assume that the Gauss-Bonnet scalar
coupling function has a linear form, that is,
\begin{equation}
\centering
\label{xiexample3}
\xi(\phi)=\frac{\phi}{f}\, ,
\end{equation}
where $[f]=$eV. This choice is extremely convenient as now
$\xi''=0$ and thus the time derivative of the scalar field scales
as $\dot\phi\sim\frac{1}{\phi}$. In turn, the continuity equation
of the scalar field is simplified and the resulting expression of
the scalar potential reads,
\begin{equation}
\centering
\label{Vexample3}
V(\phi)=V_0(\kappa\phi)^{-4\frac{(1-\beta)\kappa^2}{c}}\, ,
\end{equation}
where similar to the previous case, $[V_0]=$eV$^4$. Here, the
potential seems to have a power-law form with a fixed exponent,
which once again is not necessarily an integer. In particular,
having $c=-\frac{4}{M_P^2}$, $\beta=0.009$, $N=60$ and
$f=10^8M_P$, which is a set of parameters that produces results
that are compatible with observations, suggests that the exponent
of the potential is approximately $1$. Regarding the results, we
report that the spectral indices are $n_\mathcal{S}=0.968604$ and
$n_\mathcal{T}=-0.00666$ while the tensor to scalar ratio has the
value of $r=0.0531572$. Therefore the model predicts a red
spectrum and is in agreement with observations. In this approach,
it becomes clear that the potential has an almost linear form,
that is $V\sim\phi$. In other words, this model belongs to the
category $\xi(\phi)=\lambda\kappa^4V$.

\section{Chern-Simons Axionic Gravity}

In this section of the review we shall consider a really
interesting model that makes quite unique predictions for the
inflationary dynamics. Let us introduce a gravitational
Chern-Simons model of the form,
\begin{equation}
\centering
\label{CSaction}
\mathcal{S}=\int d^4x\sqrt{-g}\bigg(\frac{f(\phi)R}{2\kappa^2}-\frac{1}{2}\omega(\phi)g^{\mu\nu}\nabla_\mu\phi\nabla_\nu\phi-V(\phi)+\frac{\nu(\phi)}{8}R\tilde R\bigg)\, ,
\end{equation}
where $R\tilde
R=\eta^{\mu\nu\rho\sigma}R_{\mu\nu}^{\,\,\,\,\,\,\alpha\beta}R_{\rho\sigma\alpha\beta}$
is the Chern-Pontryagin density and for the sake of generality, an
arbitrary coupling between the scalar field and the Ricci scalar
is introduced. In an essence, it is reminiscing of the product
$F^*_{\mu\nu}F^{\mu\nu}$ which appears in fibre bundles and is
frequently named a Chern-Simons term because it is connected to a
3D Chern-Simons term cohomologically, that is $\nu(\phi)R\tilde
R=d(Chern-Simons)$, therefore the above characterization shall be
used. Both scalar coupling functions $f(\phi)$ and $\nu(\phi)$ in
this scenario as treated as dimensionless functions. It should be
stated that while the inclusion of $f(\phi)$ is not mandatory and
was considered only for illustrative purposes, the Chern-Simons
scalar coupling function $\nu(\phi)$ is of paramount importance,
similar to the case of the Gauss-Bonnet coupling. Here however,
the Chern-Simons term does not vanish identically in $D=4$ only
but due to the fact that it is a parity odd term, its contribution
on the energy momentum tensor vanishes as a whole. Truthfully,
even under the assumption that an arbitrary coupling between the
scalar field and the gravitational Chern-Simons term exists, the
background equations of motion remain unaffected and thus the
equations of motion remain exactly the same as in the case of the
non-minimally coupled canonical scalar field. This can easily be
ascertained from the definition of the energy stress tensor
$T^{(CS)}_{\mu\nu}=\frac{1}{4\sqrt{-g}}\frac{\delta(\sqrt{-g}\nu(\phi)R\tilde
R)}{\delta g^{\mu\nu}}$ which in this case the variation of the
aforementioned contribution yields \cite{Hwang:2005hb},
\begin{equation}
\centering
\label{stresstensoraxion}
T_{\mu\nu}^{(CS)}=\eta_\mu^{\,\,\,\alpha\beta\gamma}\bigg(\nu_{,\gamma;\delta}R^{\delta}_{\,\,\nu\alpha\beta}-2\nu_{,\gamma} R_{\nu\alpha;\beta}\bigg)\, ,
\end{equation}
where it becomes abundantly clear that indeed if the scalar
coupling function $\nu(\phi)$ is not dynamical then it does not
contribute. Note also that in this approach the field equations
obtain the following form,
\begin{equation}
\centering
\label{fieldeqaxion}
\frac{f}{\kappa^2}G_{\mu\nu}=(\nabla_\mu\nabla_\nu-g_{\mu\nu}\Box)\frac{f}{\kappa^2}+\omega\nabla_\mu\phi\nabla_\nu\phi-\bigg(\frac{1}{2}\omega g^{\alpha\beta}\nabla_\alpha\phi\nabla_\beta\phi+V\bigg)g_{\mu\nu}+T^{(CS)}_{\mu\nu}\, .
\end{equation}
For the sake of simplicity, let us assume that the spatial
curvature is equal to zero similar to the previous cases. In
consequence,  by taking into consideration that the metric
corresponds to a flat, isotropic and homogeneous background as
recalling that the scalar field is assumed to be time dependent
only, then the energy-stress tensor obtains the following form,
\begin{equation}
\centering
\label{stresstensoraxion2}
T^{(CS)i}_{\,\,\,\,\,\,\,\,\,\,\,\,\,j}=\frac{1}{a}\epsilon^{ik\lambda}\bigg[(\ddot\nu-H\dot\nu)\dot h_{jk,\lambda}-\dot\nu \Box h_{jk,\lambda}\bigg]+(i\leftrightarrow j)\, ,
\end{equation}
where recall that $h_{ij}(t,\bm x)$ describes traceless and
transverse tensor modes (\ref{perturbedmetric}) in the 3D surface
and $-\Box h_{ij}=\ddot h_{ij}+3H\dot
h_{ij}-\frac{\nabla^2}{a^2}h_{ij}$ due to the FRW metric. Now
owning to the fact that the totally antisymmetric Levi-Civita
tensor in 3D $\epsilon^{ijk}$ is present, it can easily be
inferred that the aforementioned tensor does not contribute in the
background equations of motion since the diagonal parts are needed
for the Friedmann and Raychaudhuri equations and of course the
same applies to the continuity equation of the scalar field. This
is a really interesting result since the inclusion of a new scalar
coupling function, namely $\nu(\phi)$, does not influence the
evolution of the scalar field and in a sense serves as a new
degree of freedom. Of course, it could be possible to implement
certain constraints on the free parameters of the scalar coupling
function $\nu(\phi)$ if the amplitude of tensor perturbations
$\mathcal{A}_{\mathcal{T}}(k)$ were to be computed at the pivot
scale, or in other words if the tensor spectral index was
computed, similar to how the potential amplitude in potential
driven inflation is constrained from the amplitude of scalar
curvature perturbations $\mathcal{A}_{\mathcal{S}}$. In the end,
the equations of motion coincide with the non-minimal case
\cite{Hwang:2005hb},
\begin{equation}
\centering
\label{axionmotion1}
\frac{3fH^2}{\kappa^2}=\frac{1}{2}\omega\dot\phi^2+V-\frac{3H\dot f}{\kappa^2}\, ,
\end{equation}
\begin{equation}
\centering
\label{axionmotion2}
-\frac{2f\dot H}{\kappa^2}=\omega\dot\phi^2+\frac{\ddot f-H\dot f}{\kappa^2}\, ,
\end{equation}
\begin{equation}
\centering
\label{axionmotion3}
\omega(\ddot\phi+3H\dot\phi)+V'+\frac{1}{2}\omega'\dot\phi^2-\frac{f'R}{2\kappa^2}=0\, ,
\end{equation}
where similar to the previous models, the prime denotes
differentiation with respect to the scalar field. The inclusion of
the gravitational Chern-Simons term however has a major impact on
tensor perturbations. In this case, by assuming that the
background is perturbed as,
\begin{equation}
\centering
\label{tensorperturbmetric}
ds^2=-c^2dt^2+a^2(t)(\delta_{ij}+h_{ij}(t,\bm x))dx^idx^j\, ,
\end{equation}
one can shown that the tensor mode must satisfy the following
equation in configuration space \cite{Hwang:2005hb},
\begin{equation}
\centering
\label{axionictensormode}
\frac{1}{a^3f}\frac{d}{dt}\bigg(a^3f\dot h_{ij}\bigg)-\frac{\nabla^2}{a^2}h_{ij}-\frac{2\kappa^2}{af}\epsilon_{(i}^{\,\,k\lambda}\bigg[(\ddot\nu-H\dot\nu)\dot h_{j)k}-\dot\nu \Box h_{j)k}\bigg]_{,\lambda}=0\, ,
\end{equation}
which as shown, it is greatly affected by the Chern-Simons scalar
coupling function. This can be shown explicitly by performing a
Fourier expansion,
\begin{equation}
\centering
\label{Cijexpand}
h_{ij}(t,\bm x)=\sqrt{\mathcal{V}}\int \frac{d^3\bm k}{(2\pi)^3}\sum_le_{ij}^{(l)}(\bm k)h_{l\bm k}(t)\e^{i\bm k\cdot\bm x}\, ,
\end{equation}
in order to move to momentum space and thus one can easily see
that Eq. (\ref{axionictensormode}) obtains the following form,
\begin{equation}
\centering
\label{tensormodemomentumspace}
\frac{1}{a^3Q_{t,l}}\frac{d}{dt}\bigg(a^3Q_{t,l}\dot h_{l\bm k}\bigg)+\bigg(\frac{c_\mathcal{T}}{c}\bigg)^2\frac{k^2}{a^2}h_{l\bm k}=0\, ,
\end{equation}
where $Q_{t,l}=\frac{f}{\kappa^2}+2\lambda_l\dot\nu\frac{k}{a}$ is
an auxiliary parameter denoting the shifted squared Planck mass
but obviously differing from the previous string corrections
definition. Here, $e_{ij}^{(l)}$ is the circular polarization of
tensor perturbations which in this approach is either left or
right handed. Furthermore, $h_{l\bm k}(t)$ describes tensor
perturbations in momentum space while $c_\mathcal{T}$ which stands
for the propagation velocity of tensor perturbations is
identically equal to the speed of light therefore the model is in
agreement with the GW170817 event. Finally, parameter $\lambda_l$
is used in order to distinguish between the two different
polarization states with $\lambda_L=-1$ and $\lambda_R=+1$. Hence,
in the gravitational Chern-Simons model, tensor perturbations
satisfy different differential equations in momentum space
depending on their polarization state and they are essentially
distinguished based on their chirality \cite{Odintsov:2022hxu}.
This parity-odd term can also be used in order to study
gravitational leptogenesis.  As a result, the different behavior
of modes based on their chirality has a major impact on the tensor
spectral index and the tensor-to-scalar ratio. In particular, it
turns out that \cite{Hwang:2005hb},
\begin{align}
\centering
\label{axionobservables}
n_\mathcal{S}&=1-2\frac{2\epsilon_1+\epsilon_2-\epsilon_3+\epsilon_4}{1-\epsilon_1}\, ,&r&=8|\epsilon_1+\epsilon_3|\sum_{l=L,R}\bigg|\frac{f}{\kappa^2Q_{t,l}}\bigg|\, ,&n_\mathcal{T}&=-2\frac{\epsilon_1+\epsilon_5}{1-\epsilon_1}\, ,
\end{align}
where,
\begin{align}
\centering
\label{axionindices}
\epsilon_1&=-\frac{\dot H}{H^2}\, ,&\epsilon_2&=\frac{\ddot\phi}{H\dot\phi}\, ,&\epsilon_3&=\frac{\dot f}{2Hf}\, ,&\epsilon_4&=\frac{\dot E}{2HE}\, ,&\epsilon_5&=\frac{1}{2}\sum_{l=L,R}\frac{\dot Q_{t,l}}{2HQ_{t,l}}\, ,
\end{align}
with $E=\frac{f}{\kappa^2}\bigg[1+\frac{3\dot
f^2}{2\kappa^2f\dot\phi^2}\bigg]$. It should be stated that the
factor of $\frac{1}{2}$ in front of $\epsilon_5$ is for averaging
the contribution of the two polarization states. It is worth
making a comment on the expression of the tensor spectral index
$n_\mathcal{T}$. Due to the fact that index $\epsilon_5$
participates, a blue tilted tensor spectral index can be obtained
only if $\epsilon_5<-\epsilon_1$ which can result in an
amplification of the energy spectrum of primordial gravitational
waves.

Before we proceed with any examples, it is worth making two
comments on the Chern-Simons model. Firstly, as demonstrated
before, the parity odd terms breaks chirality and therefore modes
with different circular polarization have a different behavior.
This difference is of course model dependent and is connected to
the Chern-Simons scalar coupling functions and the apparent
dominance of the factor $2\dot\nu\frac{k}{a}$ over the Planck mass
squared, recall the new expression for $Q_t$. This realization,
combined with the fact that a blue spectrum can in principle be
manifested seems to have interesting phenomenological
implications. Consider for instance high frequency modes that
become sub-horizon after inflation. Since the scalar field evolves
dynamically with respect to cosmic time, then chirality is broken
and therefore the difference between circular polarization states
should leave an imprint on the energy spectrum of primordial
gravitational waves. The fact that a blue-tilted tensor spectral
index is obtained simply implies that the amplitude is enhanced
compared to the GR predictions therefore it is easier, in
principle, to observe it however the same distinction is present
even if the spectrum is red, as most scalar-tensor theories
predict. Obviously, low frequency modes should not be affected
since chirality is restored provided that the scalar field has
reached its vacuum expectation value. Therefore, for the
Chern-Simons model and in the high frequency regime, one should
expect for a given frequency, two signals due to the fact that
chirality is broken and this is quite different compared to the
previous string inspired models. It can be shown that even though
the tensor spectral index is one, since it is averaged over the
circular polarizations exactly as the tensor-to-scalar ratio,
chirality is still broken and since different evolution laws must
be satisfied based on the circular polarization, the running
Planck mass $a_{M,l}=\frac{\dot Q_{t,l}}{HQ_{t,l}}$ differs
between modes meaning that the enhancement factor, which we shall
showcase explicitly in the following sections, also differs.

Another issue that one should keep in mind when working on
Chern-Simons models is the possibility of producing ghost
instabilities. In principle, the effective Chern-Simons mass scale
is a parameter which can be used in order to extract information
about ghosts. The aforementioned mass is defined as,
\begin{equation}
\centering
\label{mcs}
m_{CS}=\frac{2}{\kappa^2\dot\nu}\, ,
\end{equation}
and to no ones surprise is connected to the arbitrary scalar
coupling function $\nu(\phi)$, or more specifically its dynamical
evolution. An observant reader might notice that the denominator
appears in the definition of the auxiliary parameter $Q_{t,l}$
which is indeed the case and serves as an effective shift to the
Planck mass, depending on the polarization of the mode.
Primordially, the scalar field evolves with respect to cosmic
time, assuming that it is only homogeneous, therefore the
Chern-Simons mass scale has a finite value. As the Universe
expands and subsequently cools down, reaching equilibrium, the
scalar field itself decreases in magnitude, compared to the near
Planck scale values that it originally had, and tries to reach its
vacuum expectation value. When the scalar field reaches its
minimum value, it no longer evolves dynamically with respect to
time therefore $\dot\nu$ vanishes identically. Note that quantum
fluctuations around the vacuum expectation value are still present
however this is different from dynamical evolution. Hence, since
the vacuum expectation value has been reached, the denominator in
Eq. (\ref{mcs}) tends to zero and in consequence the effective
Chern-Simons mass scale explodes. Phenomenologically speaking,
this means that the impact of the parity odd term on the model is
lifted, therefore one obtains the GR description, or any other
modified expression that was accompanied by the gravitational
Chern-Simons term. This is in agreement with the previous
statement about low frequency modes on the energy spectrum of
gravitational waves. In order to effectively avoid the appearance
of ghost instabilities, once should make sure that the numerical
value of the Chern-Simons mass scale is in fact greater than a
given threshold during the inflationary era. Subsequent eras are
also affected but provided that $\dot\phi$ decreases in magnitude
with respect to time until it reaches zero, the above mass scale
increases in turn until it reaches infinity.  Any viable models
should respect these thresholds otherwise it is in peril as ghost
instabilities are manifested. Let us now focus on a toy models to
see what is the impact of the Chern-Simons term in the
inflationary era that we are interested in.

The simplest example and most interesting from a phenomenological
point of view is the case of chaotic inflation. Let us assume that
in the gravitational action (\ref{CSaction}) is described by
$f(\phi)=\alpha$ and a power-law potential
$V(\phi)=V_0(\kappa\phi)^n$. The power-law model is known for
being incompatible with observations because there exists no pair
of exponent $n$ and viable e-folding number $N$ that result in
acceptable values for the scalar spectral index and the
tensor-to-scalar ratio simultaneously, as it was also presented in
the canonical case as well. The gravitational Chern-Simons model
however has the advantage that only tensor perturbations are
affected by its inclusion. Therefore, for a viable e-folding
number $N=60$ and the Chaotic model $n=2$, irrespective of the
potential amplitude $V_0$, the new degree of freedom, meaning the
scalar coupling function $\nu(\phi)$, along with the auxiliary
parameter $\alpha$, can be chosen such that the a viable
tensor-to-scalar ratio is manifested. For instance, having a
linear coupling function $\nu(\phi)=\phi/f$ with $f=M_P$,
$V_0=M_P^4$ and $\alpha=1$ suggests that the chaotic model becomes
viable as the values $n_\mathcal{S}=0.96667$ and $r=0.00639$ seem
to be in agreement with the latest Planck data. For the sake of
generality, it should be mentioned that for the exact same set of
parameters but a quadratic Chern-Simons coupling, the tensor
spectral index is slightly positively defined as
$n_\mathcal{T}\simeq4\times 10^{-8}$ and increases with the
increase of the exponent of the Chern-Simons coupling. On the
other hand the tensor-to-scalar ratio decreases. Overall, it can
easily be inferred that the gravitational Chern-Simons model is
quite useful since due to the fact that tensor perturbations
behave differently depending on their polarization state, recall
Eq. (\ref{tensormodemomentumspace}), previously discarded models
can now be rectified.

\section{Vacuum $f(R)$ Gravity Inflation}

Now let us consider the description of inflationary dynamics in
the context of vacuum $f(R)$ gravity, for details we refer to Ref.
\cite{Odintsov:2020thl}. In the most general vacuum scenario with
a modified gravity of the $f(R)$ type, the gravitational action
reads,
\begin{equation}\label{fraction}
    \mathcal{S}= \frac{c^4}{16 \pi G} \int d^4x \sqrt{-g} f(R) ,
\end{equation}
and by varying with respect to the metric one obtains the gravitational field equations of motion,
\begin{equation} \label{fieldeqa}
    f_{R} R_{\mu \nu} - \frac{1}{2}g_{\mu \nu} f - \nabla_{\mu} \nabla_{\nu} f_{R} + g_{\mu \nu} \square f_{R} = 0 ,
\end{equation}
where $f_{R}=\frac{\partial f}{\partial R}$ and $\square = \nabla^a \nabla_a$.
The Ricci tensor is expressed in terms of the Christoffel symbols as,
\begin{equation}\label{ricci1a}
    R_{ab}=\partial_c \Gamma^{c}_{ab} - \partial_b \Gamma^{c}_{ac}+\Gamma^{c}_{dc}\Gamma^{d}_{ab}-\Gamma^{c}_{db}\Gamma^{d}_{ac} \, .
\end{equation}
We are using the flat FRW metric,
\begin{equation}\label{frwa}
    ds^2=-c^2dt^2+a^2(t)\left(dr^2+r^2 (d\theta^2 + \sin^2\theta d{\phi}^2) \right) \, ,
\end{equation}
for which the non-zero Ricci tensor components are,
\begin{align}\label{riccifrw}
     &R^{0}_{0}=\frac{3\ddot{a}}{a} \\ \notag
     &R^{1}_{1}=R^{2}_{2}=R^{3}_{3}=\frac{\ddot{a}}{a} + 2\left(\frac{\dot{a}}{a}\right)^2 \, .
\end{align}
The Ricci scalar is obtained by contracting the Ricci tensor with the metric tensor,
\begin{equation}\label{ricciscab}
    R=6\left(  \frac{\ddot{a}}{a} + \left( \frac{\dot {a}}{a}\right)^2 \right)\,=12H^2 +6 \dot{H} ,
\end{equation}
where,
\begin{equation}\label{H}
    H=\frac{\dot a}{a} \, ,
\end{equation}
is the Hubble rate and its first derivative with respect to time is found to be,
\begin{equation}\label{dotH}
    \dot {H}=\frac{\ddot a}{a}- {\left(\frac{\dot a}{a}\right)}^2=\frac{\ddot a}{a}-{H}^2 \, .
\end{equation}
Hence, the field equations (\ref{fieldeqa}) take the form,
\begin{equation}\label{frfieldeq1}
    -\frac{f}{2}+3({H}^2 + \dot{H}) f_{R}-18(4{H}^2 \dot{H} + H \ddot{H})f_{RR}=0 \, ,
\end{equation}

\begin{equation}\label{frfieldeq1}
    \frac{f}{2}-(3{H}^2 + \dot{H}) f_{R}+6(8{H}^2 \dot{H}+4{\dot{H}}^2 +6 H\ddot{H}+\dddot{H})f_{RR} + 36(4H \dot{H}+\ddot{H})^2 f_{RRR}=0 \, .
\end{equation}
Our ultimate goal is to find an expression for the spectral index
$n_\mathcal{S}$ and the tensor-to-scalar ratio $r$ in terms of the
slow-roll indices. For a theory without matter components the
latter read,
\begin{align}\label{fRslowrollindices}
\epsilon_1=-\frac{\dot{H}}{H^2},\,\,\,\epsilon_2=0,\,\,\,
\epsilon_3=\frac{\dot f_{R} }{2Hf_{R}},\,\,\,\epsilon_4=\frac{\ddot f_R}{H\dot f_R}\, .\\
\end{align}
The observational parameters $n_\mathcal{S}$ and $r$ are given by,
\begin{equation}\label{fRns}
    n_\mathcal{S}= 1 - \frac{4\epsilon_1 + 2\epsilon_2 -2 \epsilon_3 + 2 \epsilon_4}{
 1 -\epsilon_1} \, ,
\end{equation}
and
\begin{equation}\label{fRr}
    r=48\frac{{\epsilon_3}^2}{(1+\epsilon_3)^2} \, .
\end{equation}
Now, the relations (\ref{fRslowrollindices})-(\ref{fRr}) hold true for any model. If we further assume that the we have slow-roll inflation, the conditions,
\begin{equation}\label{slowrollH}
    \dot{H} \ll H^2 \ , \ \ddot{H} \ll H \dot{H} \, ,
\end{equation}
 correspond to $\epsilon_i \ll 1$, $i=1,3,4$ and we can use them to simplify the expressions of the slow-roll indices and observational parameters.
To this end, $\epsilon_1$ can be approximated as,
\begin{equation} \label{frepsilon1app}
    \epsilon_1=-\epsilon_3(1-\epsilon_4) \simeq -\epsilon_3 \, .
\end{equation}
Hence, the scalar spectral index can be written as,
\begin{equation}\label{frnsappr1}
    n_\mathcal{S} \simeq 1-6\epsilon_1-2\epsilon_4 \, ,
\end{equation}
and the tensor-to-scalar ratio,
\begin{equation}\label{frrappr1}
    r \simeq 48 {\epsilon_1}^2 \, .
\end{equation}
Now let us try to simplify the expression of $\epsilon_4$, which is,
\begin{equation}\label{frepsilon4}
    \epsilon_4=\frac{\ddot {f_R}}{H\dot{f_R}}=\frac{f_{RRR} {\dot{R}}^2 +f_{RR} \ddot{R}}{H f_{RR} \dot{R}} \, ,
\end{equation}
where the dot denotes derivative with respect to time. The time derivative of the Ricci scalar is,
\begin{equation}\label{frdorR}
    \dot{R}=24 \dot{H}H+6\ddot{H} \simeq 24 \dot{H}H=-24{H}^3 \epsilon_1 \, ,
\end{equation}
where we have made use of the slow-roll approximation. Hence, $\epsilon_4$ becomes,
\begin{equation}\label{frepsilon4approx1}
    \epsilon_4 \simeq- \frac{24 f_{RRR} {H}^2 }{f_{R R}}\epsilon_1 -3 \epsilon_1 + \frac{\dot {\epsilon_1}}{H \epsilon_1} \, .
\end{equation}
But $\dot \epsilon_1$ as seen by (\ref{fRslowrollindices}) is,
\begin{equation}\label{epsilon1dot}
    \dot \epsilon_1 = - \frac{\ddot {H} H^2 - 2\dot{H^2} H}{H^4} = -\frac{\ddot{H}}{H^2}+\frac{2\dot{H^2}}{H^3} \simeq 2H\epsilon_1^2 .
\end{equation}
So we can further approximate $\epsilon_4$ as,
\begin{equation}\label{frepsilon4approx2}
    \epsilon_4 \simeq - \frac{24 f_{RRR}H^2}{f_{RR}} \epsilon_1 - \epsilon_1 = -\frac{x}{2} \epsilon_1 - \epsilon_1 \, ,
\end{equation}
where we named the dimensionless parameter in front of $\epsilon_1$, $x/2$,
\begin{equation}\label{x}
    x = \frac{48 f_{RRR}H^2}{f_{RR}} \, .
\end{equation}
Hence, by combining (\ref{fRns}), (\ref{fRr}), (\ref{frepsilon4approx2}) we find,
\begin{equation}\label{nsf}
    n_\mathcal{S} = 1- 4\epsilon_1+x\epsilon_1 \ ,
\end{equation}
\begin{equation}\label{rf}
   r \simeq \frac{48(1-n_\mathcal{S})^2}{(4-x)^2} \, .
\end{equation}
Let us take a moment to discuss this result. The dimensionless
parameter $x = \frac{48 f_{RRR}H^2}{f_{R R}}$ does not have to be
constant, as it depends on the Hubble rate and its derivatives
both directly and through the Ricci scalar $R$ appearing in $f(R)$
and the derivatives, so its evolution can be calculated if one can
solve the Friedmann equations for the evolution of the Hubble rate
in the case of vacuum $f(R)$ gravity. Using the definition of
e-foldings number $N = \int_{t_k}^{t_{end}} H(t) dt$ and inverting
it, having found the evolution of $H$ from the Friedmann equations
we can express the initial horizon crossing time instance in terms
of the e-foldings $N$ number and the time instance $t_{end}$ that
corresponds to the end of inflation. However, the condition that
marks the end of the inflationary era $\epsilon_1(t_{end})=1$ can
be used to extract the final time $t_{end}$ in terms of the
model's free parameters, which we call $\sigma$. So, following
this path we end up with an expression of $t_k$ as a function of
$t_{end}(\sigma)$ and the e-foldings number $N$, which we can plug
in to the solution of the Hubble rate evolution and following to
the expression of $x$, so we will end up with $x=x(N,\sigma)$, an
expression dependent of the model's free parameters and the
e-foldings number which we know for inflation has to be around
$~60$.   The ``recipe'' described above is not so easy to follow
whatsoever, since solving the Friedmann equations is usually a
difficult task for the majority of cases even assuming
slow-rolling inflation. So instead of calculating the analytical
form of x, let us discuss about its estimated behavior in some
limiting cases, specifically,
\begin{align*}
  &\bullet\, |x|\ll1 \\
  &\bullet\, x\sim \mathcal{O}(4)\\
  &\bullet\, |x|\gg1
\end{align*}
The first case we will examine is $x \ll 1$, which also includes
the case $x=0$. In this scenario the expression for the
tensor-to-scalar ratio (\ref{rf}) can by a series expansion as,
\begin{equation}\label{rx1}
    r \simeq 3(1-n_\mathcal{S})^2 + \frac{3(1-n_\mathcal{S})^2}{2}x+\frac{9(1-n_\mathcal{S})^2}{16}x^2 \, ,
\end{equation}
which to leading order is given by,
\begin{equation}\label{rx2}
    r \simeq 3(1-n_\mathcal{S})^2 \, .
\end{equation}
This is also the expression obtained directly from setting $x=0$,
so at leading order the expansion of the $x\ll1$ case gives the
same $r-n_\mathcal{S}$  relation as the $R^2$ model ($f_{RRR}=0$).
Now, in the case where $x$ is in the order of magnitude of $1$,
then we cannot approximate the $r-n_\mathcal{S}$ relation further
than (\ref{rf}), so profoundly this case is more complex and also
blows up for $f(R)$ theories which yield $x\sim 4$, which are
obviously non-viable. However, there may be found cases for which
$x$ is in the vicinity of $4$ and yield a viable phenomenology.
For $x\gg1$ we can approximate (\ref{rf}) as,
 \begin{equation}\label{rxf}
   r \simeq \frac{48(1-n_\mathcal{S})^2}{x^2} \, .
\end{equation}
There is one problem in this case, recall that $
\epsilon_4=-\frac{x}{2}\epsilon_1-\epsilon_1$ and from the
slow-roll conditions we have $\epsilon_i\ll1$, so if $x\gg1$ then
the slow roll condition for $\epsilon_3$ does not hold true
anymore and the formulation we have described so far is not valid
in this case, although for a spectral index respecting the Planck
2018 constraints $x\gg1$ would probably yield a very small
tensor-to-scalar ratio and satisfy the Planck 2018 constraint. If
$x$ is constant, hence it does not depend on the e-foldings number
which quantifies the evolution, then the $r-n_\mathcal{S}$
relation takes the form,
\begin{equation}\label{rxconst}
    r \simeq 3\alpha(1-n_\mathcal{S})^2 \, ,
\end{equation}
where $\alpha=\frac{16}{(4-x)^2}$, which is the exact relation for
$\alpha$-attractors, $f(R)=\alpha R$ and some other string theory
motivated $f(R)$ gravities. However, this similarity does not
render the $x=$constant case viable, since if $x\gg1$ yields a
breakdown of the slow-roll conditions for the reason explained in
the previous paragraph. However, it can produce a viable
phenomenology for $x\ll1$ and for some $x\sim4$. Nonetheless, the
results are model dependent in any case.  Let us explore the
$x=$constant case a little further. The Planck 2018 data
constraint the scalar spectral index and the tensor-to-scalar
ratio as $0.962514 \pm 0.00406408$ , $r< 0.064$, and the slow-roll
index $\epsilon_1\sim\mathcal{O}(10^{-3})$. Additionally, for the
slow-roll condition $\epsilon_4\ll1$ to hold true, the maximum
value of $\epsilon_4$ can be $\sim \mathcal{O}(10^{-2})$, so
(\ref{frepsilon4approx2}) constraints $x$ to be
$x\sim\mathcal{O}(10)$ in order of magnitude. Combing the
restrictions for $n_\mathcal{S}$ and $r$ and using (\ref{rxconst})
there can be found the ranges of $x=$constant that can yield
viable slow-roll inflation era with $N\sim50-60$.

\section{ Scalar field Assisted f(R) Gravity}

In this section we present the case of an inflationary era
generated by a canonical scalar field in the presence of a $f(R)$
gravity. The action contains a gravitational term, the quadratic
kinetic term of the scalar field and its potential,
\begin{equation}\label{action}
\mathcal{S}=\int{d^4x\sqrt{-g}\left(\frac{
f(R)}{2\kappa^2}-\frac{1}{2}g^{\mu
\nu}\partial_{\mu}\phi
\partial_{\nu}\phi-V(\phi)\right)}\, .
\end{equation}
From the variation principle of the action $\delta \mathcal{S}=0$,
with respect to the metric we can obtain the Einstein field
equations and with respect to the field $\phi$ we obtain the
equation of motion for the field. The Einstein field equations for
the action (\ref{action}) read,
\begin{equation}\label{efe}
f_{R} R_{\mu \nu}-\frac{1}{2} g_{\mu \nu} f-\nabla_\mu \nabla_\nu f_{R}+g_{\mu \nu} \square f_{R}=\kappa^2 T^{(\phi)}_{\mu \nu} \, ,
\end{equation}
where,
\begin{equation}\label{set}
    T_{a b}^{(\phi)}=-\frac{2}{\sqrt{g}} \frac{\delta S_{\phi}}{\delta g^{a b}}=\partial_a \phi \partial_b \phi-g_{a b}\left(\frac{1}{2} \partial^c\phi \partial_c \phi+V(\phi)\right) \, .
\end{equation}
Now, taking into account (\ref{ricciscab}), $R=12H^2+6\dot{H}$,
and using it in the Einstein field equations, the latter reduce to
the Friedmann equations which along the field equation of motion
read,
\begin{equation}\label{freq1}
 3 f_{R} H^2= \frac{R f_{R}-f}{2}-3H \dot{f_R}+\kappa^2\left(\frac{1}{2} \dot{\phi}^2+V\right) \, ,
\end{equation}
\begin{equation}\label{freq2}
-2 f_{R} \dot{H}=\kappa^2 \dot{\phi}^2+\ddot{f}_{R}-H \dot{f}_{R} \, ,
\end{equation}
\begin{equation}\label{freom}
 \ddot{\phi}+3 H \dot{\phi}+V^{\prime}=0 \, ,
\end{equation}
where the dot denotes differentiation with respect to the cosmic time and the prime with respect to the field $\phi$. We assume slow-roll inflation, hence,
\begin{equation}\label{slowrollH}
    \dot{H} \ll H^2 \ , \ \ddot{H} \ll H \dot{H} \, ,
\end{equation}
but also the slow-rolling of the field demands that its potential
term is significantly larger than its kinetic term, but also the
kinetic part of the field does not change rapidly so that
inflation can last a sufficient amount of time, thus,
\begin{equation}\label{approxphi}
   \frac{1}{2} \dot \phi^2 \ll V(\phi) \ , \ \ddot \phi \ll H\dot
   \phi \, ,
\end{equation}
which immediately reduces the field equation of motion to,
  \begin{equation} \label{phidot}
   \dot \phi \simeq - \frac{V'}{3H} \, .
\end{equation}
Now that we have a complete set of equations, we can use the
Friedmann equations (\ref{freq1}), (\ref{freq2}) to express the
Hubble rate and its derivatives in terms of the potential
$V(\phi)$ and the fields derivatives $\dot \phi, \ddot \phi$,
which we can also express in terms of the potential through the
equation of motion (\ref{phidot}), so we ultimately have the
Hubble rate and its derivatives expressed only as a function of
$\phi$, the potential's free parameters and the function's
$f(\mathcal{R})$ free parameters. Having these, we can calculate
the expression for the slow-roll indices,
\begin{align}\label{slowrollindices}
\epsilon_1=-\frac{\dot{H}}{H^2},\,\,\,\epsilon_2=\frac{\ddot
\phi }{H \dot \phi},\,\,\,
\epsilon_3=\frac{\dot f_{R} }{2Hf_{R}},\,\,\,\epsilon_4=\frac{\dot E}{2HE}\, ,\\
\end{align}
where,
\begin{equation}\label{Edef}
    E=\frac{f_{R}}{\kappa^2}+\frac{3 \dot f_{R}^2}{2\kappa^4\dot
    \phi^2}\, ,
\end{equation}
which quantify the slow-roll inflationary evolution and since we
assume the slow-roll approximation, then the slow-roll indices
have to satisfy $\dot\epsilon_i \ll 1$, $i=1,2,3,4$. So now one
can find the scalar spectral index $n_\mathcal{S}$ and the
tensor-to-scalar ratio $r$ for such models, which are given by,
\begin{equation}\label{ns}
    n_\mathcal{S}= 1 - \frac{4\epsilon_1 + 2\epsilon_2 -2 \epsilon_3 + 2 \epsilon_4}{
 1 -\epsilon_1} \, ,
\end{equation}
\begin{equation}\label{r}
    r=16|\epsilon_1 + \epsilon_3| \, .
\end{equation}
The end of the inflationary era is characterized by
$\epsilon_1(\phi_{end})=1$, because at that instance the
derivative of the Hubble rate is zero, so the Hubble horizon stops
decreasing and will begin increasing again, marking the end of
this cosmic era. From $\epsilon_1(\phi_{end})=1$ we can then
express $\phi_{end}$ in terms of the model's free parameters, but
also using the definition of the e-foldings number,
\begin{equation}
    \centering\label{efold}
    N = \int_{t_k}^{t_{end}} {H} dt \, ,
\end{equation}
we can write it as a function of the potential and the field and
change the integration variable from cosmic time to the scalar
field $\phi$, and use it to solve for the initial value of the
field at the beginning of inflation in terms of the e-foldings
number $N$ and $\phi_{end}$, so ultimately have the initial
$\phi_k$ only as a function of the model's free parameters, which
can then use to find the slow-roll indices and observational
parameters for this $\phi$ only depending on these free
parameters. Then, demanding confrontation with the Planck 2018
constraints we can find for which values of the parameter space
these models are viable.  We mention two examples in this section,
one is a string-theory motivated corrected gravity,
\begin{equation}\label{fr1}
    f(R)=R+\frac{R^2}{36 M^2} \, ,
\end{equation}
 where M is a free parameter, and
 \begin{equation}\label{fr2}
    f(R)=\alpha R \, ,
\end{equation}
the latter is an effective modified Einstein-Hilbert action, motivated by the Lagrangian contained in the action,
\begin{equation}\label{effectiveaction}
\mathcal{S}=\int
d^4x\sqrt{-g}\left(\frac{1}{2\kappa^2}\left(\alpha {R}+ \frac{\gamma
^3 \lambda \Lambda ^3}{6 {R}^2}-\frac{\gamma ^2 \lambda \Lambda
^2}{2 {R}}-\frac{\Lambda}{\zeta
}\left(\frac{{R}}{m_s^2}\right)^{\delta
}+\mathcal{O}(1/{R}^3)+...\right)-\frac{1}{2}\partial_\mu{\phi}\partial^\mu{\phi}-V({\phi})\right)\,
,
\end{equation}
with $\alpha=1-\lambda$, but in the large curvature regime $R\rightarrow\ \infty$, at leading order is,
\begin{equation}
    \centering\label{act}
    \mathcal{S} = \int d^4 x \sqrt{- g}\left(\frac{\alpha R}{2 \kappa^2} - \frac{1}{2}g^{\mu \nu} \partial_\mu {\phi} \partial_\nu {\phi} - V \right) \, .
\end{equation}
Let us begin with the first example,
\begin{equation*}
    f(R)=R+\frac{R^2}{36M^2}\, ,
\end{equation*}
For the field equations we need the derivatives of
$f(\mathcal{R})$ with respect to the Ricci scalar and the
derivatives of the latter with respect to the cosmic time. They
read,
\begin{equation}\label{frr}\centering
    f_{R}=1+\frac{R}{18M^2} \, \ , \ f_{RR}=\frac{1}{18M^2} \, ,
\end{equation}
\begin{equation}\label{dotfr}\centering
    \dot {f_{R}}= f_{RR} \dot{R} \, \ , \ \ddot {f_{R}}=f_{RRR} \dot{R}^2 + f_{RR}\ddot{R} \, .
\end{equation}
Combining (\ref{ricciscab}), (\ref{fr2}), (\ref{frr}) and (\ref{dotfr}) result in (\ref{fr1}), (\ref{fr2}) for this model to be,
\begin{equation} \label{Friedman2}
    3H^2+\frac{3}{M^2}H^2 \dot{H}= \frac{\dot{H}^2}{2}-\frac{\ddot{H} H}{M^2}+\kappa^2\bigg(\frac{1}{2}\dot \phi^2 +V \bigg) ,
\end{equation}
\begin{equation} \label{Raycha2}
    -2 \dot{H}-\frac{2}{M^2}\dot{H}^2 = -\frac{\ddot{H} H}{M^2}+\kappa^2 \dot \phi^2 \, .
\end{equation}
We now make use of the slow-roll approximations (\ref{slowrollH}),(\ref{approxphi}) and we further assume that,
\begin{equation}\label{approxH}
    \frac{\dot{H}^2}{M^2} \ll H^2,\,\,\,\frac{\dot{H}^2}{M^2} \ll
    V(\phi)\, ,
\end{equation}
which is an assumption we are making to simplify the formulation ,
based on the fact that for usual quasi-de Sitter expansion this
holds true, but has to constantly be tested throughout
calculations.  So the field equations become,
\begin{equation} \label{Friedman3}
    3H^2+\frac{3H^2}{M^2} \dot{H}\simeq  \kappa^2 V \, ,
\end{equation}
\begin{equation} \label{Raycha3}
     -2 \dot{H}-\frac{2}{M^2}\dot{H}^2 \simeq \kappa^2 \dot \phi^2  \, .
\end{equation}
We can solve the second equation (\ref{Raycha3}) for $\dot{H}$, and we get,
\begin{equation}\label{dothanalyticsol}
\dot{H}=\frac{-M^2+M\sqrt{M^2-2\dot{\phi}^2\kappa^2}}{2}\,
,
\end{equation}
and by plugging it in (\ref{Friedman3}) we get,
\begin{equation}\label{friedaux}
\frac{3H^2}{2}-\frac{3H^2\sqrt{M^2-2\dot{\phi}^2\kappa^2}}{2M}\simeq
\kappa^2 V\, .
\end{equation}
At this point we will make another assumption, that the following approximation also holds true,
\begin{equation}\label{taylorphi}
    \frac{2\kappa^2 \dot \phi^2}{M^2} \ll 1\, ,
\end{equation}
the validity of which should also be tested as we move ahead to
calculations. Finally, we Taylor expand the term
$\sqrt{1-\frac{2\kappa^2\dot{\phi}^2}{M^2}}$ and we arrive at the
final forms of the Friedmann equations and field equation of
motion, which is our system of dynamical equations that describe
this model,
\begin{equation} \label{Friedman}
    H^2 \simeq \frac{\kappa^2 V}{3}+\mathcal{O}\bigg(\frac{\kappa^2 \dot{\phi}^2}{2}H^2\bigg),
\end{equation}
\begin{equation} \label{Raychad}
     \dot{H} \simeq -\frac{\kappa^2 \dot{\phi}^2}{2}  -\frac{\kappa^4
     \dot{\phi}^4}{4M^2}\, ,
\end{equation}
\begin{equation}\label{dotphisimpleq}
   \dot \phi \simeq - \frac{V'}{3H} \, .
\end{equation}
Making use of all the ingredients,   (\ref{frr}),
(\ref{dotfr}),(\ref{Friedman}), (\ref{Raychad}) and after some
algebra, one can find the expression of the slow-roll indices for
the model at hand, ${R}^2$ minimally coupled scalar field
inflation.
\begin{equation}\label{e1}
    \epsilon_1=\frac{1}{2\kappa^2} \left( \left(\frac{V'}{V}\right)^2 +\frac{1}{6M^2}{\left(\frac{V'}{V}\right)}^2\frac{V'^2}{V}\right)\, ,
\end{equation}
\begin{equation}\label{e2}
    \epsilon_2=-\frac{V''}{\kappa^2 V}+\epsilon_1 \, ,
\end{equation}
\begin{equation}\label{e3}
    \epsilon_3=\frac{\epsilon_1}{-1-\frac{3M^2}{2H^2}+\frac{\epsilon_1}{2}} \, .
\end{equation}
We quote the expressions for $E$ and $\dot E$,
\begin{equation}\label{E}
    E=1+\frac{2R}{36M^2}+\frac{8}{3\kappa^2 M^4}\frac{H^2 \dot{H}^2}{\dot \phi^2} \, ,
\end{equation}
\begin{equation}\label{dotE}
    \dot E=\frac{4 H \dot{H}}{3M^2} + \frac{16}{3 \kappa^2 M^4 \dot \phi^4}\big( H\dot{H}^3 \dot \phi^2-H^2 \dot{H}^2 \dot \phi \ddot \phi \big) \, ,
\end{equation}
but omit the full expression of $\epsilon_4$ since it is very long. The e-foldings number (\ref{efold}) for this model can easily be found to be,
\begin{equation}\label{NR2}
    N=\int_{\phi_{end}} ^{\phi_k} \kappa^2 \frac{V}{V'} d \phi \, .
\end{equation}
We pick a potential to perform a realistic test of this model and
apply our framework. The potential we choose is a simple power-law
potential, sometimes referred to as chaotic inflation, which does
not yield a viable phenomenology when considered in the context of
usual Einstein-Hilbert gravity,
\begin{equation}\label{V}
    V(\phi)=\frac{V_0}{\kappa^4}(\kappa \phi)^2 \, ,
\end{equation}
where $V_0$ is a  model's dimensionless free parameter. Using (\ref{e1})-(\ref{dotE}) we find the expressions of the slow-roll indices,
\begin{equation}\label{e1V}
    \epsilon_1=\frac{\frac{4}{\phi^2}+\frac{8 V_0}{3M^2\kappa^2
    \phi^2}}{2\kappa^2}\, ,
\end{equation}
\begin{equation}\label{e2V}
    \epsilon_2=\frac{4 V_0}{3 \kappa ^4 M^2 \phi ^2}\, ,
\end{equation}
\begin{equation}\label{e3V}
    \epsilon_3=\frac{4 V_0 \left(3 \kappa ^2 M^2+2
V_0\right)}{-27 \kappa ^4 M^4-6 \kappa ^2 M^2
V_0 \left(\kappa ^2 \phi ^2-1\right)+4
V_0^2}\, ,
\end{equation}
and from (\ref{ns}), (\ref{r}) the expressions for the
observational quantities can also be extracted. We omit the latter
as well as the expression for $\epsilon_4$ due to their length.
From $\epsilon_1(\phi_{end})=1$ we find
$\phi_{end}=\frac{\sqrt{\frac{2}{3}}\sqrt{2V_0+3M^2\kappa^2}}{M\kappa^2}$
and employing (\ref{NR2}), we calculate the integral and solve
with respect to $\phi_k$ and find
$\phi_k=\frac{\sqrt{\frac{2}{3}}\sqrt{2V_0+3M^2\kappa^2+6M^2N\kappa^2}}{M\kappa^2}$.
It is obvious that we have expressed $\phi_k$ only in terms of the
model's free parameters and the e-foldings number, which for
inflation is $N \sim 60$.  This model is a viable model, since
there can be found ranges of values of the free parameters $V_0$
and $\beta$, where $\beta=\kappa M$, and we set $\kappa=1$, that
satisfy the constraints on $n_s$ and $r$ imposed by the Planck
2018 mission,
\begin{equation}\label{constraints}
    n_{\mathcal{S}}=0.9649 \pm 0.0042 \ ,  \ r<0.064 \, ,
\end{equation}
but also comply with the observation that the amplitude of the primordial scalar perturbations is,
\begin{equation}\label{psconstraints}
    \mathcal{P}_{\zeta}(k)=2.19\pm0.02\times 10^{-9} \, .
\end{equation}
The amplitude is given by,
\begin{equation}\label{ampl}
    \mathcal{P}_\zeta(k)=\left(\frac{k\left(\left(-2 \epsilon_1-\epsilon_2-\epsilon_4\right)\left(0.57+\log \left(\left|\frac{1}{1-\epsilon_1}\right|\right)-2+\log (2)\right)-\epsilon_1+1\right)}{2 \pi z}\right)^2 \, ,
\end{equation}
where $z=\frac{(\dot{\phi} k) \sqrt{\frac{E}{f_{R} / \kappa^2}}}{H^2\left(\epsilon_3+1\right)}$.
Furthermore, the free parameters should also respect the validity of our assumption $\frac{\dot{H}^2}{M^2} \ll H^2,\,\,\,\frac{\dot{H}^2}{M^2} \ll
    V(\phi)$ and $\frac{2\kappa^2 \dot \phi^2}{M^2} \ll 1$.
Let us demonstrate an example, for $V_0=9.37 \times 10^{-13}$,
$\beta=6.8 \times 10^{-6}$ and $N=60$ we find
$n_{\mathcal{S}}=0.96611,r=0.063968$ and $\mathcal{P}_{\zeta}(k)=
2.19216 \times 10^{-9} $.  One noteworthy feature of this model is
that it sets an upper bound on the e-foldings number at $N=67$,
since for greater values of $N$ there cannot be found values for
the free parameters such that the constraints on the spectral
index $n_\mathcal{S}$, on the amplitude $\mathcal{P}_{\zeta}$ and
the approximation $\mathcal{P}_{\zeta}$ are satisfied
simultaneously.

Let us now apply our framework to yet another case, that of
\begin{equation*}
    f(R)=\alpha R \, ,
\end{equation*}
where $\alpha$ is a dimensionless parameter $0 < \alpha \leq 1$.
Using the slow-roll approximation again we obtain the field
equation of motion (\ref{phidot}), while the Einstein field
equations at leading order are,
\begin{equation}
    \centering\label{eom}
    \frac{\alpha}{\kappa^2}\left({R}_{\mu \nu} - \frac{1}{2}Rg_{\mu \nu}\right) = \partial_\mu {\phi} \partial_\nu {\phi} - g_{\mu \nu}\left(\frac{1}{2}g^{\rho \sigma}\partial_\rho {\phi} \partial_\sigma {\phi} +{V}({\phi}) \right) \, .
\end{equation}
Substituting our $f(R)$ in (\ref{fr1}) and (\ref{fr2}) we find the Friedmann and Raychaudhuri equations for this case,
\begin{equation}
    \centering\label{fr1ar}
    \frac{3\alpha}{\kappa^2} {H}^2 =\frac{1}{2}\dot{{\phi}}^2 + {V} \, ,
\end{equation}
\begin{equation}
    \centering\label{fr2ar}
    \frac{2 \alpha}{\kappa^2}\dot{{H}} =-\dot{{\phi}}^2 \, .
\end{equation}
Using the slow-roll approximation (\ref{slowrollH}) the first Friedmann equation is simplified to,
\begin{equation}
    \centering\label{h2}
    {H}^2 = \frac{\kappa^2}{3 \alpha}{V} \, .
\end{equation}
Now we have the full set of dynamical equations
(\ref{h2}),(\ref{fr2ar}), (\ref{phidot}) that describe the
inflationary era generated by a minimally coupled scalar field in
the presence of a $\alpha R$ gravity, so we charge forward to
calculating the slow-roll indices. This case is rather simple and
the only non-zero slow-roll indices are $\epsilon_1,\epsilon_2$,
\begin{equation}
    \centering\label{eps1}
     \epsilon_1 = -\frac{\dot{{H}}}{{H}^2}= \frac{\alpha}{2 \kappa^2}\frac{{V}'^2}{{V}^2} \, ,
\end{equation}
\begin{equation}\label{e2frv}
    \epsilon_2=-\frac{V''}{\kappa^2 V}+\epsilon_1 \, ,
\end{equation}
since,
\begin{equation}
    \centering\label{ddphi}
    \ddot{{\phi}} = - \frac{\dot{{H}}}{{H}}\dot{{\phi}} -{V}'' \frac{\dot{{\phi}}}{3 {H}} \, .
\end{equation}
The e-foldings number $N$ can be found using (\ref{efold}) and (\ref{h2}) and is,
\begin{equation}
    \centering\label{Nhil}
    N = \frac{\kappa^2}{\alpha} \int_{{\phi}_{end}}^{\phi_k} \frac{{V}}{{V}'}d {\phi} \, .
\end{equation}
Last, we need the expressions of the observational parameters, which are,
\begin{equation}
    \centering\label{spec}
    n_\mathcal{S} - 1 = -4\epsilon_1 -2\epsilon_2=1 + 2 \alpha \eta - 6 \alpha  \epsilon \, ,
\end{equation}
\begin{equation}
    \centering\label{ttsr}
    r = 8 \kappa^2 \frac{\dot{{\phi}}^2}{{H}^2}=16 \alpha  \epsilon \, .
\end{equation}
Let us mention a brief and simple paradigm. We study the potential,
\begin{equation}\label{ETV}
    V({\phi})=V_0(1-e^{-\kappa q {\phi} }) \ , \ 10^{-3}\leq q \leq 10^3  \, ,
\end{equation}
and the slow-roll indices are,
\begin{equation}\label{ETe1}
    \epsilon_1=\frac{\alpha  q^2}{2 \left(e^{
    \kappa q {\phi} }-1\right)^2} \, , \, \epsilon_2=\frac{\alpha  q^2 \left(2 e^{\kappa q {\phi} }-1\right)}{2 \left(e^{\kappa q {\phi} }-1\right)^2} \, .
\end{equation}
From $\epsilon_1(\phi)=1$ marking the end of inflation we find ${\phi}_f=\frac{\ln \left(\frac{1}{2} \left(\sqrt{2} \sqrt{\alpha } q+2\right)\right)}{\kappa q}$ and from (\ref{Nhil}) we get $N=\frac{\kappa {\phi} }{\alpha  q}-\frac{e^{\kappa q {\phi} }}{\alpha  q^2}$. We keep the exponential term only since ${\phi}$ is of order of $10-20M_{pl}$ so this is the leading term. Solving for $\phi_i$ we find expression for ${\phi}_i=\frac{\ln \left(\frac{1}{2} \left(2 \alpha  q^2 N+\sqrt{2} \sqrt{\alpha } q+2\right)\right)}{\kappa q}$. We can then express the slow-roll indices as well as $n_s$,$r$ in the initial state, which we quote since they are brief,
\begin{equation}\label{ETei}
    \epsilon_1=\frac{\alpha  q^2}{2 \left(\frac{1}{2} \left(2 \alpha  q^2 N+\sqrt{2} \sqrt{\alpha } q+2\right)-1\right)^2} \, , \,  \epsilon_2=\frac{\alpha  q^2 \left(2 \alpha  q^2 N+\sqrt{2} \sqrt{\alpha } q+1\right)}{2 \left(\frac{1}{2} \left(2 \alpha  q^2 N+\sqrt{2} \sqrt{\alpha } q+2\right)-1\right)^2} \, ,
\end{equation}
\begin{equation}\label{ETns}
    n_\mathcal{S}=-\frac{4 \alpha  q}{\sqrt{2} \sqrt{\alpha }+2 \alpha  q N}-\frac{12}{\left(2 \sqrt{\alpha } q N+\sqrt{2}\right)^2}+1 \, , \, r=\frac{32}{\left(2 \sqrt{\alpha } q N+\sqrt{2}\right)^2} \, .
\end{equation}
The expressions are rather simple in this case, so we can even
solve analytically the constraint expressions $0.962514 \pm
0.00406408$ , $r< 0.064$ and we find that for the free parameters
($\alpha,q$) that satisfy,
\begin{equation}\label{TEconx}
    0.993963 \leq \alpha q^2 \leq 1.011855 \, ,
\end{equation}
the latter are satisfied.

\section{Inflation in the Case of k-essence $f(R)$ Gravity }

The usual inflationary scenarios employ a scalar field $\phi$
whose potential $V(\phi)$ dominates over its kinetic energy
$\frac{\dot\phi^2}{2}$ forcing the Universe to an exponential type
of accelerating expansion. In this section we will describe a
class of models which yield an inflationary evolution in the
absence of a potential, but in the presence of non quadratic
kinetic terms along with the vacuum $f(R)$ gravity. So, the action
looks like,
\begin{equation}\label{kaction}
    \mathcal{S}=\int d^4x \sqrt{-g}\bigg[ \frac{1}{2\kappa^2}f(R)+G(X)\bigg] \, ,
\end{equation}
where $\kappa^2=\frac{8\pi G}{c^4}$ and $X=\frac{1}{2}\partial^{\mu}\phi \partial_{\mu} \phi$, and if the scalar field is homogeneous, thus depends only on the cosmic time $t$, then $X=-\frac{1}{2}\dot \phi^2$. The field equations and equation of motion for the metric (\ref{frwa}) read,
\begin{equation}\label{kfr1}
    -\frac{1}{2}(f-f_{R} R)-\frac{\kappa^2}{2}G_X\dot \phi^2 - 3 H \dot{f_{R}}=3f_{R} H^2 \, ,
\end{equation}
\begin{equation}\label{kfr2}
   \ddot{f_{R}}- H \dot{f_{R}} + 2\dot{H}f_{R}-\frac{\kappa^2}{2}G_X\dot{\phi^2}=0 \, ,
\end{equation}
\begin{equation}\label{keom}
   \frac{1}{a^3} \frac{d}{dt}(a^3 G_X\dot{\phi})=0\, ,
\end{equation}
where $f_{R}=\frac{\partial f(R)}{\partial R}$ ,
$G_X=\frac{\partial G}{\partial X}$. We set $\kappa^2=1$ for
simplicity and present the case of slow-roll inflation, hence, we
work in the regime where the slow-roll condition holds true,
\begin{equation}\label{kslowrollcondition}
    \ddot \phi \ll H \dot \phi \, .
\end{equation}
Let us note that in the case of $f(R)=R$ it reduces to the case of
k-essence inflation in the familiar Einstein-Hilbert gravity
framework. Now, the slow-roll indices which quantify the evolution
during the slow-rolling inflation era are given by,
\begin{equation}\label{slowrollindices}
    \epsilon_1 = \frac{\dot{H}}{H^2} \, , \, \epsilon_2 = \frac{\ddot \phi}{H\dot \phi} \, , \, \epsilon_3=\frac{\dot{f_{R}}}{2H f_{R}} \, , \, \epsilon_4 = \frac{\dot E}{2 H E} \, ,
\end{equation}
where,
\begin{equation}\label{E}
    E = -\frac{f_{R}}{2\kappa^2X} \bigg( X G_X+2X^2 G_{XX} +\frac{3\dot{f_{R}}^2}{2 \kappa^2f_{R}}\bigg) \, .
\end{equation}
We also need to express the observational quantities, i.e. the
spectral index of the scalar primordial curvature perturbations,
$n_\mathcal{S}$, and the ratio of tensor to scalar ratio, $r$, in
terms of the slow-roll indices. The latter are,
\begin{equation}\label{kns}
    n_\mathcal{S}-1=\frac{4\epsilon_1-2\epsilon_2+2\epsilon_3-2\epsilon_4}{1+\epsilon_1} \, ,
\end{equation}
\begin{equation}\label{kr}
    r=16|\epsilon_1-\epsilon_3|c_A \, ,
\end{equation}
where,
\begin{equation}\label{cA}
    c_A=\sqrt{\frac{X G_X + \frac{3 \dot{f_{R}}^2}{2\kappa^2f_{R}}}{X G_X + 2X^2 G_{XX}+\frac{3 \dot{f_{R}}^2}{2\kappa^2f_{R}}}} \, .
\end{equation}
Now the path to be followed in order to determine the viability of
a model of this class is one. After choosing a model, thus a
$f(R)$ gravity and a form for the function $G(X)$, one can use the
Friedmann equations and equation of motion
(\ref{kfr1})-(\ref{keom}) to calculate the evolution of the Hubble
rate, employing this and also using (\ref{ricciscab}) can then
find the evolution of the slow-roll indices and observational
parameters. Demanding that the latter satisfy the Planck 2018
constraints and also the slow-roll conditions $\epsilon_i\ll1$
hold true during Inflation, the values of the model's parameter
space for which it is viable can be determined. As examples we
will present two models,
\begin{equation}\label{kexaction}
    \mathcal{S}=\int d^4x \sqrt{-g} \bigg[ \frac{1}{2}f(R) \pm X \pm \frac{1}{2}f_1 X^m \bigg] \, .
\end{equation}
 In the first case,
\begin{equation}\label{kexaction1}
    \mathcal{S}=\int d^4x \sqrt{-g} \bigg[ \frac{1}{2}f(R) - X - \frac{1}{2}f_1 X^m \bigg] \, ,
\end{equation}
with $f_1>0$, so that the theory is free of ghost degrees of freedom. For this action (\ref{keom}) becomes,
\begin{align}\label{keom1}
     &3f_1 2^{-m}m H(t) \dot \phi(t) (\dot \phi (t)^2)^{m-2} -3H(t)\dot \phi (t) -\ddot \phi(t)-f_1 2^{1-m}m^2 \dot \phi(t)^2-\ddot \phi(t) \\
    &-(\dot \phi(t)^2)^{m-2}-f_12^{1-m}m\dot \phi(t)^2 \ddot \phi(t) (\dot \phi(t)^2)^{m-2}f_1+2^{-m}m\ddot \phi(t)(\dot \phi(t)^2)^{m-1}=0 \nonumber \, .
\end{align}
We can simplify this expression using the slow-roll condition
(\ref{kslowrollcondition}), which allows us to discard terms
containing higher powers of the first derivative and the second
derivative of the scalar field. Finally we have,
\begin{align}\label{keom2}
    & 3f_12^{-m}H(t)\dot \phi(t)^{2m}-3H(t)\dot\phi(t)=0 \Rightarrow \\
    & \dot \phi(t)=(2^{-m} m f_1)^{\frac{1}{1-2m}} \, ,
\end{align}
and the solution is simply found to be,
\begin{equation}\label{kphit}
    \phi(t)=(2^{-m} m f_1)^{\frac{1}{1-2m}} t \, .
\end{equation}
The fact that we can obtain explicitly the evolution of the
inflaton as a function of cosmic time, using the slow-roll
condition,  will assist in the calculation of the slow-roll
indices significantly. To move ahead we need to solve the
Friedmann equations to specify the evolution of Hubble rate, to
this end we need to choose the $f(R)$ model we want to study. For
the purpose of this review we will present the case of a $f(R)$
gravity which does not produce the desired results when considered
alone, to investigate the difference made by the presence of the
k-essence term $G(X)$ in the Lagrangian. The model we study is,
\begin{equation}
    f(R) = R + \alpha R^n \, ,
\end{equation}
where $\alpha>0$ and $n\in[\frac{1+\sqrt{3}}{2},2)$, so that we
have an inflationary acceleration and not superacceleration. For
$n=2$ we have the case of the so called Starobinsky model which
provides a viable phenomenological description of the inflationary
era, in contrast to the case of $n\in[\frac{1+\sqrt{3}}{2},2)$,
for which the constraints on the spectral index $n_\mathcal{S}$
and the tensor-to-scalar ratio $r$ imposed by the 2018 Planck
mission results cannot be satisfied simultaneously.  In the
slow-roll approximation, $\ddot{H} \ll H \dot{H}$, in essence
dismissing terms containing the second time derivative of the
Hubble rate and using the result for the evolution of the scalar
field $\phi(t)$ (\ref{kphit}) the first Friedmann equation
(\ref{kfr1}) reads,
\begin{align}\label{kfr11}
    &3n\alpha R^{n-1} H^2+\frac{\alpha (n-1)}{2} R^n - 3n(n-1)\alpha H R^{n-2}\dot R\\
    &- \frac{1}{2}(f_1 2^{-m} m )^{\frac{2}{1-2m}}(f_1 2^{-m} m (f_1 2^{-m} m)^{\frac{2(m-1)}{1-2m}}-1) =0 \nonumber \, .
\end{align}
The last term can be dismissed since it is subleading, considering
the fact that $n\in[\frac{1+\sqrt{3}}{2},2)$. Also, recall the
relation between the Ricci scalar and the Hubble rate is $R=12H^2
+6 \dot{H}$ and to leading order $\dot{R}= 24H\dot{H}$,
substituting these in (\ref{kfr11}) and solving the differential
equation one obtains the evolution of the Hubble rate with cosmic
time,
\begin{equation}\label{kHubble1}
    H(t)=\frac{1}{c_1 (t-\frac{t_i}{c_1})} \, ,
\end{equation}
where,
\begin{equation}\label{kc1}
    c_1 = \frac{2-n}{(n-1)(2n-1)} \, ,
\end{equation}
and $t_i$ is an initial time instance.  Using these results
(\ref{kphit}), (\ref{kHubble1}), we can now express the slow-roll
parameters $\epsilon_i$ (\ref{slowrollindices}) and subsequently
the observational quantities $n_\mathcal{S},r$ in term of the
model's parameters $t_i,f_1,m,n,\alpha$. In the following we quote
the expression for $\epsilon_1,\epsilon_2,\epsilon_3$, but omit
the ones for $\epsilon_4$ as well as $n_\mathcal{S},r$ since they
are very lengthy, but they can be found in closed form.
\begin{align}\label{kslowroll}
    &\epsilon_1 = -c_1 \nonumber \\
    &\epsilon_2=0 \\
    &\epsilon_3=-c_1(n-1) \nonumber \, .
\end{align}
Testing this model in terms of satisfaction of the Planck 2018
constraints $0.962514 \pm 0.00406408$ , $r< 0.064$, one can
recover viable phenomenology for a range of parameter space
values. Let us provide an example. For
($N,n,t_i$)=($60,1.36602,10^{-25}$), where $N$ is the e-foldings
number, then there can be found pairs of $f_1=[2.03291,204]$ and
$\alpha=[4.59837\times10^{-15},6\times10^{-15}]$ that result in
$n_\mathcal{S}=0.965$ and $r=0.06$, e.g for
($f_1,\alpha$)=($2.03291,4.59843\times 10^{-15}$). As can be seen
the values of $\alpha$ are restricted to a very narrow range. Let
us now present the second case of (\ref{kexaction}),
\begin{equation}\label{kexaction2}
    \mathcal{S}=\int d^4x \sqrt{-g} \bigg[ \frac{1}{2}f(R) + X + \frac{1}{2}f_1 X^m \bigg] \,  .
\end{equation}
Using again the slow-roll approximation on the corresponding field equation of motion (\ref{keom}) we get,
\begin{equation}\label{keom2}
    3H(t)\dot\phi(t)-3f_1 2^{-m}mH(t)\dot{\phi(t)}^{2m}=0 \, ,
\end{equation}
whose solution is the same as in the ghost-free case
(\ref{kphit}). The slow-roll indices obtained in this case are the
same as (\ref{kslowroll}) except $\epsilon_4$ and thereafter
calculate the expressions of the scalar spectral index and the
tensor to scalar ratio. For this model, viable phenomenology can
be yielded, but only with extreme fine tuning of the model's
parameters. We mention one example: for
($f_1,\alpha,n,m,t_i$)=($10^{-40},6.751\times10^{43},1.36602,1.4,10^{-10}$)
then, $n_\mathcal{S}=0.966$ and $r=0.0613$, which are compatible
with the Planck 2018 constraints. One detail about this model is
that by default, for $f_1>0$, due to the sign of the $+$ sign of
the kinetic term, ghosts occur, so even the viable theories that
can be described by this formulation, can only be effective
theories, since the physical theory has to be free of ghosts.

We will now present another approach to examining k-essence $f(R)$
gravity inflationary theories, that of reconstruction, which is a
procedure that can be used for inflationary evolution realization
in the slow-roll approximation in other cases as well. So, in this
approach we know the action at hand, for us (\ref{kaction}) and we
have yielded the field equations and equations of motion
(\ref{kfr1})-(\ref{keom}), as well as the evolution of the
inflaton with cosmic time (\ref{kphit}), which is the same for
both scenarios of our action. What we want to do now is
reconstruct the form of the $f(R)$ gravity and examine for which
parts of the corresponding parameter space the whole theory
confronts with the Planck 2018 constraints.  The desired evolution
of the Hubble rate in terms of the e-foldings number we would like
to realize is,
\begin{equation}\label{kHN}
    H(N)= \gamma e^{\frac{N}{4\sqrt{3}\beta}} \, ,
\end{equation}
where $ \beta$ and $\gamma$ are free parameters. We set $H^2(N)=A(N)$, so the Ricci scalar is given by,
\begin{equation}\label{kRicci}
    R(N)=12A(N)+3\dot A(N) \, ,
\end{equation}
and combining (\ref{kHN}) and (\ref{kRicci}) we find,
\begin{equation}\label{kNR}
    N(R)=2\sqrt{3}\beta \ln\bigg(\frac{2\beta R}{(24\beta+\sqrt{3})\gamma^2}\bigg) \,.
\end{equation}
Now, we have expressed the Hubble rate in terms of the e-foldings
number and the latter in terms of the Ricci scalar, in a backwards
sort of procedure, so next we can use this result in the first
Friedmann equation to express it only in terms of the Ricci scalar
and derivatives of the function $f(R)$ with respect to $R$, solve
the differential equation and obtain the desired form of $f(R)$.
The first Friedmann equation (\ref{kfr1}) by plugging in the
solution for the scalar field evolution (\ref{kphit}), which takes
into account the slow-roll condition, reads,
\begin{equation}\label{kfr13}
    -9A(N(R))(4A'(N(R)))+A''(N(R))f''(R)+(3A(N(R))+\frac{3}{2}A'(N(R)))f'(R)-\frac{f(R)}{2}+J_3^{\pm}=0
\end{equation}
where, $A'(N)=\frac{dA}{dN}$, $A''(N)=\frac{d^2A}{dN}$ and $f'(R)=\frac{df}{dR}$, while $J_3^{\pm}$ is,
\begin{equation}\label{kJ3}
    J_3^{\pm}=2 (2^{-\frac{2m}{1-2m}-1})m^{\frac{2}{1-2m}}f_1^{\frac{2}{1-2m}}(\pm 1- 2^{-\frac{2(m-1)m}{1-2m}-m}m^{\frac{2(m-1)}{1-2m}+1}f_1^{\frac{2(m-1)}{1-2m}+1}) \, ,
\end{equation}
with the $+$ sign corresponding to the phantom case
(\ref{kexaction2}) and the $-$ sign to the ghost-free case
(\ref{kexaction1}). The Friedmann equation the yields the
differential equation,
\begin{equation}\label{kfrdif}
    -\frac{(3(8\sqrt{3}\beta+1)R^2)f''(R)}{(24\beta+\sqrt{3})^2}+\frac{((12\beta+\sqrt{3})R)f'(R)}{2(24\beta+\sqrt{3})}+J_3^{\pm} \, ,
\end{equation}
whose solution is,
\begin{equation}\label{kfr}
f(R)=\mathcal{C}_1R^{\mu}+\mathcal{C}_2R^{\nu}+\frac{2(24\beta J_3^{\pm}+\sqrt{3}J_3^{\pm})}{24\beta+\sqrt{3}} \, ,
\end{equation}
where $\mathcal{C}_1,\mathcal{C}_2$ are constants coming from integration and the parameters $\mu,\nu$ are,
\begin{align}\label{munu}
    &\mu=\frac{96\beta^2+\frac{\sqrt{24\beta+\sqrt{3}}\sqrt{384\sqrt{3}\beta^3-912\beta^2-32\sqrt{3}\beta+1}}{3^{2}}+28\sqrt{3}\beta+3}{32\sqrt{3}\beta+4} \\
    &\nu=\frac{96\beta^2-\frac{\sqrt{24\beta+\sqrt{3}}\sqrt{384\sqrt{3}\beta^3-912\beta^2-32\sqrt{3}\beta+1}}{3^{2}}+28\sqrt{3}\beta+3}{32\sqrt{3}\beta+4} \nonumber \, .
\end{align}
Since we know the expression of the Hubble rate evolution we want
(\ref{kHN}), we can easily express the Ricci scalar in terms of
the e-foldings number using (\ref{kRicci}) and we also know the
explicit form of our $f(R)$ function, which can in turn express in
terms of $N$., we can then use this to calculate the slow-roll
indices and as well as the spectral index and the tensor to scalar
ratio. For this purpose we note that since,
\begin{equation}
    \frac{d}{dt}=H\frac{d}{dN} \, ,
\end{equation}
the slow-roll indices can be expressed in terms of the e-foldings number $N$ as,
\begin{align}\label{ksr}
    &\epsilon_1=\frac{H'(N)}{H(N)} \, , \nonumber \\
    &\epsilon_2=0 \, , \\
    &\epsilon_3=\frac{12H(N)H'(N)\frac{d^2f_{R}}{dR(R(N))^2}}{f_{R(R(N))}} \nonumber \, .
\end{align}
and,
\begin{equation}\label{knsr}
    n_\mathcal{S}-1=\frac{2(2\epsilon_1+\epsilon_3-\epsilon_4)}{\epsilon_1+1} \, , \, r= 16c_A|\epsilon_1-\epsilon_3| \, ,
\end{equation}
with
\begin{equation}
    c_A=\sqrt{\frac{\frac{864H(N)^4 H'(N)^2\left(\frac{d^2 f_{R}(R(N))}{d R^2}\right)^2}{f_{R}(N)}+J_1}{\frac{864 H(N)^4 H'(N)^2\left(\frac{d^2 f_{R}(R(N))}{d R^2}\right)^2}{f_{R}(R(N))}+J_1+J_2}} \, ,
\end{equation}
where,
\begin{align}
& J_1=f_1 2^{-m-1} m\left(\left(f_1 2^{-m} m\right)^{\frac{2}{1-2 m}}\right)^m-\frac{1}{2}\left(f_1 2^{-m} m\right)^{\frac{2}{1-2 m}} \\
& J_2=2^{-\frac{2 m^2}{1-2 m}-m}(m-1) m^{\frac{2 m}{1-2 m}+1} f_1^{\frac{2 m}{1-2 m}+1} \nonumber \, ,
\end{align}
we omit the expression of $\epsilon_4$ for brevity. Having all the
ingredients, using (\ref{kHN}), (\ref{ksr}), (\ref{kfr}) and
(\ref{knsr}) we can find the closed form for the slow-roll
parameters and the observational quantities for both of the models
at hand. The free parameters are ($f_1,\gamma,\beta,m,N$), so by
investigation on the parameter space one can determine the
viability of each model.  For the phantom model, there can be
found ranges of values of the free parameters that yield a viable
phenomenology. For example, for ($N,\beta,\gamma,f_1,m$)=($60,
1.4983, 0.001,1.2470,1.2$) we get $n_\mathcal{S}=0.964894$ and
$r=0.0179065$. However, for the ghost-free model there cannot be
found values of the free parameters that satisfy the constraints
on $n_\mathcal{S}$,$r$ simultaneously, so we cannot realize a
viable phenomenology with Hubble rate evolution of (\ref{kHN})
with this model.

\section{Generalized $f(R,\phi,X)$ gravity}

At this stage, it becomes abundantly clear that the overall
phenomenology described previously can be written in a compact
manner. This realization simplifies the analysis to quite an
extend and also paves the wave to other theoretical models that
are not presented here. Let us consider that the gravitational
action is written as,
\begin{equation}
\centering
\label{frphixaction}
\mathcal{S}=\int d^4x\sqrt{-g}\bigg(\frac{f(R,\phi,X)}{2\kappa^2}+\mathcal{L}^{(c)}\bigg)\, ,
\end{equation}
where $f(R,\phi,X)$ is an arbitrary function of the Ricci scalar,
the scalar field and the kinetic term
$X=\frac{1}{2}g^{\mu\nu}\nabla_\mu\phi\nabla_\nu\phi$ while
$\mathcal{L}^{(c)}$ is the Lagrangian density for corrections such
as string corrections and axionic strings as presented before.
This choice of action corresponds to the following set of
equations,
\begin{equation}
\centering
\label{frphixfieldeq}
f_{,R}G_{\mu\nu}=-(f_{,R}R-f)g_{\mu\nu}+f_{,R,\mu;\nu}-g_{\mu\nu}\Box f_{,R}-\frac{1}{2}f_{,X}\nabla_\mu\phi\nabla_\nu\phi+\kappa^2T^{(c)}_{\mu\nu}\, ,
\end{equation}
\begin{equation}
\centering
\label{frphixKG}
\nabla_\mu(f_{,X}\nabla^\mu\phi)-f_{,\phi}+T^{(c)}=0\, .
\end{equation}
As an example, one can obtain the canonical scalar field case for
$f(R,\phi,X)=R-2\kappa^2(X+V)$, the k-essence case for
$f(R,\phi,X)=R-2\kappa^2(X+\omega X^2)$. Also, for the
Chern-Simons model, one simply needs to replace
$T^{(c)}=0=T^{(c)\mu}_{\,\,\,\,\,\,\,\,\mu}$. In this unified way,
the background equations read,
\begin{equation}
\centering
\label{Friedmannfrphix}
\frac{3f_{,R}H^2}{\kappa^2}=\frac{f_{,X}X}{\kappa^2}+\frac{f_{,R}R-f}{2\kappa^2}-\frac{3H\dot f_{,R}}{\kappa^2}-T^{(c)0}_{\,\,\,\,\,\,\,\,0}\, ,
\end{equation}
\begin{equation}
\centering
\label{Raychaudhurifrphix}
-\frac{2f_{,R}\dot H}{\kappa^2}=\frac{f_{,X}X}{\kappa^2}+\frac{\ddot f_{,R}-H\dot f_{,R}}{\kappa^2}+\frac{1}{3}T^{(c)\alpha}_{\,\,\,\,\,\,\,\,\alpha}\, ,
\end{equation}
and,
\begin{equation}
\centering
\label{conteqfrphix}
\frac{1}{a^3\kappa^2}\frac{d}{dt}\bigg(a^3f_{,X}\dot\phi\bigg)+\frac{f_{,\phi}}{\kappa^2}=T^{(c)}\, ,
\end{equation}
where the components of the energy stress tensor for string
corrections are given by the expression
$T^{(c)}_{\mu\nu}=\frac{1}{2\sqrt{-g}}\frac{\delta(\sqrt{-g}\mathcal{L}^{(c)})}{\delta
g^{\mu\nu}}$ and it could be either the contribution from string
corrections studied in (\ref{Tabstring}) or from the gravitational
Chern-Simons case (\ref{stresstensoraxion}). In the latter case,
the equation of motion remain unaffected obviously and only tensor
perturbations are affected as showcased before. In this approach,
the $f(R,\phi,X)$ model can be used in order to study non
canonical theories such as Dirac-Born-Infeld or the inclusion of
higher order kinetic terms. This description is convenient when
studying primordial scalar non Gaussianities under the
constant-roll assumption since the sound wave velocity in this
case, which reads,
\begin{equation}
\centering
\label{cafrphix}
c_\mathcal{S}^2=c^2\frac{Xf_{,X}+\frac{3\dot f^2}{2\kappa^2f}}{Xf_{,X}+2X^2f_{,XX}+\frac{3\dot f^2}{2\kappa^2 f}}\, ,
\end{equation}
could obtain a quite small value and thus the equilateral
nonlinear term $f_{NL}^{eq}$ can be enhanced to the point where it
may be detectable in subsequent experiments. For an arbitrary
model, the sound wave velocity, as mentioned before, should be
well-behaved, meaning that it satisfies the condition
$0<c_\mathcal{A}\leq1$. In addition, the slow-roll indices that
are required in order to study the inflationary era are,
\begin{align}
\centering
\label{slowrollindicesfrphix}
\epsilon_1&=-\frac{\dot H}{H^2}\, ,&\epsilon_2&=\frac{\ddot\phi}{H\dot\phi}\, ,&\epsilon_3&=\frac{\dot f_{,R}}{2Hf_{,R}}\, ,&\epsilon_4&=\frac{\dot E}{2HE}\, ,&\epsilon_5&=\frac{\dot Q_t}{2HQ_t}\, ,
\end{align}
where all the auxiliary parameters can be taken from Eq.
(\ref{auxiliarystringterms}) with the only change being $f\to
f_{,R}$. Note that for the $f(R,\phi,X)$ model without string
corrections, the auxiliary parameter $E$ that participates in the
4-th slow-roll index is given by the following expression,
\begin{equation}
\centering
\label{Efrphix}
E=-\frac{f_{,R}}{2\kappa^2X}\bigg[Xf_{,X}+2X^2f_{,XX}+\frac{3\dot f_{,R}^2}{2\kappa^2f_{,R}}\bigg]\, ,
\end{equation}
which, for the case of $f(R,\phi,X)=f(R,\phi)-2\kappa^2(X+\omega
X^2+V)$, it obviously coincides with the result obtained from
action (\ref{Stringaction}) for $c_1=0=c_2=c_3$ and
$\xi(\phi)c_4\to\frac{\omega}{4}$, with $\omega$ being $\phi$
independent. Furthermore, if the Chern-Simons model is considered,
then the slow-roll index $\epsilon_5$ should be replaced as
$\epsilon_5=\frac{1}{2}\sum_{l=L,R}\frac{\dot Q_{t,l}}{2HQ_{t,l}}$
with $Q_{t,l}$ now being equal to
$Q_{t,l}=\frac{f_{,R}}{\kappa^2}+2\lambda_l\dot\nu\frac{k}{a}$. In
the end, the spectral indices have the following form ( for the
tensor-to-scalar-ratio see Eq.(\ref{observablesstring}) for the
former case or Eq.(\ref{axionobservables}) for the latter),
\begin{align}
\centering
\label{generalobservables}
n_\mathcal{S}&=1-2\frac{2\epsilon_1+\epsilon_2-\epsilon_3+\epsilon_4}{1-\epsilon_1}\, ,&n_\mathcal{T}&=-2\frac{\epsilon_1+\epsilon_5}{1-\epsilon_1}\, .
\end{align}
In the following we shall present certain examples based on this
formalism, describe in detail the results obtained and showcase
certain differences between the models at hand.

\subsection{Kinetic Axion $f(R)$ Model}

Let us start by presenting the results from the kinetic axion
$f(R)$ model. This model has an intriguing phenomenology due to
the axion dynamics and its impact on both the inflationary era and
subsequent cosmological eras. For this model, the gravitational
action reads,
\begin{equation}
\centering
\label{kineticactionmodel1}
\mathcal{S}=\int d^4x\sqrt{-g}\bigg(\frac{f(R)}{2\kappa^2}-\frac{1}{2}g^{\mu\nu}\nabla_\mu\phi\nabla_\nu\phi-V(\phi)\bigg)\, ,
\end{equation}
where the functions of the model are specified as,
\begin{align}
\centering
\label{kineticaxionfunctions}
f(R)&=R+\frac{R^2}{6M^2}\, ,&V(\phi)&=m_\alpha^2f_\alpha^2\bigg(1-\cos\bigg(\frac{\phi}{f_\alpha}\bigg)\bigg)\, ,
\end{align}
with $M$ being a mass scale indicative of the energy scale beyond
which the $R^2$ term dominates, $m_\alpha$ stands for the axion
mass and finally $f_\alpha$ serves as the axion decay constant. In
this approach, it is assumed that the $f(R)$ gravity dominates
primordially which is a reasonable assumption considering that the
Ricci scalar $R=12H^2+6\dot H$ has a quite large value. In turn,
the canonical scalar field can safely be considered as subleading
and thus the background equations of motion
(\ref{Friedmannfrphix}) and (\ref{Raychaudhurifrphix}) are
equivalent to the ones obtained in the pure $f(R)$ case, that is,
\begin{equation}
\centering
\label{kineticaxionexample1motion1}
3f_{,R}H^2\simeq\frac{f_{,R}R-f}{2}-3H\dot f_{,R}\, ,
\end{equation}
\begin{equation}
\centering
\label{kineticaxionexample1motion2}
-2f_{,R}\dot H\simeq\ddot f_{,R}-H\dot f_{,R}\, .
\end{equation}
Therefore, the background equations coincide with the vacuum $R^2$
model which is known for admitting the following solution for the
Hubble rate expansion,
\begin{equation}
\centering
\label{kineticaxionexample1Hubble}
H(t)=H_I-\frac{M^2}{6}t\, .
\end{equation}
Here, it should be stated that apart from the linear dependence on
cosmic time $t$, which in turn implies an exponential expansion
(note that the condition $\ddot a>0$ is satisfied), it also
implies that the (quasi) de Sitter expansion is fully determined
by the $f(R)$ gravity and in particular the effective mass scale.
In order to be in agreement with observations, the mass scale is
considered to be approximately $M=\frac{150}{2N}10^{-5}M_P$ where
$N$ stands for the e-folding number. In addition, parameter $H_I$
is assumed to be approximately of order $\mathcal{O}(10^{13})$ GeV
so that a viable inflationary era is obtained. Regarding the
scalar field, for the kinetic axion model, it is assumed that
$\phi$, which at this stage evolves dynamically, is inferior to
the axion decay constant, that is $\phi\ll f_\alpha$. As a result,
the canonical potential can be well approximated by a quadratic
potential,
\begin{equation}
\centering
\label{kineticaxionpotential}
V(\phi)\simeq\frac{1}{2}m_\alpha^2\phi^2\, ,
\end{equation}
which, according to the latest estimates for the axion mass which
predict an upper bound $m_\alpha\leq\mathcal{O}(10^{-12})$ eV, it
can easily be inferred that the scalar potential primordially is
negligible. Note that while it would be acceptable to further
approximate the potential with the inclusion of a quartic term,
since it would describe fundamental interactions between the
axions as mentioned in the canonical scalar field case previously,
the interaction coupling in this case is given by the ratio
$\bigg(\frac{m_\alpha}{f_\alpha}\bigg)^2$ which is extremely
small, given that the axion decay constant is quite large, thus
the mass term suffices. As a result, the Klein-Gordon equation
(\ref{sscFE}) can be approximated as,
\begin{equation}
\centering
\label{kineticaxionexample1conteq}
\ddot\phi+3H\dot\phi\simeq0\, ,
\end{equation}
which in turn implies that the axion, primordially, behaves as a
stiff matter perfect fluid. This can easily be inferred from the
EoS which reads,
\begin{equation}
\centering
\label{kineticaxionexample1eos}
\omega=\frac{\dot\phi^2-2V}{\dot\phi^2+2V}\simeq1\, ,
\end{equation}
therefore the energy density scales as $\rho_\alpha\sim
a^{-6}(t)$. As mentioned before, this value for the EoS is the
maximally allowed value that is in agreement with causality and in
short states that pressure is directly proportional to the energy
density, that is $P=\rho c^2$, therefore the deceleration
parameter in this case has the largest value possible, that is
$q=2$. The reader should also keep in mind that while this is
indeed the behavior of the axion primordially, the canonical
potential becomes important subsequently when the scalar field
approaches its vacuum expectation value $\varphi=\theta_\alpha
f_\alpha$, with $\theta_\alpha$ being the misalignment angle
ranging between $0<\theta_\alpha<1$, and this in turn implies that
the axion redshifts as dark matter since
$\rho_\alpha=\frac{1}{2}\dot\phi^2+V\sim a^{-3}(t)$, making it a
perfect candidate for successfully describing cold dark matter but
for the time being, we shall focus only on the primordial
phenomenology. Since the kinetic term dominates and the effective
EoS at the end of inflation is specified by a stiff matter perfect
fluid, the inflationary era is in this case further prolonged
depending on the reheating temperature. Before we showcase this
however, an important feature should be addressed. Due to the fact
that the axion mass is approximately
$m_\alpha\sim\mathcal{O}(10^{-12})$eV, thus rendering the scalar
potential negligible, the second index is completely specified by
the continuity equation of the scalar field
(\ref{kineticaxionexample1conteq}) and thus it becomes of order
$\mathcal{O}(1)$, that is $\epsilon_2\simeq-3$. This is a prime
example as to why the indices in Eq. (\ref{slowrollindicesfrphix})
should not be regarded blindly as slow-roll indices since their
numerical value may be important. This may seem as a problem given
that a quite large value for $\epsilon_2$ accompanied by typical
values for the rest indices $\epsilon_i$ produce a scalar spectral
index which is at variance with reality, as it was shown in Refs.
\cite{Odintsov:2020sqy}, however for the case at hand, it is the
$f(R)$ part that actually saves the model given that at leading
order, $\epsilon_4=-\epsilon_1-\epsilon_2$, thus resulting in an
elegant cancellation between the large indices. As a result, for
$N=60$ and $M=1.25\times 10^{-5}M_P$, the results obtained
coincide with the vacuum $R^2$ model. The only constraint that
needs to be considered is the initial value of the scalar field
during the first horizon crossing relative to the axion decay
constant. More specifically, in order for the expansion
(\ref{kineticaxionpotential}) to be valid, one requires
$f_\alpha\gg\phi_k$. In addition, the fact that
$\epsilon_2\simeq-3$ primordially implies that during the first
horizon crossing the scalar field obeys a constant-roll condition.
This is connected to the appearance of scalar non-Gaussianities in
the CMB however due to the fact that the $R^2$ term dominates, the
usual result is obtained which implies that the equilateral
nonlinear term is of order $\mathcal{O}(10^{-2})$ so no
significant prediction is made.

Let us return to the kinetic axion dynamics and focus particularly
on the EoS. As briefly mentioned before, the dominance of the
kinetic term of the axion compared to its scalar potential implies
that the axion behaves as stiff matter which increases the
duration of inflation depending on the reheating temperature. More
specifically, in Ref. \cite{Adshead:2010mc}, it was shown that the
duration of inflation has the following form,
\begin{equation}
\centering \label{kinetixaxionexample1efolds}
N_k=56.12-\ln\bigg(\frac{k}{k_*}\bigg)+\frac{1}{3(1+\omega)}\ln\bigg(\frac{2}{3}\bigg)+\ln\bigg(\frac{\rho_k^{\frac{1}{4}}}{\rho_{end}^{\frac{1}{4}}}\bigg)+\frac{1-3\omega}{3(1+\omega)}\ln\bigg(\frac{\rho_{reh}^{\frac{1}{4}}}{\rho_{end}^{\frac{1}{4}}}\bigg)+\ln\bigg(\frac{\rho_k^{\frac{1}{4}}}{10^{16}GeV}\bigg)\,
,
\end{equation}
with $k_*=0.05$Mpc$^{-1}$ being the pivot scale and subscripts
$k$, $end$ and $reh$ are used in order to denote the moment where
inflation starts and modes become superhorizon, the ending stage
of inflation and the ending of the reheating era or in other words
the start of the radiation domination era respectively. Note that
at the end of the reheating era, the radiation fluid is assumed to
be in thermal equilibrium therefore its energy density is
specified by the relation
$\rho_{reh}=\frac{\pi^2}{30}g_*T_{reh}^4$ with $g_*$ being the
relativistic degrees of freedom at that instance hence the reason
why we stated that extension of inflation is affected by the
reheating temperature. Note that in principle the degrees of
freedom differ between different cosmological eras therefore in a
sense they also scale with temperature however for the case at
hand one can safely assume that $g_*\sim\mathcal{O}(100)$. In a
sense, the vacuum $R^2$ model dominates inflation however when it
reaches its end, the kinetic term of the axion dominates,
therefore an intermediate era of stiff matter manifests between
the inflationary era and the radiation domination era, meaning
during the reheating era. One can understand this as a result of
the stability of the de Sitter fixed point that emerges when an
autonomous dynamical analysis is performed for the power-law
scalar field assisted $R^2$ models. In short, the de Sitter fixed
point is nonhyperbolic and after a few e-folds have elapsed,
depending on the initial conditions for the dynamical parameters
used, the trajectories on the phase space are driven away however
instead of reaching the expected EoS of $\omega=\frac{1}{3}$
corresponding to radiation, the kinetic axion kicks in and results
in an intermediate stiff matter era. Now if the EoS in Eq.
(\ref{kinetixaxionexample1efolds}) is replaced as $\omega=1$ and
the reheating temperature is specified, then depending on the
value of the temperature, the e-folding number is increased by
$\mathcal{O}(1)$. The change may seem not extensive however the
impact of this change has a significant effect on the
observational quantities. We report that according to the findings
of Ref. \cite{Oikonomou:2023kqi}, a really high reheating
temperature of approximately $T_R\sim\mathcal{O}(10^{12})$GeV
results in an increase of the e-folding number by approximately
$5$ e-folds, making it equal to $N=65.3439$ and in consequence the
scalar spectral index now reads $n_\mathcal{S}=0.969393$ while the
tensor-to-scalar ratio becomes equal to $r=0.00281042$. While the
tensor-to-scalar ratio can be decreased by approximately 15$\%$,
the scalar spectral index lies outside the area of viability,
recall that $n_\mathcal{S}=0.9649\pm0.0042$ with a 68$\%$ C.L.
This implies that extremely large reheating temperatures of order
$\mathcal{O}(10^{12})$GeV and beyond should not be considered in
the kinetic axion $f(R)$ model since the following predictions are
not in agreement with observations. On the other hand, a typical
value of $T_{reh}=10^7$GeV affects mildly the results as now
$n_\mathcal{S}=0.967483$ and $r=0.00317206$. Overall, large
reheating temperatures increase the scalar spectral index of
primordial curvature perturbations while the tensor-to-scalar
ratio, and as a result the tensor spectral index, decrease.
Obviously, the opposite applies to the case of small reheating
temperatures but a lower bound for the reheating temperature
around the MeV scale should be considered since subsequent
cosmological eras cannot follow for smaller values of $T_{reh}$.

\subsection{Kinetic Axion Gauss-Bonnet Model}

Having introduced a phenomenologically interesting scalar field
assisted $f(R)$ model, it stands to reason that higher
contribution of the scalar field as presented before can affect
the inflationary era.  For the second model let us consider that
the gravitational action contains also string corrective terms and
in particular a non-minimal coupling between the scalar field and
curvature through the inclusion of the Gauss-Bonnet topological
density \cite{Oikonomou:2023kqi},
\begin{equation}
\centering
\label{kineticactionmodel2}
\mathcal{S}=\int d^4x\sqrt{-g}\bigg(\frac{f(R)}{2\kappa^2}-\frac{1}{2}g^{\mu\nu}\nabla_\mu\phi\nabla_\nu\phi-V(\phi)-\xi(\phi)\mathcal{G}\bigg)\, ,
\end{equation}
where now Eq. (\ref{kineticaxionfunctions}) is still valid, along
with the expansion (\ref{kineticaxionpotential}) and in addition,
the Gauss-Bonnet scalar coupling function is assumed to be linear,
that is $\xi(\phi)=\frac{\phi}{f}$. Of course, all the previous
statements are still valid, meaning that the $R^2$ dominates the
background equations and the scalar field affects the
observational quantities though the second and fourth indices
$\epsilon_2$ and $\epsilon_4$, not to be considered as slow-roll
indices. The model is chosen for two reason. Firstly, the linear
Gauss-Bonnet scalar coupling function is obviously the simplest
model that one can consider. Secondly, it can be shown that if the
constraint on the propagation velocity of tensor perturbations
$\ddot\xi=H\dot\xi$ is imposed on the case of linear coupling, the
constant-roll condition $\ddot\phi=H\dot\phi$ that emerges is in
fact too strong and spoils the viability of the model regardless
of the scalar potential that is chosen. In the previous example
however we showcased explicitly that while $\epsilon_2$ is quite
large, $\epsilon_4$ turns out to be proportional to $-\epsilon_2$
and thus they cancel when the scalar spectral index is computed,
leaving us only with the rest slow-roll indices which can be
considered as slow-roll given that they are approximately of order
$\mathcal{O}(10^{-3})$. In particular, without imposing any
approximations, one can show that for the $R^2$ model, index
$\epsilon_4$ reads,
\begin{equation}
\centering
\label{index4kineticaxionfr}
\epsilon_4=\epsilon_3+\bigg[1-\frac{x(1+2y)}{x(1+2y)+(\epsilon_3+y)^2}\bigg]\bigg[\frac{-\epsilon_1\epsilon_3+y(1-2\epsilon_1)}{\epsilon_3+y}-\epsilon_2-\epsilon_5\bigg]\, ,
\end{equation}
where $\epsilon_5=\frac{\epsilon_3+y(1-\epsilon_1)}{1+2y}$ and for
simplicity, the auxiliary variables that were introduced are equal
to $x=\frac{\kappa^2\dot\phi^2}{6f_{,R}H^2}$ and
$y=\frac{\kappa^2Q_a}{2f_{,R}H}$. Obviously, in the limit of $y\to
0$ and for a negligible $x$, the previous result
$\epsilon_4=-\epsilon_1-\epsilon_2$ is extracted. It should also
be stated that a viable inflationary model should predict quite
small values for the auxiliary parameters $x$ and $y$, something
that can easily be inferred from the background equations
(\ref{Friedmannfrphix})-(\ref{Raychaudhurifrphix}) since the
$f(R)$ term cannot dominate the equations of motion if $x$ and $y$
are comparable to it. In consequence, even though string
corrections are present, the condition $\epsilon_5<-\epsilon_1$
cannot be satisfied and thus the model at hand is incapable of
producing a blue tilted tensor spectral index unless string
corrections dominate over the $f(R)$ part. Tensor perturbations
are most likely to be affected solely by the $f(R)$ part, meaning
that the consistency relation $r=-8n_\mathcal{T}$ is still
satisfied exactly as in the canonical scalar field case. In
particular, one can show that the tensor-to-scalar ratio for the
constrained Gauss-Bonnet model has the following form,
\begin{equation}
\centering
\label{rkineticaxionfr}
r=16\bigg|\bigg(3\epsilon_1+2y\bigg)\frac{\epsilon_1c_\mathcal{S}^3}{1+2y}\bigg|\, ,
\end{equation}
however for the constrained Gauss-Bonnet model,
$c_\mathcal{S}\simeq c$ even in the presence of an $f(R)$ gravity
and provided that $y\ll\epsilon_1$, the vacuum $R^2$ result is
extracted. Let us now proceed with the derivation of $\dot\phi$.
According to the continuity equation of the scalar field
(\ref{conteqstring}), assuming that the canonical potential is
inferior and that the scalar field satisfies the constant-roll
condition $\ddot\phi=H\dot\phi$, one can easily see that,
\begin{equation}
\centering
\label{dotphikineticaxionfr2}
\dot\phi=-\frac{2M^3(1-\epsilon_1)(N+0.5)^{\frac{3}{2}}}{\sqrt{3}f} \, ,
\end{equation}
from which one can see that in this approach,
$\dot\phi(t_{end})=0$ when inflation stops. As shown, the overall
phenomenology is quite different compared to the previous case not
only due to the inclusion of the Gauss-Bonnet density but also due
to the constant-roll condition imposed from the propagation
velocity of tensor perturbations. Let us see the impact of this
result on the inflationary era. For typical values for the free
parameters as $M=1.25\times 10^{-5}M_P$, $N=60$ and
$f=10^{-5}M_P$, one can show that $n_\mathcal{S}=0.96686$,
$r=0.00327847$ and $n_\mathcal{T}=-0.000413283$ which are
obviously in agreement with experimental data. In addition, the
tensor spectral index and the tensor-to-scalar ratio coincide with
the vacuum $R^2$ result which in turn implies that parameters $x$
and $y$ are not so important which should be the case in order for
the $R^2$ to dominate the background equations. On the other hand,
when $x$, depending on the value of $f$, increases quite close to
$\epsilon_3^2$, then the scalar spectral index can be affected
though index $\epsilon_4$. Note also that parameter $f$ is not
necessarily the axion decay, here it was simply treated as an
additional degree of freedom however if one identifies it as the
axion decay rate, then this in turn implies that
$f_\alpha\sim\mathcal{O}(10^{13})$GeV and thus the value of the
scalar field during the first horizon crossing needs to be lesser
than this in order for the expansion of the scalar potential to be
valid.

Concerning the form of $\dot\phi$ in Eq.
(\ref{dotphikineticaxionfr2}), one can easily deduce that at the
ending stage of inflation where $\epsilon_1=1$, the kinetic term
of the axion vanishes identically. Obviously this is not exactly
true but the idea that the kinetic term now becomes inferior is
correct. To showcase this, let us include the scalar potential in
the continuity equation and go to the limit where $\epsilon_1=1$.
In this case, the time derivative of the scalar field reads,
\begin{equation}
\centering
\label{finaldotphikineticaxionfr2}
\dot\phi=-\frac{m_\alpha^2\phi}{4H}\, ,
\end{equation}
and thus, the ratio between the kinetic term and the scalar
potential at that time instance is,
\begin{equation}
\centering
\label{ratiokineticaxionfr}
\frac{\dot\phi^2}{2V}\bigg|_{end}=\frac{3}{8}\bigg(\frac{m_\alpha}{M}\bigg)^2\ll1\, ,
\end{equation}
and thus no extension of the inflationary era occurs. This is a
drastic change between the canonical case and the Gauss-Bonnet
model. It should be stated that the constraint is really powerful
since it alters the continuity equation of the scalar field from a
second order to a first order differential equation and thus the
solution $\dot\phi(\phi)$ can be found algebraically. In fact,
including additional string corrective terms does not seem to
affect the outcome since either
$c\xi(\phi)G^{\mu\nu}\nabla_\mu\phi\nabla_\nu\phi$ or
$\omega\bigg(\frac{1}{2}g^{\mu\nu}\nabla_\mu\phi\nabla_\nu\phi\bigg)^2$
will still result in the condition $\dot\phi(t_{end})=0$.
Obviously the phenomenology is completely altered if the
constraint on the propagation velocity of tensor modes is not
initially imposed.

\subsection{Kinetic Axion Chern-Simons Model}

For the final model we shall consider the kinetic axion model
studied in the previous two cases however now a parity odd term
will be included. Suppose that the gravitational action reads
\cite{Oikonomou:2023kqi},
\begin{equation}
\centering
\label{kineticaxionfrcs}
\mathcal{S}=\int d^4x\sqrt{-g}\bigg(\frac{f(R)}{2\kappa^2}-\frac{1}{2}g^{\mu\nu}\nabla_\mu\phi\nabla_\nu\phi-V(\phi)+\frac{\nu(\phi)}{8}R\tilde R\bigg)\, ,
\end{equation}
where the Chern-Simons scalar coupling function $\nu(\phi)$ is
assumed to be quadratic, that is
$\nu(\phi)=\bigg(\frac{\phi}{f}\bigg)^2$ with $f$ being an
auxiliary parameter with mass dimensions of $[f]=$eV, not to be
confused with the axion decay constant necessarily. In this
approach, phenomenologically speaking, there exist not many
differences. In fact, regarding index $\epsilon_4$, the correct
expression is extracted from Eq. (\ref{index4kineticaxionfr}) in
the limit of $y\to0$. In addition, the scalar spectral index is
expected to be the same as in the previous cases if $N=60$ is
selected. The only element that is affected non trivially is the
description for tensor perturbations and more specifically the
tensor spectral index and the tensor-to-scalar ratio.  For
simplicity, let us designate two auxiliary dimensionless
parameters as $y_l=\frac{2\lambda_l\kappa^2\dot\nu H}{F}$ and
$z=\frac{\nu''\dot\phi}{H\nu'}$. In consequence, the fifth index
(\ref{axionindices}) now reads,
\begin{equation}
\centering
\label{kineticaxionfrindex5}
\epsilon_5=\epsilon_3+\frac{1}{2}\bigg[\frac{\epsilon_2-\epsilon_1+z}{2}-\epsilon_3\bigg]\sum_{l=L,R}\frac{y_l}{1+y_l}\, ,
\end{equation}
and thus the tensor observable, which for the model at hand are
given by the following expressions,
\begin{align}
\centering
\label{kineticaxionfrcstensorobservables}
n_\mathcal{T}&=-2\frac{\epsilon_1+\epsilon_5}{1-\epsilon_1}\, ,&r&=8|\epsilon_1+\epsilon_3|\sum_{l=L,R}\frac{1}{|1+y_l|}\, .
\end{align}
Let us now see how the results can differ from the previous cases.
For the same set of values as before and for $f=10^{-9}M_P$,
$\phi_k=10^{-10}M_P$, one finds that while the scalar spectral
index and the tensor-to-scalar ratio are not affected and in fact
coincide with the vacuum $R^2$ model, the tensor spectral index
obtains the value $n_\mathcal{T}=0.0132771$ thus the Chern-Simons
model can in fact yield a blue-tilted tensor spectral index, in
contrast to the previous two cases. This is because the new degree
of freedom does not participate in the background equations and
thus a large value of $\dot\nu$ during the first horizon crossing
is possible without spoiling the dominance of the $f(R)$ gravity
at the level of the equations of motion. Furthermore, due to the
fact that the gravitational Chern-Simons term does not influence
the background equations, recall Eq. (\ref{stresstensoraxion2}),
the continuity equation of the scalar field suggests that the
apparent dominance of the kinetic term of the axion over its
scalar potential results in the appearance of a stiff matter era
during reheating, therefore the duration of the inflationary era
is prolonged depending on the reheating temperature, exactly as
was the case with the kinetic axion $f(R)$ model. In addition,
since the energy density during the first horizon crossing and at
the end of inflation is not affected by the gravitational
Chern-Simons term, the increase in the e-folds is exactly the
same. We report that for high values of the reheating temperature
as $T_{reh}=10^{12}$GeV and further, where it was shown that the
scalar spectral index is at variance with observations, the tensor
spectral index decreases in value and now reads
$n_\mathcal{T}=0.0117172$, roughly speaking a 12\% decrease while
typical values for the reheating temperature as $T_{reh}=10^7$GeV
suggest that $n_\mathcal{T}=0.0128038$,, only a 3.6\% decrease. In
short, this model proves that the tensor spectral index can easily
be manipulated by the assumption that the aforementioned parity
odd term participates in the gravitational action
(\ref{kineticaxionfrcs}) without spoiling scalar perturbations or
even the background equations, in other words the phase space has
exactly the same fixed points whether the Chern-Simons term is
considered or not, the only change lies with the behaviour of
tensor perturbations depending on their polarization state. This
is also an interesting outcome since it showcases that
scalar-field assisted $f(R)$ gravity models can in fact result in
the amplification of the energy spectrum of primordial
gravitational waves, which is an exciting result that may explain
a possible signal from a SGWB provided by LISA in the next decade
or so.

\section{Energy Spectrum Of Primordial Gravitational Waves}

In the last section of this review, we shall briefly discuss the
impact of modified theories of gravities in the energy spectrum of
primordial gravitational waves. This is because certain models
manage to predict quite different results compared to the general
relativistic description, at least in the high frequency regime.
Hence, making comparisons with a future signal detected by a third
generation detector, provided that it is attributed to a
stochastic gravitational wave background, can result in further
constraints on parameters used to quantify inflation. Even to
date, the NANOGrav 2023 detection \cite{NANOGrav:2023gor} of a
stochastic signal of primordial gravitational waves can be
attributed to cosmological sources and even to some modified
gravities with blue tilted tensor spectral index
\cite{Oikonomou:2023qfz,Vagnozzi:2023lwo} In order to showcase the
impact that modified theories of gravity have on the energy
spectrum of gravitational waves, it is preferable to firstly study
the general relativistic predictions.

Let us commence our study by considering only tensor perturbations
in the perturbed metric Eq. (\ref{perturbedmetric}), that is,
\begin{equation}
\centering
\label{perturbedmetricGW}
ds^2=a^2(\tau)(-c^2d\tau^2+(\delta_{ij}+h_{ij}(\tau,\bm x))dx^idx^j\, ,
\end{equation}
where the conformal time $\tau$ is used for simplicity and $x^i$
describes the spatial coordinates in a comoving frame. Here,
$h_{ij}$ denotes the gauge-invariant perturbed metric which
describes symmetric and traceless transverse models, meaning that
$h_{ij}$ satisfies the following conditions,
\begin{align}
\centering
\label{hijconditions}
h_{ij}&=h_{ji}\, ,&\delta^{ij}h_{ij}&=0\, ,&\partial_j h^{ij}&=0\, .
\end{align}
These conditions are implemented in order to properly describe
gravitational waves. Now for the sake of simplicity, let us work
with Einstein-Hilbert gravity such that,
\begin{equation}
\centering
\label{GRaction}
\mathcal{S}=\int d^4x\sqrt{-g}\bigg(\frac{c^4R}{16\pi G}+\mathcal{L}_{matter}\bigg)\, ,
\end{equation}
with $G$ being the gravitational constant, $c$ stands for the
speed of light and $\mathcal{L}_{matter}$ denoting the Lagrangian
density for matter. Using the perturbed metric in Eq.
(\ref{perturbedmetricGW}), the gravitational action can be written
up to second order as,
\begin{equation}
\centering
\label{perturbedaction}
\mathcal{S}=\int d\tau d^3\bm x\sqrt{-g}\bigg[-\frac{c^4g^{\mu\nu}}{64\pi G}\partial_\mu h_{ij}\partial_\nu h^{ij}+\frac{1}{2}\Pi_{ij}h^{ij}\bigg]\, ,
\end{equation}
where the anisotropic stress tensor is defined as
$\Pi^i_j=T^i_j-p\delta^i_j$ while it satisfies the traceless and
transverse conditions $\delta_{ij}\Pi^{ij}=0=\partial_j\Pi^{ij}$.
Now by varying the perturbed action (\ref{perturbedaction}) with
respect to the gauge-invariant field $h_{ij}$, one finds that,
\begin{equation}
\centering
\label{GEeq1}
h''_{ij}+2\frac{a'}{a}h_{ij}-\nabla^2h_{ij}=\frac{16\pi G}{c^4}a^2\Pi_{ij}\, ,
\end{equation}
where hereafter the prime is used in order to denote
differentiation with respect to conformal time $\tau$. Now in
order to proceed, it is convenient to work in momentum space by
performing a Fourier transformation and also treat $h_{ij}$ and
$\Pi_{ij}$ as operators describing conjugate variables. In
particular, let the Fourier expansion of the fields be,
\begin{equation}
\centering
\label{hijexpansion}
\hat h_{ij}(\tau,\bm x)=\sum_r\sqrt{\frac{16\pi G}{c^4}}\int\frac{d^3\bm k}{(2\pi)^{3}}e^r_{ij}(\bm k)\hat h^r_{\bm k}(\tau)\e^{i\bm k\cdot\bm x}\, ,
\end{equation}

\begin{equation}
\centering
\label{Pijexpansion}
\hat \Pi_{ij}(\tau,\bm x)=\sum_r\sqrt{\frac{16\pi G}{c^4}}\int\frac{d^3\bm k}{(2\pi)^{3}}e^r_{ij}(\bm k)\hat\pi^r_{\bm k}(\tau)\e^{i\bm k\cdot\bm x}\, ,
\end{equation}
where $e^r_{ij}$ is the polarization operator for tensor
perturbations describing states $+$ and $\times$. For the sake of
consistency, the polarization operator must satisfy the traceless
and transverse conditions $\delta^{ij}e_{ij}=0=k^ie_{ij}$ since
the same requirements appear in momentum space as well. Here,
$\hat h_{ij}^r(\tau,\bm x)$ is treated as an operator since
perturbations in momentum space can be interpreted as a creation
of a mode with momentum $-\bm k$ or an annihilation of a mode with
momentum $\bm k$, that is $\hat h^r_{\bm k}(\tau)=h_k(\tau)\hat
a^r_{\bm k}+h^*_k(\tau)\hat a^{r\dagger}_{-\bm k}$ in order for
$\hat h^r_{ij}$ to be a Hermitian operator. Also, in order to
quantize the theory, the following equal time commutation
relations are imposed,
\begin{align}
\centering
\label{quantization}
[\hat h^r_{\bm k}(\tau),\hat\pi_{\bm k'}^s(\tau)]&=i\delta^{rs}\delta^{(3)}(\bm k-\bm k')\, ,&[\hat h^r_{\bm k}(\tau),\hat h_{\bm k'}^s(\tau)]&=0\, , &[\hat\pi^r_{\bm k}(\tau),\hat\pi_{\bm k'}^s(\tau)]&=0\, ,
\end{align}
where $\hat\pi^r_{\bm k}(\tau)=a^2(\tau)\hat h^{r'}_{-\bm
k}(\tau)$, or using the creation/annihilation operators,
\begin{align}
\centering
\label{quantization2}
[\hat a^r_{\bm k},\hat a^{s\dagger}_{\bm k'}]&=(2\pi)^3\delta^{rs}\delta^{(3)}(\bm k-\bm k')\, ,&[\hat a^r_{\bm k},\hat a^{s}_{\bm k'}]&=0\, ,&[\hat a^{r\dagger}_{\bm k},\hat a^{s\dagger}_{\bm k'}]&=0\, .
\end{align}
Note that the factor of $(2\pi)^3$ appears in the commutation
relation between creation and annihilation operator of tensor
modes for consistency only because the Fourier transformation
considered for tensor perturbations in Eq. (\ref{hijexpansion})
has a power of $(2\pi)^3$ in the denominator. Now, in order to
derive an expression for the energy spectrum of primordial
gravitational waves, we first need information about the tensor
power spectrum. Consider the vacuum expectation value of the field
operator $\hat h_{ij}(\tau)$ which reads,
\begin{equation}
\centering \label{GWcorrelation1} \langle\hat h_{ij}(\tau,\bm
x)\hat h^{ij}(\tau,\bm
x)\rangle=\int_0^\infty\frac{dk}{k}\frac{64\pi
G}{c^4}\frac{k^3}{2\pi^2}|h_k(\tau)|^2\, ,
\end{equation}
from which the tensor power spectrum can be extracted by
considering the scale independent derivative as,
\begin{equation}
\centering \label{tensorpowerspectrum}
\Delta^2_h(k,\tau)=\frac{d\langle\hat h^2_{ij}\rangle}{d\ln k}\, .
\end{equation}
Now the energy density of gravitational waves can be extracted
from the temporal component of the energy-stress tensor, that is
$\rho_{GW}=-T^0_0$ with $T_{\mu\nu}=\bar
g_{\mu\nu}\mathcal{L}-2\frac{\delta\mathcal{L}}{\delta\bar
g^{\mu\nu}}$ being the energy stress tensor for the Lagrangian
density presented in the perturbed action (\ref{perturbedaction})
and defined by making use of the background metric. Hence, in the
end one can show that the energy spectrum in general relativity is
given by the following expression \cite{Odintsov:2022cbm},
\begin{equation}
\centering \label{Grenergyspectrum}
\Omega_{gw}(k,\tau)=\frac{1}{\rho_{crit}(\tau)}\frac{d\langle\hat\rho_{gw}(\tau)\rangle}{d\ln
k}=\frac{1}{12}\frac{k^2\Delta_h^2(k,\tau)}{H_0^2(\tau)}\, ,
\end{equation}
where $\rho_{crit}(\tau)$ is the critical energy density and $H_0$
is the current value of the Hubble rate expansion\footnote{Given
that in modern cosmology the $H_0$ tension remains unresolved, the
result is in principle affected by the exact value chosen but this
applies to the amplitude of the energy spectrum of tensor
perturbations and not on the scaling with respect to frequency.}.
Note that in the above computation, the current scale factor is
assumed to be equal to unity for simplicity. When working with the
energy spectrum of gravitational waves, one is interested in high
frequency modes since they re-enter the horizon in the early era.
Therefore, the high frequency regime in principle carries
information about the early Universe such as the reheating era,
hence the reason why the study of gravitational waves is
important. In the end, by capitalizing on the fact that the
Universe is adiabatic with the temperature scaling as $T\sim(1+z)$
and the fact that the degrees of freedom as mentioned before
change with the cosmological era, one can show that the tensor
power spectra reads \cite{Odintsov:2022cbm},
\begin{equation}
\centering
\label{Grtensorpowerspectrafinal}
\Delta^2_h(k,\tau)=\Delta^{(p)2}_h(k)\bigg(\frac{\Omega_m}{\Omega_\Lambda}\bigg)^2 \bigg(\frac{g_*(T_{in})}{g_{*0}}\bigg)\bigg(\frac{g_{*s0}}{g_{*s}(T_{in})}\bigg)^{\frac{4}{3}}\bigg(\overline{\frac{3j_1(k\tau_0)}{k\tau_0}}\bigg)^2T^2_1(x_{eq})T^2_2(x_R)\, ,
\end{equation}
where $x_A=\frac{k}{k_A}$ is the fraction between a wavenumber $k$
with a specific wavenumber $k_a$ corresponding to a cosmological
instance of interest, $j_1$ is the spherical Bessel of first
kind\footnote{Note that the average over many periods can be
approximated as $j_1(x)=\frac{1}{\sqrt{2}x}$}, $T$ stands for
transfer function for the matter-radiation equivalence and the
reheating era respectively while $\Omega_\Lambda$ is the density
parameter for dark energy. In principle, its presence does not
require the existence of a cosmological constant since modified
theories of gravity that replicate the $\Lambda$CDM results can
work without the need for $\Lambda$. We shall return to this later
when the energy spectrum for modified theories of gravity is
discussed. In addition, $\Delta^{(p)2}_h(k)$ denotes the
primordial tensor power spectrum that carries information about
the inflationary era and is in fact model dependent, hence the
reason why analyzing the gravitational wave energy spectrum at
high frequencies is important. Model dependence arises from the
fact that the primordial tensor power spectrum depends on the
tensor spectral index as,
\begin{equation}
\centering
\label{GRprimordialtensorpowerspectrum}
\Delta^{(p)2}_h(k)=\mathcal{A}_\mathcal{T}(k_*)\bigg(\frac{k}{k_*}\bigg)^{n_{\mathcal{T}}}\, ,
\end{equation}
therefore there exists a clear distinction between models which
produce a positive or negative power-spectra. Here, $k_*$ denotes
the CMB pivot scale and $\mathcal{A}_\mathcal{T}$ denotes the
amplitude of tensor perturbations which, in contrast to the
amplitude of scalar curvature perturbations
$\mathcal{A}_{\mathcal{S}}$, it has yet to be determined. A direct
measurement of the amplitude of tensor perturbations implies that
B-modes are actually measured in the CMB or in other words, both
the tensor spectral index and the tensor-to-scalar ratio are
numerically computed. In the GR limit, the amplitude of tensor
perturbations should constrain a parameter similar to how the
amplitude of scalar perturbations imposes limits on the strength
of the potential amplitude when potential driven inflationary
models are studied. Note also that the tensor spectral index
should in principle be a scale dependent object, meaning that it
changes with scale as,
\begin{equation}
\centering
\label{nTscaling}
n_\mathcal{T}(k)=n_\mathcal{T}(k_*)+\sum_{n=1}^{\infty}\frac{d^nn_{\mathcal{T}}}{d\ln^nk}\bigg|_{k_*}\frac{\ln ^n\frac{k}{k_*}}{(n+1)!}\, ,
\end{equation}
however due to the fact that a nearly scale invariant result is
expected according to the latest Planck observations, one can
focus on the first order running if not the pivot scale tensor
spectral index. In other words one can assume that,
\begin{equation}
\centering
\label{nTform}
n_\mathcal{T}(k)=n_\mathcal{T}(k_*)+\frac{a_\mathcal{T}(k_*)}{2}\ln \frac{k}{k_*}\, ,
\end{equation}
and substitute in Eq. (\ref{GRprimordialtensorpowerspectrum}). In
addition, referring to the amplitude of tensor perturbations, it
is known that the amplitude is connected to the tensor-to-scalar
ratio as,
\begin{equation}
\centering
\label{ATamplitude}
A_{\mathcal{T}}(k_*)=r\mathcal{A}_\mathcal{S}(k_*)\, ,
\end{equation}
and thus, by combing all the above expressions, one can show that
the energy spectrum for gravitational waves is
\cite{Odintsov:2022cbm},
\begin{equation}
\centering
\label{finalOmegaGW}
\Omega_{GW}=\frac{k^2}{12H_0^2}r\mathcal{A}_\mathcal{S}(k_*)\bigg(\frac{k}{k_*}\bigg)^{n_\mathcal{T}(k_*)+\frac{a_\mathcal{T}(k_*)}{2}\ln\frac{k}{k_*}}\bigg(\frac{\Omega_m}{\Omega_\Lambda}\bigg)^2\bigg(\frac{g_*(T_{in})}{g_{*0}}\bigg)\bigg(\frac{g_{*s0}}{g_{*s}(T_{in})}\bigg)^{\frac{4}{3}}\bigg(\overline{\frac{3j_1(k\tau_0)}{k\tau_0}}\bigg)^2T_1^2(x_{eq})T_2^2(x_R)\, .
\end{equation}
This result applies to general relativity and it should be stated
that due to the fact that for a canonical scalar field that
dominates during inflation the tensor spectrum is red, it can
easily be inferred that a suppression of modes is expected at high
frequencies. Let us now see how modified theories of gravity can
in principle affect the overall procedure. By following the same
reasoning as before, one needs to extract information about the
behavior of tensor perturbations. In previous sections, a detailed
analysis was provided, see for instance Eq.
(\ref{tensormodestring}) for string corrections. Overall, the main
difference between GR and modified gravity lies with the running
of the Planck mass as now, for an arbitrary modified scalar-tensor
theory, traceless and transverse modes satisfy the following
equation \cite{Odintsov:2022cbm},
\begin{equation}
\centering
\label{modifiedmodeeq}
\ddot h_{ij}+(3+a_M)H\dot h_{ij}-\frac{\nabla^2}{a^2}h_{ij}=0\, ,
\end{equation}
where $a_M=\frac{\dot Q_t}{HQ_t}$ as mentioned before is the
running Planck mass and parameter $Q_t$ is the shifted Planck mass
square and has a unique form for every model. Note that not all
modified theories of gravity affect the aforementioned mode
equation, see for instance k-essence models. In this case,
modifications simply alter numerically the tensor-to-scalar ratio
and the tensor spectral index by means of the first slow-roll
index $\epsilon_1$ as stated before. Now in order to solve this
equation, one can incorporate the WKB approximation, therefore the
solution can be written with respect to the GR solution as,
\begin{equation}
\centering
\label{hwkb}
h=\e^{-\mathcal{D}}h_{GR}\, ,
\end{equation}
where $h_{ij}=he_{ij}$ and the exponent is equal to
$\mathcal{D}=\frac{1}{2}\int_0^zdz'\frac{a_M}{1+z'}$. In the end,
since the tensor power spectrum (\ref{tensorpowerspectrum}) is
proportional to the vacuum expectation value of the contraction of
a transverse traceless mode with itself, it becomes clear that the
gravitational wave energy spectrum is affected in a similar manner
by modified theories of gravity through this exponent and in the
end, the final expression that takes into consideration the effect
of modified theories of gravity reads \cite{Odintsov:2022cbm},
\begin{equation}
\centering
\label{finalOmegaGWWKB}
\Omega_{GW}=\e^{-2\mathcal{D}}\frac{k^2}{12H_0^2}r\mathcal{P}_\zeta(k_*)\bigg(\frac{k}{k_*}\bigg)^{n_\mathcal{T}(k_*)+\frac{a_\mathcal{T}(k_*)}{2}\ln\frac{k}{k_*}}\bigg(\frac{\Omega_m}{\Omega_\Lambda}\bigg)^2\bigg(\frac{g_*(T_{in})}{g_{*0}}\bigg)\bigg(\frac{g_{*s0}}{g_{*s}(T_{in})}\bigg)^{\frac{4}{3}}\bigg(\overline{\frac{3j_1(k\tau_0)}{k\tau_0}}\bigg)^2T_1^2(x_{eq})T_2^2(x_R)\, .
\end{equation}
In principle, it should be stated that the effects of modified
theories do not lie only on the aforementioned exponent
$\mathcal{D}$ since different considerations can in principle
impose certain constraints on the reheating era for instance and
therefore the reheating temperature which participates in the
second transfer function affects the energy spectrum, see for
instance the kinetic axion Chern-Simons model. In this approach,
models with a blue spectrum are actually favored since they can
result in an apparent amplification of the energy spectrum at high
frequencies compared to the result of GR. In consequence, a
potential signal in the near future from second and third
generation detectors of gravitational waves  can be explained by
modified gravity models that predict a blue-tilted tensor spectral
index. On the other hand, if a signal is detected near the GR
prediction then this could still be explained by modified
theories, for instance power-law $f(R)$ models that a small,
insignificant enhancement is extracted. Finally, models with a
heavy red spectrum can suppress the signal thus making the search
of primordial gravitational waves more difficult. Overall,
modified theories seem to perplex the results however they stand
strong in explaining a potential signal that is attributed to a
stochastic gravitational wave background. It should also be stated
that as long as the running of the Planck mass $a_M$ for a given
model is nontrivial, then the amplitude of tensor perturbations in
the energy spectrum differs from the GR prediction, hence the
reason why the high frequency regime is studied since in this
regime the inflaton has yet to reach its vacuum expectation value.
This however does not mean that high frequencies are only
important since the case of an $f(R)$ gravity could in principle
have interesting implications in low frequencies as well.

As a final note, let us present in detailed the running of the
Planck mass for various models that were previously studied. For
the models considered in the present article, it should be stated
that only a few affect the energy spectrum of gravitational waves
and are presented in Table. \ref{planckmasses}. In order to
understand which models affect the spectrum, one needs to consider
the mode equation and the effect that terms included in the
gravitational action have on the running Planck mass. The simplest
example is the $f(R)$ gravity for which it can easily be inferred
that $a_M=\frac{\dot F}{HF}$ where $F=\frac{df}{dR}$ for
simplicity. The same expression applies to general $f(R,\phi)$
models since only the derivative with respect to the Ricci scalar
appears. In consequence, the numerator that contains information
about the time derivative gives rise to a term proportional to
$\dot\phi$ and affects the results for as long as the scalar field
evolves dynamically. In other words, when the vacuum expectation
value is reached, only the contribution from the Ricci scalar
remains, as stated previously. This description is of course model
dependent. The second model that we considered here is the
Einstein-Gauss-Bonnet gravity. As shown, it is one of the terms
that affect the propagation velocity of tensor perturbations and
in general the behavior of tensor modes. By recalling the form of
$Q_t$ from (\ref{auxiliarystringterms}) for the case of a general
$f(R,\phi)$ model, it becomes clear that $a_M$ is proportional to
$H\ddot\xi+\dot H\dot\xi$. This term, similar to the previous
case, appears when the scalar field evolves dynamically with
respect to time. It is interesting to mention that under the
assumption that tensor perturbations propagate through spacetime
with the velocity of light, the Gauss-Bonnet scalar coupling
function satisfies the differential equation $\ddot\xi=H\dot\xi$
which upon solving, it becomes clear that the contribution of the
Gauss-Bonnet term is proportional to the scale factor and thus
$\dot\xi=\frac{\lambda}{1+z}$ with $\lambda$ being an auxiliary
parameter with mass dimensions of eV. In addition, for
$\omega_{eff}=-\frac{1}{3}$, or in other words $\ddot a=0$, the
contribution of $\ddot\xi H+\dot H\dot\xi$ vanishes identically.
Hence, regardless of the coupling function, the scaling with
redshift is the same and therefore the results are in fact
universal. Now in Eq. (\ref{auxiliarystringterms}), it becomes
clear that the running of the Planck mass is affected also by the
inclusion of the kinetic coupling
$\xi(\phi)G^{\mu\nu}\nabla_\mu\phi\nabla_\nu\phi$ however now the
results are not universal, in contrast to the Gauss-Bonnet case,
as a different coupling results in a different evolution for
$\dot\phi$. Note that is the kinetic coupling is considered on its
own without the Gauss-Bonnet density, then while it is possible to
extract information about the running Planck mass and in
consequence the conditions under which a blue spectrum is
extracted, if they are not strictly model dependent, compatibility
with the GW170817 event cannot be extracted as now the description
for primordial massless gravitons requires a vanishing kinetic
coupling. Finally, the Chern-Simons model takes into consideration
the different circular polarizations as shown in Eq.
(\ref{tensormodemomentumspace}). This model is quite interesting
since for a given frequency, two signals are expected with a
different amplitude. This can be inferred from the fact that modes
satisfy a different differential equation based on their circular
polarization and therefore a difference is expected, depending on
the choice of the Chern-Simons scalar coupling function it may be
insignificant or on the contrary, quite important. This result
appears regardless of the sign of the tensor spectral index and is
indicative of the chirality that gravitational waves posses in the
Chern-Simons model since it violates parity. For this model it is
expected that for a viable inflationary era, high frequency modes
are separated in the energy spectrum based on their chirality with
the difference tending to zero as the frequency decreases. This is
similar to having the scalar field reach its vacuum expectation
value. When the minimum is reached and no dynamical evolution is
present, chirality is restored in the limit of the Chern-Simons
mass scale reaching infinity and therefore no split is expected
between modes any longer.
\begin{table}
\centering
\begin{tabular}{|c|c|}
\hline
Model&$a_M$\\ \hline
$f(R,\phi)$&$(F_{,R}\dot R+F_{,\phi}\dot\phi)/(HF)$\\ \hline
$f(R,\phi)$ Gauss-Bonnet& $(F_{,R}\dot R+F_{,\phi}\dot\phi-8\kappa^2(H\ddot\xi+\dot H\dot\xi))/(H(F-8\kappa^2\dot\xi H))$\\ \hline
$f(R,\phi)$ Constrained Gauss-Bonnet& $(F_{,R}\dot R+F_{,\phi}\dot\phi+8\kappa^2\dot\xi H^2q)/(H(F-8\kappa^2\dot\xi H))$\\ \hline
$f(R,\phi)$ String Corrections&$(F_{,R}\dot R+F_{,\phi}\dot\phi-8c_1\kappa^2\dot\xi H^2(\frac{\ddot\xi}{H\dot\xi}+\frac{\dot H}{H^2})+c_2\kappa^4\xi H\dot\phi^2(\frac{\dot\xi}{H\xi}+2\frac{\ddot\phi}{H\dot\phi}))/(H(F-8\kappa^2c_1\dot\xi H+c_2\kappa^4\xi\dot\phi^2))$ \\ \hline
$f(R,\phi)$ Constrained String Corrections&$(F_{,R}\dot R+F_{,\phi}\dot\phi-8c_1\kappa^2\dot\xi H^2+2c_2\kappa^4\xi H\dot\phi^4(1+\frac{\dot\xi}{2H\xi}+\frac{\ddot\phi}{H\dot\phi}))/(H(F-8\kappa^2c_1\dot\xi H+c_2\kappa^4\xi\dot\phi^2))$ \\ \hline
$f(R,\phi)$ Chern-Simons&$(( F_{,R}\dot R+F_{,\phi}\dot\phi)/(HF)+(2\lambda_l\kappa^2\dot\nu/F)(\frac{\ddot\nu}{H\dot\nu}-1)k/a)/(1+\frac{2\lambda_l\kappa^2\dot\nu}{F}\frac{k}{a})$\\ \hline
\end{tabular}
\label{planckmasses}
\caption{Factor $a_M$ for various modified theories of gravity. Here, $q=-1-\frac{\dot H}{H^2}$ is as usual the deceleration parameter.}
\end{table}

\section{Conclusions}

In this review we presented the most recent trends for
inflationary dynamics in terms of modified gravity theories. The
motivation for using modified gravity theories in order to
describe inflation is multi-fold since the standard single scalar
field description of inflation has many shortcomings, among which,
there must be a  large number of couplings to the standard model
particles in order for the Universe to be reheated, or if
inflation can explain the 2023 NANOGrav stochastic gravitational
background observation the tensor spectral index has to be
significantly blue tilted and so on. Apart from the inflationary
era motivation, the late-time era also must be described by
modified gravity since the possibility that the total background
equation of state parameter is slightly phantom, cannot be
described by the standard single scalar field description of
general relativity without resorting to phantom scalar fields.
Thus we provided a modern text that describes the most timely
extensions of general relativity for describing the inflationary
era. Note that the general relativistic descriptions of the
inflationary era are basically provided by single scalar field and
k-essence descriptions.  We provided an informative overview of
the inflationary paradigm in which we pointed out the shortcomings
of the standard hot Big Bang scenario, and we explained how the
inflationary paradigm may theoretically solve these problems. Also
we emphasized how important is the inflationary paradigm
theoretically and why it should eventually be the correct
description of nature regarding the early time era, since it is
the only scenario which provides a nearly scale invariant power
spectrum of primordial scalar curvature fluctuations, which are
necessary ingredients for a successful large scale matter
structures existence. We provided a detailed overview of how the
inflationary era may be generated by a single scalar field theory
with minimal and non-minimal couplings. We calculated in some
detail the necessary slow-roll indices, and we provided and proved
in some detail several well-known formulas regarding the single
scalar field inflationary theories. After we briefly discussed the
swampland criteria, and the constant-roll evolution which serves
as an alternative to the standard slow-roll evolution, we studied
and analyzed several string motivated models of inflation, which
involve Gauss-Bonnet couplings of the scalar field, higher order
derivatives of the scalar field, and some subclasses of viable
Horndeski theories. We also presented and analyzed inflation in
the context of Chern-Simons theories of gravity, including various
subcases and generalizations of string corrected modified
gravities which also contain Chern-Simons correction terms, with
the scalar field being identified with the invisible axion, which
is the most viable dark matter candidate to date. We also provided
a detailed account of vacuum $f(R)$ gravity inflation, and also
inflation in $f(R,\phi)$ and kinetic-corrected $f(R,\phi)$
theories of gravity. In the end of the review we discussed how the
gravitational waves evolve in the context of modified gravity and
we quantified the effect of modified gravity on the general
relativistic waveform, quantifying the overall effect in a single
parameter which must be evaluated for all the evolutionary eras of
our Universe.

In the next decade, several experiments will provide concrete
evidence of inflation, like the stage 4 CMB experiments in 2027
and the interferometric gravitational wave experiments like LISA
and the Einstein telescope in 2035. The smoking gun for the
existence of inflation is the actual observation of CMB B-modes
(curl modes). In such a case, the interferometric experiments will
provide hints on which model may be the correct description of the
inflationary era. Indeed, if the NANOGrav signal originates from
an inflationary era, then the theories that can describe such an
era have specific characteristics. If the NANOGrav signal is
combined with other future gravitational waves experiments, this
could be useful for determining the actual theory behind
inflation. For example a signal detectable in more than two
distinct frequency ranges, could for example indicate a flat
energy spectrum for gravitational waves, or if the signal is
detectable by some experiments, this could point out specific
characteristics of the theory that describes inflation and the
subsequent reheating era. Thus the necessity for a multi-frequency
study of gravitational waves is compelling. In addition, the
observation of scalar modes of gravitational waves will provide
direct evidence for a modified gravity theory describing nature,
although such a task is challenging due to the sensitivity of the
detectors. In all cases, the next decade belongs to the quest of
the inflationary era and the early Universe probes.

\section*{Acknowledgments}

This work was partially supported by MICINN (Spain), project
PID2019-104397GB-I00  and by the program Unidad de Excelencia
Maria de Maeztu CEX2020-001058-M, Spain (SDO). Research at
Perimeter Institute is supported in part by the Government of
Canada through the Department of Innovation, Science and Economic
Development and by the Province of Ontario through the Ministry of
Colleges and Universities (Ifigeneia Giannakoudi). This work was
supported by BCGS (Eirini Lymperiadou).

\end{document}